\documentclass[a4paper,11pt]{article}
\usepackage{jheppub} 
\usepackage[T1]{fontenc}
\usepackage{lmodern}
\usepackage{makecell}
\usepackage{graphicx} %
\usepackage{tikz}
\usepackage{float}
\usepackage{hyperref}
\usepackage{simpler-wick}
\usepackage{bm}
\usepackage{amssymb}
 \usepackage{braket}
 \usepackage{amsmath}
 \allowdisplaybreaks[4]
 \usepackage{bbding}
 \usepackage{xcolor}
 \usepackage{multirow}
 \usepackage{bbding}
 \usepackage{graphicx} %
 \usepackage{subfigure} 
   \usepackage{color}  
   \usepackage{indentfirst}
   \usepackage{ytableau}

\newcommand{\mc}[1]{\mathcal{#1}}

\title{\boldmath Positivity from J-Basis Operators in the Standard Model Effective Field Theory}

\author[c,a]{Chengjie Yang,}
\author[a,b]{Zhe Ren,}
\author[a,b,d,e,f]{Jiang-Hao Yu,}

\affiliation[a]{CAS Key Laboratory of Theoretical Physics, Institute of Theoretical Physics, Chinese Academy of Sciences, Beijing 100190, China}
\affiliation[b]{School of Physical Sciences, University of Chinese Academy of Sciences, Beijing 100049, P.\ R.\ China}
\affiliation[a]{Theory Division, Institute of High Energy Physics, Chinese Academy of Sciences, Beijing 100190, China}
\affiliation[c]{Center for High Energy Physics, Peking University, Beijing 100871, China}
\affiliation[d]{School of Fundamental Physics and Mathematical Sciences, Hangzhou Institute for Advanced
Study, UCAS, Hangzhou 310024, China}
\affiliation[e]{International Centre for Theoretical Physics Asia-Pacific, Beijing/Hangzhou, China}

\emailAdd{(corresponding authors) renzhe@itp.ac.cn}
\emailAdd{jhyu@itp.ac.cn}

\abstract{
In the effective field theory (EFT), the positivity bound on dim-8 effective operators tells us that the $s^2$ contribution in the scattering amplitude of 2-to-2 process geometrically corresponds to the convex cone composed of the ultraviolet (UV) states as the external rays. The J-Basis method can provide a complete group theory decomposition of the scattering amplitude on the direct product of the gauge group and the Lorentz group, thus to search for all UV states. Compared to previous methods, which can only perform direct product decomposition on the gauge groups, the J-Basis method greatly improves the strictness of the restrictions and also provides a systematic scheme for calculating the positivity bounds of the dim-8 operators.
}

\begin{document}
\maketitle
\flushbottom

\section{Introduction}
The Standard Model Effective Field Theory (SMEFT) framework provides a systematic approach to parameterize new physics (NP) effects at high energy by using low energy degrees of freedom.
As a non-renormalizable theory, SMEFT Lagrangian contains many operators with higher mass dimension written as
\begin{equation}
\mathcal{L}=\mathcal{L}_{\mathrm{SM}}+\sum_{n=5}^{\infty} \frac{1}{\Lambda^{(n-4)}} \sum_i C_i^{(n)} O^{( n)}\,,
\end{equation}
where $C^{(n)}$ and $O^{(n)}$ are Wilson Coefficients (WCs) and effective operators respectively of mass dimension $n$. These effective operators are written based on the standard model field building blocks, following the Lorentz and gauge symmetries \cite{Weinberg:1979sa,Buchmuller:1985jz,Grzadkowski:2010es}. They are enumerated order by order via the canonical mass dimension and form the complete and independent basis up to dimension 8 and higher in Ref.~\cite{Henning:2015alf,Lehman:2014jma,Liao:2016hru,Li:2020gnx,Murphy:2020rsh,Li:2020xlh,Liao:2020jmn,Harlander:2023psl}, with generalization to any mass dimension in Ref.~\cite{Li:2022tec, Harlander:2023ozs}.
The WCs parameterize ultraviolet (UV) information from NP theory. In the top-down approach, once the heavy states of a UV theory are integrated out, effective operators at the low energy scale can be obtained , called the matching procedure. 
Since the WCs comprise the information from UV theory, if the experimental data shows deviation from Standard Model (SM) prediction, the WCs can be determined. 

Given the null signal of NP, the WCs can only be restricted by current data or bounded theoretically. Using various processes, it is possible to restrict the WCs by experimental data via global fits    
\cite{Barklow:2017suo,Ellis:2018gqa,Durieux:2018tev,Durieux:2018ggn,Falkowski:2019hvp,Durieux:2019rbz,Hartland:2019bjb,deBlas:2019rxi,DeBlas:2019qco}. On the other hand, positivity bound was proposed \cite{Adams:2006sv,Pham:1985cr,Pennington:1994kc,Ananthanarayan:1994hf,Comellas:1995hq,Dita:1998mh} to constrain WCs based on
the unitarity, analyticity, locality properties of quantum field theory. There are many works to discuss positivity restriction of SMEFT operator coefficients. The earliest work of the positivity bound can be traced back to Ref.~\cite{Adams:2006sv}, which established a positivity bound in the forward scattering limit of 2-to-2 elastic scattering (see also \cite{Pham:1985cr,Pennington:1994kc,Ananthanarayan:1994hf,Comellas:1995hq,Dita:1998mh} for earlier discussions and applications in strong dynamics). The main idea of the elastic positivity bound is using unitary and analyticity characters to point out that 2-to-2 elastic forward scattering amplitude is non-negative. Recent literatures use the mathematical concept \textit{Arc} to give positivity bound as a semi-positive Hankel matrix filled by the WCs linked to involving effective operators at different mass dimensions
\cite{Bellazzini:2020cot,Arkani-Hamed:2020blm}.
The partial wave analysis and unitary are also used to restrict the dim-6 operators' WCs
\cite{Remmen:2020uze} and various motivation for going beyond dim-6 have been discussed in the Ref.~\cite{Liu:2016idz,Azatov:2016sqh,Ellis:2018cos,Hays:2018zze,Alioli:2020kez,Xu:2023lpq}.

Since WCs contain UV information, it's possible to enumerate possible NP particles based on effective operators, which is called the inverse problem~\cite{Berger:2007yu,Arkani-Hamed:2005qjb}. The top-down approach is a well-studied and systematized  procedure via matching and running
\cite{Fuentes-Martin:2016uol,Cohen:2020fcu,Henning:2014wua,Cepedello:2022pyx,Banerjee:2023iiv,Guedes:2023azv,Fuentes-Martin:2020udw,Fuentes-Martin:2022jrf,Chala:2021wpj,Dawson:2020oco}.
The bottom-up inverse problem \cite{Peskin123}, however, has been rarely discussed in literature. The main difficulty is that each effective operator can be mapped to infinitely many UV theories. This
case is referred to as “degeneracy”. Some articles propose to search for the possible UV states based on group representation decomposition \cite{deBlas:2017xtg,Zhang:2018shp,Li:2022abx,Li:2023cwy,Gargalionis:2020xvt,Arzt:1994gp,Adams:2006sv,Li:2023pfw,Gu:2020thj}.
The positivity can also be used to find possible UV states in the bottom-up way by combining theoretical bounds in the SMEFT and its UV states. The theoretical framework of positivity is that from a geometric perspective, the $s^2$ contribution of SMEFT amplitude exists in a silent cone formed by the external rays linked to the corresponding UV completion with different quantum numbers~\cite{Zhang:2018shp,Zhang:2020jyn,Li:2021lpe,Fuks:2020ujk}.
Thus, the whole procedure only relies on principles of quantum field theory: i.e. unitarity and UV's locality, thus the positivity framework is quite universal.

In this work, a local UV quantum field theory (QFT) is assumed in order to link the $s^2$-order contribution of the scattering amplitude to the convex geometry, and thus the positivity bound is linked to the cone space shaped by the UV particles, as discussed in Ref.~\cite{Zhang:2018shp,Zhang:2020jyn,Li:2021lpe,Chen:2023bhu,Hong:2023zgm,Trott:2020ebl}. Starting with the analyticity behavior of the forward scattering amplitude $M_{ij\rightarrow kl}(s)$ and the generalized optical theorem, the dispersion relation can be derived as
\begin{equation}
\begin{aligned}
M^{i j k l} & =\int_{(\epsilon \Lambda)^2}^{\infty} \frac{\mathrm{d} \mu \operatorname{Disc} M_{i j \rightarrow k l}(\mu)}{2 i \pi\left(\mu-\frac{1}{2} M^2\right)^3}+(j \leftrightarrow l)+c . c . \\
& =\sum_X^{\prime} \int_{(\epsilon \Lambda)^2}^{\infty} \sum_{K=R, I} \frac{\mathrm{d} \mu m_K^{X,i j} m_K^{X,k l}}{\pi\left(\mu-\frac{1}{2} M^2\right)^3}+(j \leftrightarrow l)\,.
\end{aligned}
\end{equation}
Here $i,j,k,l$ means the color and polarization of 4 outer legs while $X$ stands for the heavy states. 
By applying the convex hull theory, one shows that the silent cone which contains $s^2$ contribution of amplitude has the form $ \mathcal{C}=\operatorname{cone}\left(\left\{m^{i(j} m^{|k| l)} \mid m^{i j} \in \mathbb{R}^{n^2}\right\}\right)$ which sum over all the possible UV amplitude products $m_K^{X,i j} m_K^{X,k l}$($i,j$ means external particles while $X$ means the heavy state, $K=R,I$ means the real and the imaginary part of the amplitude), while every UV state stands for the possible external ray of the cone which provides geometric perspective on the UV physics of the SMEFT operators.

From the geometry perspective, it is essential to find a complete list of the UV states in a systematic way. In previous works
\cite{Zhang:2020jyn,Li:2021lpe,Li:2022rag,Bellazzini:2014waa},
the gauge group projectors formed by Clebsch-Gordan (CG) coefficients is utilized and UV states are enumerated to form the cone to obtain bounds for scattering processes in the SMEFT. This is called the projection method. However, this method can not guarantee finding all the possible UV states without a systematic program on the UV completion searching.
In recent work~\cite{Li:2020gnx,Li:2020xlh,Li:2022abx,Li:2022tec,Jiang:2020rwz,Li:2020zfq,Li:2023pfw}, the Pauli-Lubanski operator $\mathcal{W}^2$ and Casimir operator are introduced to decompose contact scattering amplitude to different eigenstates with specific quantum numbers. By identifying these eigenstates as the UV particles with corresponding quantum numbers, our work provides a systematic method to exhaust all the possible UV states for the effective operators in the SMEFT, which is called the J-Basis method~\cite{Li:2022abx,Li:2022tec,Li:2023pfw}.

In this work, both the convex geometry and the J-Basis method are applied in the dispersion relation to derive the positivity bounds in the SMEFT.   
After utilizing the J-Basis method to find the complete UV completion, the previous positivity bound based on the complete UV states according to the silent cones formed by external rays are updated. By comparing our results with the previous projection method \cite{Zhang:2021eeo}, we point out that the previous method of searching UV states ignores some Lorentz structures in the group decomposition, so that it exists defects. From the comparison of the results, a more complete UV completion for the specific 2-to-2 scattering process at the Lagrangian level can be obtained so that our bounds are more precise than before.

The paper is organized as follows. In Sec.~\ref{sec2}, we derive the dispersion relation for 2-to-2 forward scattering amplitude and show how to use the dispersion relation to give a geometry perspective of amplitudes. In Sec.~\ref{sec3}, we introduce relevant the Pauli-Lubanski operator for the momentum and the Casimir operator for the gauge structure. Then we show how to build a set of amplitudes representing possible UV states with definite angular momenta $J$ and gauge quantum number $R$, that is, the J-Basis method. In Sec.~\ref{sec4}, for some typical scattering processes discussed in previous works, we show our bounds by using the J-Basis method and the UV selection to search for more complete UV completion at tree level and compare ours with previous bounds to show the 
rigour of the J-Basis. 


\section{Positivity Bounds Based on External Rays}\label{sec2}
\subsection{Dispersion Relation}\label{subsec21}

\begin{figure}[htb]
    \centering
    \includegraphics[width=0.6\textwidth]{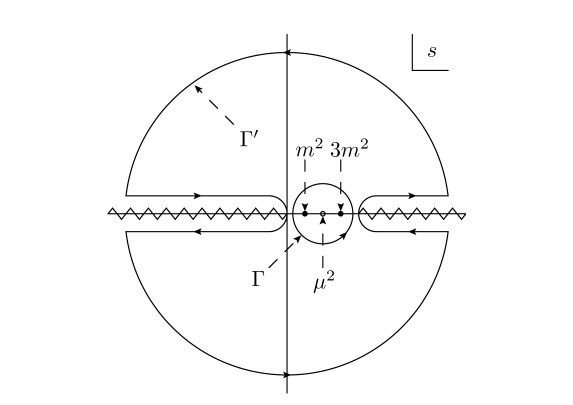}
    \caption{Diagram of the analytic structure of the forward amplitude in the complex $s$ plane in the case $m_1=m_2=m$. The simple poles at $s = m^2$ and $3m^2$ and the branch cuts starting at $s = 4m^2$ and 0 correspond to resonances and multi-particle thresholds in the $s$- and $u$-channels, respectively.} \label{fig:contour}
\end{figure}

Any 2-to-2 forward scattering amplitude $M_{ij\rightarrow kl}(s,t)$ for the full UV theory can be written as
\begin{equation}
    M_{ij\rightarrow kl}(s,t)=c_{0}+c_{2}s^2+c_{2,1}s^2 t+....+c_{n,m}s^n t^m \,.
\end{equation}
By taking derivatives on the amplitude, applying analyticity of amplitudes and considering the contour integral as shown in Fig.~\ref{fig:contour}, the dispersion relation can be obtained (e.g. Ref.~\cite{Cheung:2016yqr} by replacing $0$, $4m^2$ in the contour $\Gamma$ with $m_{-}^2$ and $m_{+}^2$) by defining  $m_{ \pm} \equiv m_1 \pm m_2$,
\begin{equation}
\begin{aligned}
c_{2}&=\left.\frac{d^2}{d s^2} M_{ij\rightarrow kl}(s, t=0)\right|_{s=\mu^2}=\frac{1}{2 \pi i} \oint_{\Gamma^{\prime}} \mathrm{d} s \frac{M_{ij\rightarrow kl}(s, 0)}{\left(s-\mu^2\right)^3}\\
&=\frac{1}{2 \pi i}\left(\int_{-\infty}^0+\int_{m_{+}^2}^{\infty}\right) \mathrm{d} s \frac{\operatorname{Disc} M_{ij\rightarrow kl}(s, 0)}{\left(s-\mu^2\right)^3}\,.
\end{aligned}
\end{equation}
Here $\operatorname{Disc} M(s, 0)=M(s+i \epsilon, 0)-M(s-i \epsilon)$. After setting $t=0$ and applying the variable replacement $u=m_{+}^2-s$, we obtain
\begin{equation}
\begin{aligned}
c_{2} & =\frac{1}{2 \pi i} \int_{m_{+}^2}^{\infty} \mathrm{d} u \frac{\operatorname{Disc} M_{ij\rightarrow kl}\left(m_{+}^2-u, 0\right)}{\left(m_{+}^2-u-\mu^2\right)^3}+\frac{1}{2 \pi i} \int_{m_{+}^2}^{\infty} \mathrm{d} s \frac{\operatorname{Disc} M_{ij\rightarrow kl}(s, 0)}{\left(s-\mu^2\right)^3} \\
& =\frac{1}{2 \pi i} \int_{m_{+}^2}^{\infty} \mathrm{d} s\left[\frac{1}{\left(s-\mu^2\right)^3}+\frac{1}{\left(s+\mu^2-m_{+}^2\right)^3}\right] \operatorname{Disc} M_{ij\rightarrow kl}(s, 0) \\
& =\frac{1}{\pi} \int_{m_{+}^2}^{\infty} \mathrm{d} s\left[\frac{1}{\left(s-\mu^2\right)^3}+\frac{1}{\left(s+\mu^2-m_{+}^2\right)^3}\right] \operatorname{Im} M_{ij\rightarrow kl}(s, 0) .
\label{eq:elasticdisperse}
\end{aligned}
\end{equation}
The above discussion is quite general: the second derivative of the low energy scattering amplitude is related to the imaginary part of the high energy scattering amplitude in the forward limit. This statement applies to both the elastic and the inelastic scatterings.

For the elastic scattering $ij\rightarrow ij$, by further applying the optical theorem, the positivity dispersion relation Eq.~\ref{posi} can be obtained,
\begin{equation}
\begin{aligned}
\textrm{Im} M_{ij \rightarrow ij}(s,0) 
&= \sum_X \int d \Pi_X |M_{ij\rightarrow X}(s,0)|^2 (2\pi)^4 \delta^4(ij-X) \\
&= [(s-\left.\left.m_{-}^2\right)\left(s-m_{+}^2\right)\right]^{1 / 2} \sigma_{\rm tot} > 0,
\end{aligned}
\label{posi}
\end{equation}
where $\sigma_{t}$ is total scattering cross section of process $ij\rightarrow X$. 
Further, taking $m_{ \pm}<\epsilon \Lambda<\Lambda$ to subtract the SM contribution, the general expression on elastic positivity bound has the form $c_2 > 0 $~\cite{Remmen:2019cyz,Remmen:2020vts}.


For the inelastic scattering $ij \rightarrow kl$, to utilize the more general optic theorem, we need to do a little more work. By adding conjugate term on $M_{ij\rightarrow kl}(s,t)$, $\mathcal{M}_{i j \rightarrow k l}\left(s=m_{+}^2 / 2\right)$ is defined as the real part of the derivative of the forward amplitude $M_{ij\rightarrow kl}(s,t)$ for scattering $ij\rightarrow kl$ process. 
By applying $M_{k l \rightarrow i j}^*(s+i \varepsilon)=M_{i j \rightarrow k l}(s-i \varepsilon)$ to connect the time reversal $i j \rightarrow k l$ and its conjugate terms, Eq.~\ref{eq:elasticdisperse} becomes 
\begin{equation}
\begin{aligned} \mathcal{M}_{i j \rightarrow k l}(\frac{m_{+}^2}{2})  \equiv & \frac{1}{2} \frac{\mathrm{d}^2}{\mathrm{~d} s^2} M_{i j \rightarrow k l}\left(s=m_{+}^2 / 2,0\right)+c . c . \\ =&\int_{m_{+}^2}^{\infty} \frac{\mathrm{d} s}{2 i \pi} \frac{\operatorname{Disc} M_{i j \rightarrow k l}(s,0)}{\left(s-\frac{m_{+}^2}{2}\right)^3}   +\int_{m_{+}^2}^{\infty} \frac{\mathrm{d} s}{2 i \pi} \frac{\operatorname{Disc} M_{i l \rightarrow k j}(s,0)}{\left(s-\frac{m_{+}^2}{2}\right)^3}+c . c .\,,
\end{aligned} \label{eq:inelasticformDisc}
\end{equation}
From the above equation, we note that in the forward limit, a twice-subtracted dispersion relation can be derived for $M_{ij\rightarrow kl}(s,t)$, assuming that a UV completion exists and is consistent with the fundamental unitary principles of the QFT.

In above Eq.~\ref{eq:inelasticformDisc}, the contributions of the kinematic poles are subtracted out \cite{deRham:2017avq,Bellazzini:2015cra,Du:2021byy,Alberte:2020jsk}. Furthermore, by assuming the $\Lambda$ is the scale of the UV theory, we can compute the amplitude in the IR to a desired accuracy
within the EFT in the energy scale within $-(\epsilon\Lambda)^2\leq s\leq (\epsilon\Lambda)^2(\epsilon\leq 1)$. Then we choose the lower limit of the integral of Eq.~\ref{eq:inelasticformDisc} turn to the value $\epsilon\Lambda$ larger than $m_{+}$, so that we can subtract out the low energy parts of the dispersion relation integrals corresponding to the EFT theory and keep the denominator of the integrands positive. Besides, the SM contribution of the Eq.~\ref{furthe} will be suppressed by inverse powers of $\epsilon \Lambda$ in Ref.~\cite{deRham:2017zjm}.

The above dispersion relation can be much simplified to
\begin{equation}
\mathcal{M}_{i j \rightarrow k l}\left((\epsilon \Lambda)^2\right)=\int_{(\epsilon \Lambda)^2}^{\infty} \frac{\mathrm{d} \mu \operatorname{Disc}M_{i j \rightarrow k l}(s)}{2 i \pi {\left(s-\frac{m_{+}^2}{2}\right)^3}}+(j \leftrightarrow l)\,.\label{furthe}
\end{equation}
This equation can be traced back to the \textit{improved positivity bounds} discussed in Ref.~\cite{Zhang:2018shp,PhysRevD.100.095003,Cheung:2016wjt}, and can also be regarded as the {\it Arc} defined in Ref.~\cite{Arkani-Hamed:2020blm,Bellazzini:2020cot}, with a radius $(\epsilon\Lambda)^2$. Now by applying the more general form of the optical theorem,
\begin{equation}
M_{i j \rightarrow k l}-\tilde{M}_{k l \rightarrow i j}^*=i \sum_X M_{i j \rightarrow X} M_{k l \rightarrow X}^*,
\end{equation}  
the dispersion relation can be written as
\begin{equation}
\mathcal{M}_{i j\rightarrow k l}(({\epsilon \Lambda})^2)=\frac{1}{2 \pi} \int_{(\epsilon \Lambda)^2}^{\infty} \frac{d s}{\left(s-\frac{m_{+}^2}{2}\right)^3} \sum_X\left[M_{i j \rightarrow X} M_{k l \rightarrow X}^*+(j \leftrightarrow l)\right]\,.
\label{eqdis}
\end{equation}

The power of analyticity is that the EFT and UV amplitude can be connected \cite{Bi:2019phv}. Considering the $s^2$ contribution corresponding to the dim-8 effective operators $\mathcal{O}_i^{(8)}$, we obtain that 
\begin{equation}
    \begin{aligned}
        &\text{EFT:}~\mathcal{L}=C_i \mathcal{O}_i^{(8)}\,,\\
        &\mathcal{M}_{i j\rightarrow k l}({\epsilon \Lambda})=\partial_s^2\mathcal{M}_{i j\rightarrow k l}^\text{EFT}(\epsilon\Lambda)=C_i\frac{\partial}{\partial s}\mathcal{O}_i^{(8)}|_{s=\epsilon\Lambda}\,
    \end{aligned}
\end{equation}
which means we establish the link between the dispersion relation of the full theory and the EFT theory to obtain the convex geometry of the EFT.

Several comments are in order. 
First, if choosing $ij=kl$, it recovers the elastic bounds.
Second, the sum in the integrand on the r.h.s. is over all the intermediate states, denoted by $X$, which might contain infinite states. Thus it provides a geometric perspective that the UV physical amplitudes $\sum_X M_{i j \rightarrow X \rightarrow k l}$ exist in a cone $\mathbf{C}$ spanned by many rays in which each ray represents contributions from UV particles $X$ with certain quantum numbers. 
Taking the shorthand notation $M_{i j \rightarrow X} \rightarrow m^{i j}$, all the UV amplitudes constitute the cone with
\begin{equation}
\begin{aligned}
\mathbf{C}^{n^4} \equiv \operatorname{cone}\left(\left\{m^{i j} m^{* k l}+m^{i \bar{l}} m^{* k \bar{j}}\right\}\right)\,.
\end{aligned}
\end{equation}
To find the boundary of the cone, it is necessary to find all the possible immediate states with certain quantum numbers. 
So the problem becomes how to find all the possible UV states for a scattering process.  


\subsection{Cone Construction}\label{subsec22}
From above, we notice that the $s^2$ contribution of the 2-to-2 amplitude should stay in the cone formed by the UV states. Now the problem becomes how to find all possible UV states: one way is the projection method by using the Irrep's (irreducible representation) projectors formed by CG coefficients and another one is the J-Basis method which is discussed in Sec.~\ref{sec3}. Here we focus on introducing the projection method and show its incompleteness in searching for UV completions.

If we don't know all possible UV states, naturally, we can use the CG coefficients to establish projectors that can expand the EFT operators \cite{Zhang:2020jyn,Bellazzini:2014waa,Remmen:2019cyz,Trott:2020ebl,Yamashita:2020gtt}, for the dim-$n$ Irrep $X$ which comes from the direct product of the two basic representation, the projectors can be written as follow,
\begin{align}
    P^{X}_{ijkl}=\sum^{n}m_n^{X,i j}m_n^{X,k l}+{j\leftrightarrow l}\,.
\end{align}
Here $m_n^{X,i j}$ is the CG coefficient where $X$ represents Irrep with different quantum number, $n$ represents the dimension of the Irrep $X$, the indices $i,j$ represent the component of the two basic representation, and $j\leftrightarrow l$ represents that the crossing symmetry~\cite{Bellazzini:2016xrt,Tolley:2020gtv} is imposed to the projectors.

Taking the $4H$ scattering as an example to show concrete steps to search for all projectors, 
The $H$ is a complex field with the $SU(2)_w$ symmetry, which can be written as  $H=(H_{2}+i H_{1},H_{4}-i H_{3})$. Thus, by considering the direct product of
\begin{equation}
\begin{split}
HHX&:\mathbf{2}\otimes\mathbf{2}=\mathbf{1}\oplus\mathbf{3}\,,\\
HHX^{\dagger}&:\mathbf{2}\otimes\mathbf{\bar{2}}=\mathbf{1}\oplus\mathbf{3}\,,\\
H^{\dagger}HX&:\mathbf{\bar{2}}\otimes\mathbf{2}=\mathbf{1}\oplus\mathbf{3}\,,\\
H^{\dagger}HX^{\dagger}&:\mathbf{\bar{2}}\otimes\mathbf{\bar{2}}=\mathbf{1}\oplus\mathbf{3}\,.\\
\end{split}
\end{equation}
Here $X$ is the heavy state while the indices of the Lorentz and the gauge group are omitted for the simplification of marking.
We can obtain the projectors listed in Table~\ref{table:4H} for expanding the $4H$ scattering amplitudes.

However, in Ref.~\cite{Murphy:2020rsh}, there are only six projectors. Once supercharge is considered in, $HHX$ and $H^{\dagger}H^{\dagger}X$'s same dimension Irreps should be merged, so the number of the UV states standing for $HHX$ and $H^{\dagger}H^{\dagger}X$ is only 2, so the number of projectors reduces to 6.
 $M^{X,n}_{k l \rightarrow X}$ is the matrix formed by CG coefficients for Irrep $X$ and its component $n$, $k$, $l$, while $i(j|k|l)$ means that crossing symmetry in QFT is imposed to the projectors.
\begin{table}[ht]
\centering
\begin{tabular}{|c|c|c|c|}
\hline Particles & Irrep($SU(2)_w$) & CG coefficients matrix & Projector \\ \hline
\multirow{2}{*}{$HH+H^{\dagger}H^{\dagger}$}& $\mathbf{1}$ & $\mathcal{C},\mathcal{C}\gamma_{4}$ & $\mathcal{C}^{i(j}\mathcal{C}^{|k|l)}+(\mathcal{C}\gamma_{4})^{i(j}(\mathcal{C}\gamma_{4})^{|k|l)}$\\ \cline{2-4}
& $\mathbf{3}$ &  $\mathcal{C}\gamma_{I},\mathcal{C}\gamma_{4}\gamma_{I}$ & $(\mathcal{C}\gamma_{I})^{i(j}(\mathcal{C}\gamma_{I})^{|k|l)}+(\mathcal{C}\gamma_{4}\gamma_{I})^{i(j}(\mathcal{C}\gamma_{4}\gamma_{I})^{|k|l)}$\\ \hline
\multirow{4}{*}{$HH^{\dagger}$ or $H^{\dagger}H$}& $\mathbf{1}_S$ & 1 & $1^{i(j}1^{|k|l)}$\\ \cline{2-4}
& $\mathbf{3}_S$ & $\gamma_{4}\gamma_{I}$ &$(\gamma_{4}\gamma_{I})^{i(j}(\gamma_{4}\gamma_{I})^{|k|l)}$\\ \cline{2-4}
& $\mathbf{1}_A$ & $\gamma_{4}$ &$\gamma_{4}^{i(j}\gamma_{4}^{|k|l)}$\\ \cline{2-4}
& $\mathbf{3}_A$ & $\gamma_{I}$ &$\gamma_{I}^{i(j}\gamma_{I}^{|k|l)}$\\ \cline{2-4}\hline
\end{tabular}
\caption{Projectors represent $4H$ scattering process.}
\label{table:4H}
\end{table}
However, by using the J-Basis method and the UV selection, nine UV states can be found in Table~\ref{table:full4H}. This shows that finding UV states by decomposing the gauge group direct product miss the spin-2 UV states in that case.

\begin{table}[ht]
\renewcommand\arraystretch{1.5}  
\centering
\begin{tabular}{|c|c|c|c|c|c|}
\hline \text { Particle } & \text { Spin } & \text {$SU(2)_w$/$U(1)_y$}  & \text { Interaction } & $\vec{c}(\text {M})$ & $\vec{c}(p) $\\
\hline $\mathcal{H}_1$ & 2 & $3_1$ & $g M^{-1} \mathcal{H}_1^{\mu \nu I \dagger}\left(\partial_\mu H^T \epsilon \tau^I \partial_\nu H\right)+h . c . $& $(3,-2,3)$ &$ (1,6,6)$ \\ \hline
$\Xi_1$ & 0 & $3_1$ & $g M \Xi_1^{I \dagger}\left(H^T \epsilon \tau^I H\right)+h . c . $& $(0,1,0)$ & $(1,0,0) $\\ \hline
$\mathcal{B}_1$ & 1 & $1_1$ & $g \mathcal{B}_1^{\mu \dagger}\left(H^T \epsilon i \stackrel{\leftrightarrow}{D_\mu} H\right)+h . c . $& $(1,0,-1) $& $(1,0,2)$ \\ \hline
$\mathcal{H}_0$ & 2 & $3_0(S)$ & $g M^{-1} \mathcal{H}_0^{\mu \nu I}\left(\partial_\mu H^{\dagger} \tau^I \partial_\nu H\right)$ & $(-7,3,8)$ & $(-4,1,-14)$ \\ \hline
$\mathcal{W}$ & 1 & $3_0(A)$ & $g \mathcal{W}_0^{\mu I}\left(H^{\dagger} \tau^I i \stackrel{\leftrightarrow}{D}_\mu H\right)$ & $(1,1,-2)$ & $(2,-1,2)$ \\ \hline
$\Xi_0$ & 0 & $3_0(S)$ & $g M \Xi_0^I\left(H^{\dagger} \tau^I H\right)$ &$ (2,0,-1) $& $(2,1,4)$ \\ \hline
$\mathcal{G}$ & 2 & $1_0(S)$ & $g M^{-1}$ $\mathcal{G}^{\mu \nu}\left(\partial_\mu H^{\dagger} \partial_\nu H\right)$ & $(3,3,-2)$ & $(6,1,6)$ \\ \hline
$\mathcal{B}_0$ & 1 & $1_0(A)$ & $g \mathcal{B}_0^\mu\left(H^{\dagger} i \stackrel{\leftrightarrow}{D}_\mu H\right)$ & $(-1,1,0)$ & $(0,-1,-2)$ \\ \hline
$\mathcal{S}$ & 0 & $1_0(S)$ & $g M \mathcal{S}\left(H^{\dagger} H\right)$ & $(0,0,1)$ & $(0,1,0)$ \\ \hline
\end{tabular}
\caption{Tree level UV completion in $4H$ scattering process. The $(A)$ and $(S)$ after $SU(2)_w/Y$ means anti-symmetry and symmetry for the amplitude $ij \rightarrow X$ under the $ij$ exchanges. In this paper, $\vec{c}(\text {M})$ is the UV-EFT matching results in the basis defined in Ref.~\cite{Murphy:2020rsh}, while $\vec{c}(p)$ is the UV-EFT matching results in the Partial Wave (P-)Basis defined in Ref.~\cite{Li:2020gnx}.}
\label{table:full4H}
\end{table}

Except the spin-2 states, the rest UV states can be checked in Ref.~\cite{Zhang:2021eeo}. Similarly, for $4W$ scattering, we obtain projectors as follow,
\begin{equation}
\begin{aligned}
P_{\alpha \beta \gamma \sigma}^1 & =\frac{1}{N} \delta_{\alpha \beta} \delta_{\gamma \sigma}, P_{\alpha \beta \gamma \sigma}^2=\frac{1}{2}\left(\delta_{\alpha \gamma} \delta_{\beta \sigma}-\delta_{\alpha \sigma} \delta_{\beta \gamma}\right) \,,\\
P_{\alpha \beta \gamma \sigma}^3 & =\frac{1}{2}\left(\delta_{\alpha \gamma} \delta_{\beta \sigma}+\delta_{\alpha \sigma} \delta_{\beta \gamma}\right)-\frac{1}{N} \delta_{\alpha \beta} \delta_{\gamma \sigma}\,.
\end{aligned}
\label{SO2}
\end{equation}

With $N=3$, these above projectors represent $SU(2)$ adjoint representation decompositions, while $N=2$ stands for decompositions in $SO(2)$ or spin space.
After imposing the crossing symmetry on these projectors, as what we did in Sec.~\ref{subsec21}. We reach the conclusion that for tree-level UV completion of $4W$ scattering, there are 9 possible UV states. However, in the tree level, we point out that the old framework of searching UV completion may cause a  mistake. By applying the UV selection analysis in the vector boson scattering (VBS) case, we find that some UV states in the tree level completion corresponding to projectors couldn't exist because their Lagrangian is zero or they are eliminated by the equation of motion (EOM), i.e. UV state corresponding to such projector doesn't exist.

Besides, the construction of the projectors for 4 fermions scattering amplitudes is a little more complicated
\cite{Li:2022rag}.
The crossing symmetry $j\leftrightarrow l$ changes to $ik\leftrightarrow\bar{k}\bar{i}$ into consideration in this case so that the projectors of the 4 fermion scattering can be written as,
\begin{equation}
\begin{aligned}
P_{i j k l}^{X}  =\frac{1}{2} \sum_\alpha\left(m_\alpha^{X,i j} m_\alpha^{X,k l}+m_\alpha^{X,i \bar{l}} m_\alpha^{X,k \bar{j}}+m_\alpha{ }^{X,\bar{k} j} m_\alpha^{X,\bar{i} l}+m_\alpha{ }^{X,\bar{k} \bar{l}} m_\alpha^{X,\bar{i}\bar{j}}\right)\,.
\end{aligned}
\end{equation}
Easily, the cone for the 4 fermions scattering can be defined as follows,
\begin{equation}
\begin{aligned}
\mathbf{C}=&\operatorname{cone}\left(\left\{m_\alpha^{X,i j} m_\alpha^{X,k l}+m_\alpha^{X,i \bar{l}} m_\alpha^{X,k \bar{j}}+m_\alpha{ }^{X,\bar{k} j} m_\alpha^{X \bar{i} l}+m_\alpha{ }^{X,\bar{k} \bar{l}} m_\alpha^{X,\bar{i}\bar{j}}+ (i \leftrightarrow j, k \leftrightarrow l) \mid m \in \mathbb{C}^{2 n \times 2 n}\right\}\right)\,.
\end{aligned}
\end{equation}.
\subsection{Cone Calculation and Obtaining Bounds}
Now we know how to construct projectors which represents UV states. Then the projectors can used to expand corresponding EFT amplitudes, and we can calculate positivity bounds.
First, we need to determine the dimension of projectors, then choose a set of basis $B^{Y}_{ijkl}$ to expand projectors and EFT amplitudes to acquire a group of vectors $\left\{c_{XY} \right\}$ for different UV states $X$ in the basis space by applying Eq.~\ref{expand}.
\begin{equation}
    P^{X}_{ijkl}=c_{XY}B^{Y}_{ijkl}\,.
     \label{expand}
    \end{equation}
For example, in Table~\ref{table:full4H}, the corresponding P-Basis EFT operators $O_{n,ijkl}$ is chosen as the basis $B^{Y}_{ijkl}$ to obtain the $\left\{\vec{c}(p) \right\}$.
 \begin{equation}
  O_{n,ijkl}=c_{nY}B^{Y}_{ijkl}\,.
    \label{expand0}
    \end{equation}
    If other $B^{Y}_{ijkl}$ rather than the operators $O_{n,ijkl}$ are chosen as basis, these can be linked according to the basis transformation relationship Eq.~\ref{expand0}.
    \begin{equation}
   M_{ijkl}=C_{n}O_{n,ijkl}=C_{n}c_{nY}B^{Y}_{ijkl}\,.
    \label{expand1}
    \end{equation}
Then, the amplitude $M_{ijkl}$ is expanded by applying Eq.~\ref{expand1} to obtain the corresponding vector $C_{n}c_{nY} $.
    \begin{equation}
    N^{m}_{Y}\cdot (C_{n}c_{nY})\geq 0\,,
    \label{expand2}
    \end{equation}
where $C_{n}c_{nY}$ is the $\vec{c}$ while $N^{m}_{Y}$ is the $\vec{n}$ that we need to search for.
 Finally we obtain the cone spanned by a set of vectors $\left\{c_{XY} \right\}$ representing UV states in the $B^{Y}_{ijkl}$ space while the EFT amplitude represented by the vector $C_{n}c_{nY}$ exists in the inner of the cone. According to the character of the cone mathematically, for any vectors $\vec{c}$ in the cone, the dot between $\vec{c}$ and every normal vectors $\vec{n}$ corresponding to the faces of the cone is larger than $0$. 

Since vectors representing UV states form the cone, naturally, we can search for faces (dim $n-1$) of the cone to describe the cone. The unique feature of a face is its normal vector. In fact, if we choose the inward direction as the positive direction for normal vectors, the dot product of every normal vector and any vectors in the cone is always positive. This is essentially the positivity bound which we search for.
For a simple linear cone, once we acquire $\vec{c}$, it’s easy to obtain all normal vectors of the cone by using the specialized mathematical calculation program like \texttt{polymake} \cite{Assarf2014ComputingCH}. 

In conclusion, we know every facet of the cone can be characterized by its normal vectors. For specific 2-2 forward scattering with determined particle types, by using group decomposition to search all projectors forming the cone which contains EFT amplitudes, then we can find all subsets $A_{i}(\vec{c})$ satisfy $Length\left(A_{i}(\vec{c})\right)=n-1$ and $Rank\left(A_{i}(\vec{c})\right)=n-1$. The collection of $A_{i}(\vec{c})$ must contain all faces of the cone, equally, we can calculate the normal vector $n_{i}(\vec{c})$ for every $A_{i}(\vec{c})$ to select $n_{i}(\vec{c})$ satisfied Eq.~\ref{expand2} to obtain positivity bound.

In the Sec.~\ref{subsec42}, we give detail calculation for the bounds of the operators involved in the $4H$ scattering by the steps introduced above.
In some more complicated cases, for the 2-to-2 scattering involving $W$ and $B$ in Sec.~\ref{sec4}, intermediate states coupling with different external particles may have a degeneracy relationship measured by a parameter $x$ like $WWX+xBBX$ where $X$ is UV state, the Lorentz and the gauge indices are omitted for the convenience of marking. It means the cone has curved surfaces parameterized by $x$. Similarly, the normal vectors corresponding to the surfaces are also parameterized by $x$. Finally, by solving positive value conditions for these multivariate quadratic polynomials, the positivity bounds with roots can be obtained.

\section{J-Basis Theory Framework}\label{sec3}
\subsection{Poincare Casimir and Partial Wave Basis}


For the Lorentz structures, we briefly introduce the Poincare Casimir operator which has been elaborated in Ref.~\cite{Li:2022abx,Li:2022tec,Jiang:2020rwz,Li:2020zfq,Li:2023pfw}. When the Poincare Casimir operator $\mathcal{W}^2$ acts on an eigenstate of spin $J$ and momentum $P$, we obtain the following equation,
\begin{equation}
    \mathcal{W}^2|P, J, \sigma\rangle=-P^2 J(J+1)|P, J, \sigma\rangle,
\end{equation}
where $\mathcal{W}^{\mu}$ is the Pauli-Lubanski operator. 


Our framework was established in spinor notation. The specific $\mathcal{W}^2$ form is introduced in Ref.~\cite{Li:2022abx},
\begin{equation}
\mathcal{W}^2=\frac{1}{8} P^2\left(\operatorname{Tr}\left[M^2\right]+\operatorname{Tr}\left[\widetilde{M}^2\right]\right)-\frac{1}{4} \operatorname{Tr}\left[P^{T} M P \widetilde{M}\right]\,.
\end{equation}
Here $P=P_\mu \sigma_{\alpha \dot{\alpha}}^\mu, P^{T}=P_\mu \bar{\sigma}^{\mu \dot{\alpha} \alpha}$ and $M,\widetilde{M}$ are chiral components of the Lorentz generator $M_{\mu \nu} \sigma_{\alpha \dot{\alpha}}^\mu \sigma_{\beta \dot{\beta}}^\nu=\epsilon_{\alpha \beta} \widetilde{M}_{\dot{\alpha} \dot{\beta}}+\tilde{\epsilon}_{\dot{\alpha} \dot{\beta}} M_{\alpha \beta}$. More specifically,

\begin{equation}
\begin{aligned}
M_{\alpha \beta}= & i \sum_{i=1}^N\left(\lambda_{i \alpha} \frac{\partial}{\partial \lambda_i^\beta}+\lambda_{i \beta} \frac{\partial}{\partial \lambda_i^\alpha}\right), \quad \widetilde{M}_{\dot{\alpha} \dot{\beta}}=i \sum_{i=1}^N\left(\tilde{\lambda}_{i \dot{\alpha}} \frac{\partial}{\partial \tilde{\lambda}_i^{\dot{\beta}}}+\tilde{\lambda}_{i \dot{\beta}} \frac{\partial}{\partial \tilde{\lambda}_i^{\dot{\alpha}}}\right), \\
\left(M^2\right)_\alpha^\beta \equiv & M_\alpha^\gamma M_\gamma^\beta=-\sum_i\left(3 \lambda_{i \alpha} \frac{\partial}{\partial \lambda_{i \beta}}+\delta_\alpha^\beta \lambda_i^\gamma \frac{\partial}{\partial \lambda_i^\gamma}\right) \\
& -\sum_{i, j}\left(\lambda_{i \alpha} \lambda_{j \gamma} \frac{\partial}{\partial \lambda_{i \gamma}} \frac{\partial}{\partial \lambda_{j \beta}}-\langle i j\rangle \frac{\partial}{\partial \lambda_i^\alpha} \frac{\partial}{\partial \lambda_{j \beta}}-\lambda_{i \alpha} \lambda_j^\beta \frac{\partial}{\partial \lambda_{i \gamma}} \frac{\partial}{\partial \lambda_j^\gamma}+\lambda_i^\gamma \lambda_j^\beta \frac{\partial}{\partial \lambda_i^\alpha} \frac{\partial}{\partial \lambda_j^\gamma}\right), \\
\left(\widetilde{M}^2\right)_{\dot{\beta}}^{\dot{\alpha}} \equiv & \widetilde{M}_{\dot{\gamma}}^{\dot{\alpha}} \widetilde{M}_{\dot{\beta}}^{\dot{\gamma}}=-\sum_i\left(3 \tilde{\lambda}^{i \dot{\alpha}} \frac{\partial}{\partial \tilde{\lambda}^{i \dot{\beta}}}+\delta_{\dot{\beta}}^{\dot{\alpha}} \tilde{\lambda}_{i \dot{\gamma}} \frac{\partial}{\partial \tilde{\lambda}_{i \dot{\gamma}}}\right) \\
& -\sum_{i, j}\left(\tilde{\lambda}_i^{\dot{\alpha}} \tilde{\lambda}_j^{\dot{\gamma}} \frac{\partial}{\partial \tilde{\lambda}_i^{\dot{\gamma}}} \frac{\partial}{\partial \tilde{\lambda}_j^{\dot{\beta}}}-[i j] \frac{\partial}{\partial \tilde{\lambda}_{i \dot{\alpha}}} \frac{\partial}{\partial \tilde{\lambda}_j^{\dot{\beta}}}-\tilde{\lambda}_i^{\dot{\alpha}} \tilde{\lambda}_{j \dot{\beta}} \frac{\partial}{\partial \tilde{\lambda}_i^{\dot{\gamma}}} \frac{\partial}{\partial \tilde{\lambda}_{j \dot{\gamma}}}+\tilde{\lambda}_{i \dot{\gamma}} \tilde{\lambda}_{j \dot{\beta}} \frac{\partial}{\partial \tilde{\lambda}_{i \dot{\alpha}}} \frac{\partial}{\partial \tilde{\lambda}_{j \dot{\gamma}}}\right) .
\end{aligned}
\end{equation}

Now we consider how the $\mathcal{W}^2$ acts on the scattering amplitude. When the $\mathcal{W}^2_\mathcal{I}$ acts on a process $\mathcal{I} \rightarrow \mathcal{I}^{\prime}$, we obtain
\begin{equation}
\begin{aligned}
\mathcal{W}_{\mathcal{I}}^2 \mathcal{M} & \equiv \sum_J \mathcal{M}_{a b}^J\left(s_{\mathcal{I}}\right)\left[\mathcal{W}^2 \mathcal{C}_a^J(\mathcal{I})\right] \cdot \mathcal{C}_b^J\left(\mathcal{I}^{\prime}\right) \\
& =\sum_J \mathcal{M}_{a b}^J\left(s_{\mathcal{I}}\right)\left[-s_{\mathcal{I}} J(J+1) \mathcal{C}_a^J(\mathcal{I})\right] \cdot \mathcal{C}_b^J\left(\mathcal{I}^{\prime}\right) \\
& =-\sum_J J(J+1) \times s_{\mathcal{I}} \mathcal{M}_{a b}^J\left(s_{\mathcal{I}}\right) \overline{\mathcal{B}}_{a b}^J\left(\mathcal{I} \mid \mathcal{I}^{\prime}\right)\,,
\end{aligned}
\end{equation}
where $\mathcal{C}_{N}^{J}$ is the C-G coefficient corresponding to the intermediate state of $N$ particles with total angular $\mathcal{J}$, and $s_{\mathcal{I}}=(\sum_{i \in \mathcal{I}} p_i)^2$ is the Mandelstam variable in the scattering channel.

\subsection{Gauge Eigen-basis and \texorpdfstring{$SU(N)$}~~Casimir}
In the previous subsection, we introduced how to construct the partial wave basis by using Poincare Casimir operators. Moreover, the decomposition of the gauge structure need to be considered in.

In fact, the projection framework 
\cite{Bellazzini:2014waa}
enumerate possible UV states by CGC. They wrote projectors $P_{\mathcal{I}\rightarrow \mathcal{I}^{\prime}}$ to expand amplitudes $\mathcal{W}_{\mathcal{I}\rightarrow \mathcal{I}^{\prime}}$. It equals to search all Invariant Subspaces of the direct product of gauge groups.
Despite having a similar principle, we introduce a more systematic tool: the $SU(N)$ Casimirs from
\cite{Li:2020gnx,Li:2020xlh,Li:2022abx,Li:2022tec}. 
First we introduce the $SU(2)$ and $SU(3)$ Casmirs as\\
\begin{equation}
\begin{aligned}
& \mathbb{C}_2=\mathbb{T}^a \mathbb{T}^a, \text { for both } S U(2) \text { and } S U(3), \\
& \mathbb{C}_3=d^{a b c} \mathbb{T}^a \mathbb{T}^b \mathbb{T}^c, \text { for } S U(3) \text { only. }
\end{aligned}
\end{equation}

In positivity, we consider the multi-states for the external particles, accordingly, we should write $T$ for the direct product representations as
\begin{equation}
\mathbb{T}_{\otimes\left\{\mathbf{r}_i\right\}}^A=\sum_{i=1}^N E_{\mathbf{r}_1} \times E_{\mathbf{r}_2} \times \cdots \times T_{\mathbf{r}_i}^A \times \ldots E_{\mathbf{r}_N}\,,
\end{equation}
with $\mathbb{T}_{\otimes\left\{\mathbf{r}_i\right\}}$ and $E_{\mathbf{r}_i}$ being the generator and identity matrix for different Irreps. The acting of $\mathbb{T}$ on a state $\Theta_{I_1 I_2 \ldots I_N}$ can be written as
\begin{equation}
\Theta_{I_1 I_2 \ldots I_N}^{\prime}=\mathbb{T}_{\otimes\left\{\mathbf{r}_i\right\}}^a \circ \Theta_{I_1 I_2 \ldots I_N} \equiv \sum_{i=1}^N\left(T_{\mathbf{r}_i}^a\right)_{I_i}^Z \Theta_{I_1 \ldots I_{i-1} Z I_{i+1} I_N}\,.
\end{equation}
Let's take the scattering of $\pi \pi$ as an example. Noticing that $\pi$ with the generator $T^{A}_{IJ}=i\epsilon_{AIJ}$~isn't basic representation of $SU(2)$ group
and considering the decomposition of $T^{m}_{\left \{12\right\}}$, firstly, we can find all independent color tensors as \begin{equation}
\mathcal{T}_1^m=\delta_{I_1 I_3} \delta_{I_2 I_4}, \mathcal{T}_2^m=\delta_{I_1 I_2} \delta_{I_3 I_4}, \mathcal{T}_3^m=\delta_{I_1 I_4} \delta_{I_2 I_3}\,.
\end{equation}
By applying properties of the \textbf{Levi-Civita} symbol $ \varepsilon_{i j k} \varepsilon^{i m n}=\delta_j{ }^m \delta_k{ }^n-\delta_j{ }^n \delta_k{ }^m  \varepsilon_{j m n} \varepsilon^{i m n}=2 \delta_j{ }^i$, we obtain\\
\begin{equation}
    \begin{aligned}
      \left(T^A T^A\right)_{\left\{1 2\right\}} \delta_{I_1 I_3} \delta_{I_2 I_4}&=\left(T^A\right)_{\left\{12\right\}}\left(T_{I_1 l_3}^A \delta_{I_2 I_4}+T_{I_2 I_4}^A \delta_{L_1 I_3}\right) \\
 &=-4 \delta_{I_1 I_3}\delta_{I_2 I_4}+2 \delta_{I_{1} I_2} \delta_{I_3 I_4}-2 \delta_{L_1 I_4} \delta_{I_2 I_3} \,,\\
\\
    \left(T^A T^A\right)_{\{12\}} \delta_{I_1 I_2} \delta_{I_3 I_4}&=T_{I_1 z}^A \delta_{z I_2} \delta_{I_{2} I_4}+T_{I_2 z}^A \delta_{I_1 z} \delta_{I_3 I_4}=T_{I_1 I_2}^A \delta_{I_3 I_4} + T_{I_2 I_1}^A \delta_{I_3 L_4}=0\,,\\
\\
 \left(T^A T^A\right)_{\{12\}} \delta_{I_1 I_4} \delta_{I_2 I_3}&=\left(T^A\right)_{\{12\}}\left(T_{I_1 I_4}^A \delta_{I_2 I_3}+\delta_{I_1 I_4} T_{I_2 I_3}^A\right)\\
& =2 \delta_{I_1 I_2} \delta_{I_3 I_4}-2 \delta_{I_1 I_3} \delta_{I_2 I_4}+4 \delta_{I_1 I_4} \delta_{I_2 I_3}\,,
    \end{aligned}
\end{equation}
i.e.
\begin{equation}
\underset{\{12\}}{\mathbb{C}_2} \circ \mathcal{T}_i^m=\underset{\{12\}}{\left(C_2^{\mathrm{T}}\right)_{i j} \mathcal{T}_j^m}=\left(\begin{array}{ccc}
4 & -2 & 2 \\
0 & 0 & 0 \\
-2 & -2 & 4
\end{array}\right)\left(\begin{array}{c}
\mathcal{T}_1^m \\
\mathcal{T}_2^m \\
\mathcal{T}_3^m
\end{array}\right)\,.
\end{equation}

After diagonalization, three eigenstates in the m-Basis can be obtained as follow,
\begin{equation}
\begin{aligned}
& \mathcal{T}(\mathbf{3})_{I_1 I_2 I_3 I_4}=-3 \mathcal{T}_1^m+2 \mathcal{T}_2^m-3 \mathcal{T}_3^m=-3 \delta_{I_1 I_3} \delta_{I_2 I_4}+2 \delta_{I_1 I_2} \delta_{I_3 I_4}-3 \delta_{I_1 I_4} \delta_{I_2 I_3} \,,\\
& \mathcal{T}(\mathbf{2})_{I_1 I_2 I_3 I_4}=\mathcal{T}_1^m-\mathcal{T}_3^m=\delta_{I_1 I_3} \delta_{I_2 I_4}-\delta_{I_1 I_4} \delta_{I_2 I_3} \,,\\
& \mathcal{T}(\mathbf{1})_{I_1 I_2 I_3 I_4}=\mathcal{T}_2^m=\delta_{I_1 I_2} \delta_{I_3 I_4}\,.
\end{aligned}
\end{equation}

\subsection{ Lorentz Eigen-Basis Construction}
Now we show $\mathcal{W}^2$ is appropriate to construct the Lorentz Eigen-Basis with angular momentum decompositions.

\subsubsection{Amplitude Operator Correspondence}

First, according to the spinor notation, the relationship between the spinor block and the operator block are obtained~\cite{Shadmi:2018xan,Ma:2019gtx,Li:2020gnx,Li:2020xlh,Arkani-Hamed:2017jhn} as follows,
\begin{table}[H]
\centering
\begin{tabular}{|l|c|c|c|c|c|}
\hline Amplitude Blocks & $\lambda_i^n \tilde{\lambda}_i^n$ & $\lambda_i^{n+1} \tilde{\lambda}_i^n$ & $\lambda_i^n \tilde{\lambda}_i^{n+1}$ & $\lambda_i^{n+2} \tilde{\lambda}_i^n$ & $\lambda_i^n \tilde{\lambda}_i^{n+2}$ \\
\hline Operator Blocks & $D^n \phi_i$ & $D^n \psi_i$ & $D^n \psi_i^{\dagger}$ & $D^n F_{L i}$ & $D^n F_{R i}$ \\
\hline
\end{tabular}
\end{table}
Taking the amplitude $\langle 12\rangle[23][24] s_{14}$ as an example:
\begin{equation}
\begin{aligned}
\langle 12\rangle[23][24] s_{14} & =\left(\lambda_1^2 \tilde{\lambda}_1\right)_{\dot{\alpha}}^{\alpha \beta}\left(\lambda_2 \tilde{\lambda}_2^2\right)_{\alpha \dot{\beta} \dot{\gamma}} \tilde{\lambda}_3^{\dot{\beta}}\left(\lambda_4 \tilde{\lambda}_4^2\right)_\beta^{\dot{\alpha} \dot{\gamma}} \\
& \stackrel{\text { amplop }}{\longrightarrow}\left(D \psi_1\right)_{\dot{\alpha}}^{\alpha \beta}\left(D \psi_2^{\dagger}\right)_{\alpha \dot{\beta} \dot{\gamma}}\left(\psi_3^{\dagger}\right)^{\dot{\beta}}\left(D \psi_4^{\dagger}\right)_\beta^{\dot{\alpha} \dot{\gamma}} \\
& \sim\left(D_\mu \psi_1\right)^\beta\left(D_\nu \psi_2^{\dagger}\right)_{\dot{\beta}}\left(\psi_3^{\dagger}\right)^{\dot{\beta}}\left(D_\rho \psi_4^{\dagger}\right)^{\dot{\alpha}}\left(\sigma^\rho \bar{\sigma}^\mu \sigma^\nu\right)_{\beta \dot{\alpha}} \\
& =\left(D_\mu \psi_1\left(g^{\rho \mu} \sigma^\nu-i \epsilon^{\mu \nu \rho \lambda} \sigma_\lambda\right) D_\rho \psi_4^{\dagger}\right)\left(D_\nu \psi_2^{\dagger} \psi_3^{\dagger}\right)+\text { EOM. }
\end{aligned}
\end{equation}

From the above, we find that the spinor notation may not be equal to operator monomials, and different operator form choices are related by the EOMs. Anyway, we see the possibility of constructing local operators through polynomials of amplitudes in the spinor notation.

According to Ref.~\cite{Li:2022abx}, multiplied by the particular Mandelstam variables in the scattering channel doesn't alter the angular momentum of the scattering states, so we can get a general form of operators corresponding to different angular momentum.
\begin{equation}
\mathcal{B}^{J, d=N+2 k}=s^k \overline{\mathcal{B}}^J \sim \mathcal{O}^{J, d}\,.
\end{equation}

Now let's take a process $\psi_1^{\dagger}, \psi_2 \rightarrow \psi_3^{\dagger}, \psi_4$ as an example:
\begin{equation}
\mathcal{B}^{J=1, d=4+2 k} \sim s^{k-2}[12]\langle 23\rangle^2[34]\,.
\end{equation}
It corresponds to operator form:
\begin{equation}
\mathcal{O}^{J=1, d=4+2 k} \sim\left(D_{\rho_1, \ldots, \rho_{k-2}}^{k-2} \psi_1^{\dagger} \bar{\sigma}_\mu D_\nu \psi_3\right)\left(D^{k-2, \rho_1, \ldots, \rho_{k-2}} D^\mu \psi_2 \sigma^\nu \psi_4^{\dagger}\right) .
\end{equation}

 \subsubsection{Poincare Casimir and Lorentz Eigen-Basis}
 
 Now we have introduced the correspondence between amplitudes and operators. Naturally, for operators of a specific category, finding its complete spinor amplitude basis to construct eigenstates for $\mathcal{W}^2$ is what we discuss in this section. 
 
\begin{figure}
    \centering
    \includegraphics[width=0.6\textwidth]{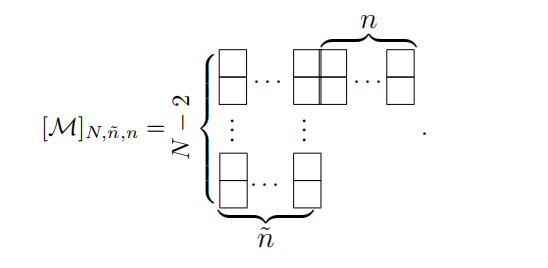}
    \caption{Young Tableau for amplitude-basis corresponding.}
    \label{fig:Young}
\end{figure}

 In Ref.~\cite{Li:2022tec,Li:2022abx,Li:2021tsq,Li:2020zfq,Li:2020tsi}, a complete basis of local amplitudes and the corresponding operators are defined as the Young Tableau (Y-)Basis. The name comes from the construction based on a Young-Tableau of the $SU(N)$ group~\cite{Henning:2019enq, Henning:2019mcv,Li:2020gnx,Li:2020xlh}, where $N$ is the number of particles involved in the amplitude.
 For the type of operators we are interested in, we define the relevant parameters of Young-Tableau as
 \begin{equation}
n=\frac{k}{2}+\sum_{h_i<0} h_i, \quad \tilde{n}=\frac{k}{2}+\sum_{h_i>0} h_i\,.
\end{equation}
Here $k$ is the number of derivatives in operator type, while  $h_i$ is the helicity of particle $i$.
The above parameters give such a Young Tableau in Fig.~\ref{fig:Young}.
Next, we just need to fill labels $1$ to $N$ into Young Tableau to acquire the basis represented by a specific Young Diagram, while the number of each label (particle $i$) is given by $\#i = \tilde{n}-2h_{i}$ for the particular class of scattering state to satisfy it's a Semi-Standard Young Tableau (SSYT): in each row the labels are non-decreasing from left to right; in each column, the labels are increasing from top to bottom.

 For example, once we consider dim-8 $4H$ operators, the Young diagram's rows and columns are equal to $2*4$  with every label $i's$ number is two. The number of its SSYT is three. After considering the gauge tensor, we can get a Y-Basis as follows,
 \begin{equation}
     \begin{aligned}  & \mathcal{M}_{D^4 H^2 H^{+2}, 1}^{(y)}=\delta_k^i \delta_l^j\langle 34\rangle^2[34]^2, \\ & \mathcal{M}_{D^4 H^2 H^{\dagger 2}, 2}^{(y)}=\delta_k^i \delta_l^j\langle 24\rangle^2[24]^2, \\ & \mathcal{M}_{D^4 H^2 H^{\dagger} 2,3}^{(y)}=-\delta_k^i \delta_l^j\langle 24\rangle\langle 34\rangle[24][34], \\ & \mathcal{M}_{D^4 H^2 H^{\dagger 2}, 4}^{(y)}=\delta_l^i \delta_k^j\langle 34\rangle^2[34]^2, \\ & \mathcal{M}_{D^4 H^2 H^{\dagger 2}, 5}^{(y)}=\delta_l^i \delta_k^j\langle 24\rangle^2[24]^2, \\ & \mathcal{M}_{D^4 H^2 H^{+2}, 6}^{(y)}=-\delta_l^i \delta_k^j\langle 24\rangle\langle 34\rangle[24][34] . \\ & \end{aligned}
 \end{equation}
 
Finally, we claim that all other bases can be reduced into the Y-Basis through the Schouten identity, the momentum conservation, and the on-shell conditions. In fact, this is a simplified approach to searching for the amplitude basis in the spinor notation.
\subsection{Gauge J-Basis from Gauge Casimir}
The correspondence of the gauge structures between operators and amplitudes is simple. The invariant tensors of group factors in the amplitudes exactly correspond to the invariant tensors that are used to contract the fields in operators to form gauge singlets.

The gauge factors were not considered in the last section, so that the Y-Basis may become polynomials when it's acted by Casimir operators. However, a complete and independent monomial basis called the gauge m-Basis, can always be calculated from these polynomials by linear transformations. An efficient algorithm to find the gauge m-Basis has been proposed in 
\cite{Li:2022tec}.

We can achieve it in two steps: First, we need to determine the Young Tableaux of the particle we consider, then use the Littlewood-Richardson (L-R) rule to find all direct products expressed by the group structure constants. Finally, we use the gauge Casimir operator to find its all invariant subspaces, just as we did in section 3.2.

\section{The updated External Ray Positivity Bounds}\label{sec4}

In this section, we consider the UV completion in the tree level for the external ray positivity bound by J-basis method. Not only the J-basis can obtain the UV states but also any decompositions of the amplitudes with specific angular momentum $\mathcal{J}$ and quantum number. Hence, the J-basis method can be applied to analyze the positivity bound for the loop-level scattering amplitudes, which can be decomposed into several angular momentum combinations.

\begin{figure}[ht]
    \centering
    \includegraphics[width=0.9\linewidth]{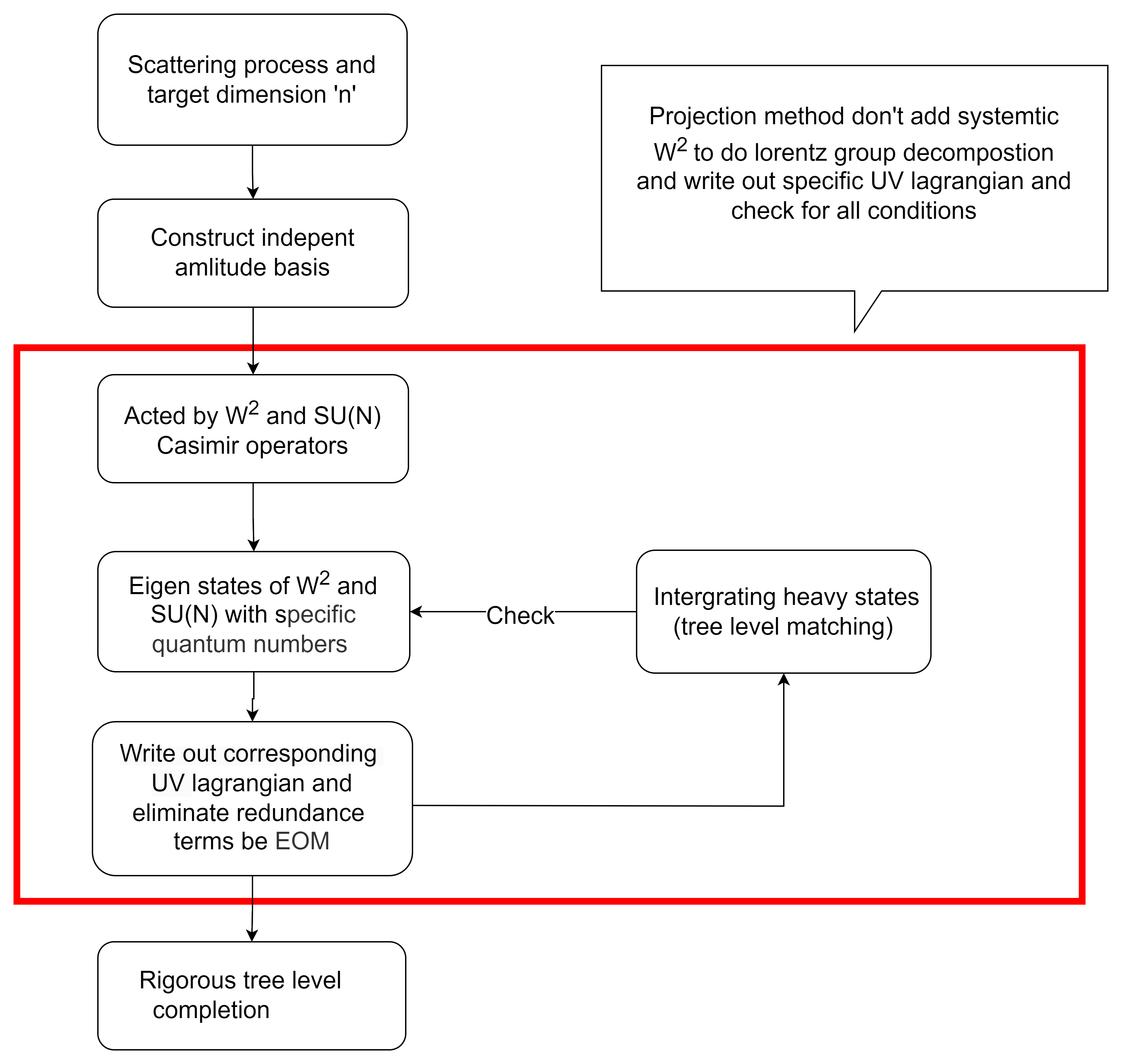}
    \caption{Flow chart of the J-Basis method to obtain the UV states corresponding to the possible external rays.}\label{fig:flow}
\end{figure}

The whole procedure for the J-basis method is described in the following.
First we process the amplitude decomposition for the amplitude basis of the specific process matching the $s^2$ contribution in the SMEFT to obtain the possible UV states. Then we use the UV selection based on repeat field, the EOM and other redundancy to obtain the UV states in the tree level formally. We present a flow chart Fig.~\ref{fig:flow} to the whole procedure and compare this method with the projection method.
Then we discuss several typical $2\rightarrow2$ scattering processes following the procedure in flowchart and show differences in the results.

\subsection{4 SM Higgs scattering}\label{subsec42}
The 2 Higgs to 2 Higgs scattering is a typical example discussed in Ref.~\cite{Helset:2018fgq}. 
It involves several dim-8 operators as
\begin{equation}
\begin{aligned}
& Q_{H^4}^{(1)}=\left(D_\mu H^{\dagger} D_\nu H\right)\left(D^\nu H^{\dagger} D^\mu H\right) \,,\\
& Q_{H^4}^{(2)}=\left(D_\mu H^{\dagger} D_\nu H\right)\left(D^\mu H^{\dagger} D^\nu H\right) \,,\\
& Q_{H^4}^{(3)}=\left(D^\mu H^{\dagger} D_\mu H\right)\left(D^\nu H^{\dagger} D_\nu H\right)\,.
\end{aligned}
\end{equation}
In the external ray method, first in Ref.~\cite{Zhang:2020jyn}, the gauge group $SU(2)_w$ CG-coefficients of the $SU(2)_w$ gauge group are used to form projectors in Table~\ref{table:4H}.
Projectors in Table~\ref{table:4H} match the UV states $\mathcal{B}_1,\mathcal{S},\mathcal{B},\mathcal{W},\Xi_0,\Xi_1$ in Table~\ref{table:full4H}. After utilizing the J-Basis method, we find extra new spin-2 UV states $\mathcal{G},\mathcal{H}_0,\mathcal{H}_1$.

Here we present the details of the J-Basis method applying to the $4H$ scattering.
First, we list the 6 P-Basis operators for the type $D^{4}H^{4}$ involved in the $4H$ scattering,
\begin{equation}
\ytableausetup{smalltableaux}
\begin{aligned} & \mathcal{O}_1^f=\frac{1}{4} \mathcal{Y}[
\begin{ytableau}
p & r 
\end{ytableau}]
\mathcal{Y}[\begin{ytableau}
s & t \end{ytableau}] H_{pi} H_{rj}\left(D_\mu D_\nu H^{\dagger^{i}}_s\right)\left(D^\mu D^\nu H^{\dagger^j}_t\right)\,, \\ & \mathcal{O}_2^f=\frac{1}{4} \mathcal{Y}[\begin{ytableau}
p & r 
\end{ytableau}]
\mathcal{Y}[\begin{ytableau}
s & t \end{ytableau}] H^{\dagger^i}_p H_{ri}\left(D_\mu D_\nu H_{js}\right)\left(D^\mu D^\nu H^{\dagger^j}_t\right)\,, \\ & \mathcal{O}_3^f=\frac{1}{4} \mathcal{Y}[\begin{ytableau}
p & r 
\end{ytableau}]
\mathcal{Y}[\begin{ytableau}
s & t \end{ytableau}] H_{pi}\left(D_\mu H_{rj}\right)\left(D_\nu H^{\dagger^i}_s\right)\left(D^\mu D^\nu H^{\dagger^j}_t\right)\,,\\
& \mathcal{O}_4^f=\frac{1}{4} \mathcal{Y}[
\begin{ytableau}
p \\ r 
\end{ytableau}]
\mathcal{Y}[\begin{ytableau}
s & t \end{ytableau}] H_{pi} H_{rj}\left(D_\mu D_\nu H^{\dagger^{i}}_s\right)\left(D^\mu D^\nu H^{\dagger^j}_t\right)\,, \\ & \mathcal{O}_5^f=\frac{1}{4} \mathcal{Y}[\begin{ytableau}
p \\ r 
\end{ytableau}]
\mathcal{Y}[\begin{ytableau}
s \\ t \end{ytableau}] H^{\dagger^i}_p H_{ri}\left(D_\mu D_\nu H_{js}\right)\left(D^\mu D^\nu H^{\dagger^j}_t\right)\,, \\ & \mathcal{O}_6^f=\frac{1}{4} \mathcal{Y}[\begin{ytableau}
p \\ r 
\end{ytableau}]
\mathcal{Y}[\begin{ytableau}
s \\ t \end{ytableau}] H_{pi}\left(D_\mu H_{rj}\right)\left(D_\nu H^{\dagger^i}_s\right)\left(D^\mu D^\nu H^{\dagger^j}_t\right)\,
.\end{aligned}
\end{equation}
Acting the Poincare Casimir operator $\mathcal{W}^2$ on these P-Basis operators, we obtain the eigenstates and the eigenvalues of J-Basis in Table~\ref{Table:4HJ}. In detail, we process these steps by using the program \texttt{ABC4EFT} in Ref.~\cite{Li:2022tec}.
Then we transform the P-Basis to the basis in Ref.~\cite{Murphy:2020rsh}. By applying the UV selection, all the possible UV states that match nine eigenstates are written out. So we can obtain the Table~\ref{table:full4H} in Sec.~\ref{subsec22} corresponding to Table~\ref{Table:4HJ}.
\begin{table}[H]
 \centering
\begin{tabular}{|c|c|}
\hline channel (Spin,$SU(3)_c$,$SU(2)_w$,$U(1)_y$)& P-Basis\\ \hline
$\left\{H_1, H_2\right\},\left\{H_3^{\dagger}, H_4^{\dagger}\right\}$ & $\mathcal{O}_{i}$ \\
\hline$(2,1,3,1)$ & $-8 \mathcal{O}_1^f-48 \mathcal{O}_2^f-48 \mathcal{O}_3^f$\\
$(0,1,3,1) $&$8 \mathcal{O}_1^f $\\
$(1,1,1,1) $&  $8 \mathcal{O}_1^f+16 \mathcal{O}_3^f  $\\
\hline$\left\{H_1, H^{\dagger}\right\},\left\{H_2, H_4^{\dagger}\right\}$ & $\mathcal{O}^j$ \\
\hline$(2,1,3,0)$ & $16 \mathcal{O}_1^f-4 \mathcal{O}_2^f+56 \mathcal{O}_3^f$\\
$(1,1,3,0)$ & $8 \mathcal{O}_1^f-4 \mathcal{O}_2^f+8 \mathcal{O}_3^f$ \\
$(0,1,3,0)$ & $8 \mathcal{O}_1^f+4 \mathcal{O}_2^f+16 \mathcal{O}_3^f $ \\
$(2,1,1,0) $& $-24 \mathcal{O}_1^f-4 \mathcal{O}_2^f-24 \mathcal{O}_3^f $\\
$(1,1,1,0)$ & $-4 \mathcal{O}_2^f-8 \mathcal{O}_3^f$\\ \hline
\end{tabular}
\caption{J-Basis analysis results for the $4H$ scattering.}
\label{Table:4HJ}
\end{table}
After obtaining all the UV states, we can apply Eq.~\ref{expand0}, Eq.~\ref{expand1} and  Eq.~\ref{expand2} to obtain positivity bounds. More detailed, we choose the EFT operators $\mathcal{O}_{n,ijkl}$ as basis $B_{ijkl}$ to expand the UV amplitude. So we just need to search for all the normal vectors of the cone constructed by all the matching results from the fifth column of Table~\ref{table:full4H}~directly.

\begin{figure}[H]
    \centering %
     \includegraphics[width=0.7\textwidth]{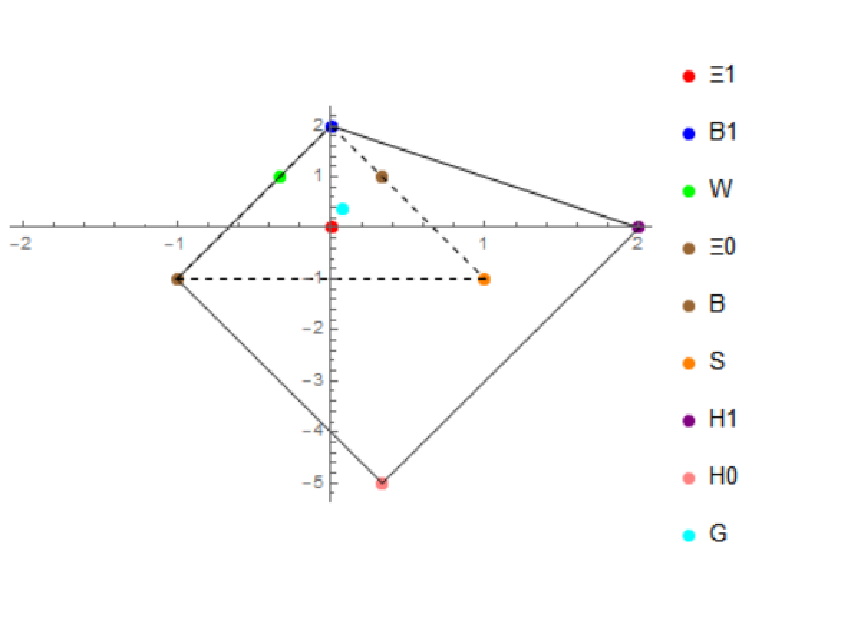}
    \caption{ The positivity cone for the 4-Higgs operators, with the corresponding generators. The x-axes represents $(C_1+C_3)/(2C_1+3C_2+C_3)$, the y axes represents $(C_1-C_3)/(2C_1+3C_2+C_3)$, the dashed line presents the cone we obtained before in Ref.~\cite{Zhang:2020jyn}, and the new cone is the solid line.}
    \label{fig:Higgs}
\end{figure}

The number of rank-2 subsets of \{$\vec{c}(p)$\} is $C_9^2=36$. Thus we could obtain 36 normal vectors corresponding to every rank-2 subset which represents the corresponding possible facet of the cone. To select the correct facets of the cone, we need to select the normal vectors which satisfy the positivity argument Eq.~\ref{expand2}. However only the following 4 normal vectors $\vec{n}(p)$ in these 36 normal vectors which are listed Eq.~\ref{nr} satisfying that for every $\vec{c}(p)_{i}$,
\begin{equation}
    \vec{c}(p)_{i}\cdot \vec{n}(p)\geq 0\,.
\end{equation}
The normal vectors that satisfies the positivity argument Eq.~\ref{nr} are
\begin{equation}
    \begin{aligned}
        (1,1,1),~(1,1,\frac{1}{2}),~(5,9,1),~(1,3,2)\,.
    \end{aligned}
    \label{nr}
\end{equation}
The EFT amplitude $(C_1,C_2,C_3)$ should exist in the cone, so we obtain new positivity bounds,
\begin{equation}
 \begin{aligned}
    & C_1+C_2+C_3\geq 0 \,, \quad \quad C_1+C_2+\frac{1}{2}C_3\geq 0 \,,\\
     & 5C_1+9C_2+C_3\geq 0 \,, \quad \quad C_1+3C_2+C_3\geq 0 \,.
 \end{aligned}   
\end{equation}
Thus external rays are changed to $\mathcal{H}_{1},\mathcal{H}_{0},\mathcal{B}_{1},\mathcal{B}_{0}$. In the perspective of the cone's bottom, we obtain Fig~.\ref{fig:Higgs}.
Based on Fig.~\ref{fig:Higgs}, the Monte Carlo Sampling shows that the allowed area of the WC space is larger than the one obtained by the projection method, and the cone is a quadrangular pyramid actually.

By applying the J-Basis method in the SM Higgs sector we find that in Ref.~\cite{Zhang:2020jyn} the projection to the UV states representing potential external ray bounds provides tighter bounds.

\subsection{\texorpdfstring{$4W$}~ Scattering}
\subsubsection{Amplitude Analysis and redundancy}\label{WWWW}
By applying the J-Basis method in $4W$ scattering, we obtain Table~\ref{table:4HW} listing all the possible UV states.
\begin{table}[H]
    \centering
\begin{tabular}{|c|c|c|}
\hline \multicolumn{3}{|c|}{ group: $\left(\operatorname{Spin}, S U(3)_c, S U(2)_w, U(1)_y)\right.$} \\
\hline \multicolumn{3}{|c|}{$\begin{aligned} \mathcal{O}_1^{(p)} & =W_{\mathrm{L} \mu \nu}^I W_{\mathrm{L}}^{I \nu \rho} W_{\mathrm{L} \lambda \rho}^J W_{\mathrm{L}}^{J \lambda \mu} \\
\mathcal{O}_2^{(p)} & =W_{\mathrm{L} \lambda \rho}^I W_{\mathrm{L} \mu \nu}^I W_{\mathrm{L}}^{J \lambda \rho} W_{\mathrm{L}}^{J \mu \nu}\end{aligned}$} \\
\hline$\left\{W_{\mathrm{L} 1}, W_{\mathrm{L} 2}\right\},\left\{W_{\mathrm{L} 3}, W_{\mathrm{L} 4}\right\}$ & $\mathcal{O}_j^{(m)}$ & $\mathcal{O}_j^{(p)}$ \\
\hline$(2, \mathbf{1}, \mathbf{5}, 0)$ & $(144,-12,0,-96,8,0,144,-12,0)$ & $8(44,3)$ \\
\hline$(1, \mathbf{1}, \mathbf{5}, 0)$ & $(48,-12,24,-32,8,-16,48,-12,24)$ & 0 \\
\hline$(0, \mathbf{1}, \mathbf{5}, 0)$ & $(0,12,0,0,-8,0,0,12,0)$ & $8(4,3)$ \\
\hline$(2, \mathbf{1}, \mathbf{1}, 0)$ & $(0,0,0,48,-4,0,0,0,0)$ & $8(-4,-3)$ \\
\hline$(1, \mathbf{1}, \mathbf{1}, 0)$ & $(0,0,0,16,-4,8,0,0,0)$ & 0 \\
\hline$(0, \mathbf{1}, \mathbf{1}, 0)$ & $(0,0,0,0,4,0,0,0,0)$ & $8(-2,0)$ \\
\hline$(2, \mathbf{1}, \mathbf{3}, 0)$ & $(-48,4,0,0,0,0,48,-4,0)$ & 0 \\
\hline$(1, \mathbf{1}, \mathbf{3}, 0)$ & $(-16,4,-8,0,0,0,16,-4,8)$ & $8(4,1)$ \\
\hline$(0, \mathbf{1}, \mathbf{3}, 0)$ & $(0,-4,0,0,0,0,0,4,0)$ & 0 \\
\hline \multicolumn{3}{|c|}{$\begin{array}{c}\mathcal{O}_1^{(p)}=W_{\mathrm{L} \mu \nu}^I W_{\mathrm{L} \lambda \rho}^J W_{\mathrm{R}}^{I \nu \rho} W_{\mathrm{R}}^{J \lambda \mu} \\
\mathcal{O}_2^{(p)}=W_{\mathrm{L} \lambda \rho}^I W_{\mathrm{L} \mu \nu}^I W_{\mathrm{R}}^{K \lambda \mu} W_{\mathrm{R}}^{K \nu \rho}\end{array}$} \\
\hline$\left\{W_{\mathrm{L} 1}, W_{\mathrm{L} 2}\right\},\left\{W_{\mathrm{R} 3}, W_{\mathrm{R} 4}\right\}$ & $\mathcal{O}_j^{(m)}$ & $\mathcal{O}_j^{(p)}$ \\
\hline$(0, \mathbf{1}, \mathbf{5}, 0)$ & $(-48,32,-48)$ & $32(-3,1)$ \\
\hline$(0, \mathbf{1}, \mathbf{3}, 0)$ & $(16,0,-16)$ & 0 \\
\hline$(0, \mathbf{1}, \mathbf{1}, 0)$ & $(0,-16,0)$ & $16(0,-1)$ \\
\hline$\left\{W_{\mathrm{L} 1}, W_{\mathrm{R} 3}\right\},\left\{W_{\mathrm{L} 2}, W_{\mathrm{R} 4}\right\}$ & $\mathcal{O}_j^{(m)}$ & $\mathcal{O}_j^{(p)}$ \\
\hline$(2, \mathbf{1}, \mathbf{5}, 0)$ & $(32,-48,-48)$ & $16(-1,-3)$ \\
\hline$(2, \mathbf{1}, \mathbf{3}, 0)$ & $(0,16,-16)$ & $16(-1,1)$ \\
\hline$(2, \mathbf{1}, \mathbf{1}, 0)$ & $(-16,0,0)$ & $16(-1,0)$ \\
\hline
\end{tabular}
\caption{J-Basis analysis results for the $4W$ scattering. Column of $\mathcal{O}_j^{(m)}$ represents the m-Basis results, and the column of $\mathcal{O}_j^{(p)}$ represents the P-Basis results. The combination of groups is defined as as $\left(\operatorname{Spin}, S U(3)_c, S U(2)_w, Y)\right.$}
\label{table:4HW}
\end{table}
In Table~\ref{table:4HW}, the 6 involved operators in the P-Basis are listed as follows,
\begin{equation}
\begin{aligned}
& \mathcal{O}_{W_{\mathrm{L}, 1}^4}^{(p)} =W_{\mathrm{L} \mu \nu}^I W_{\mathrm{L} \lambda \rho}^J W_{\mathrm{L}}^{I \nu \rho} W_{\mathrm{L}}^{J \lambda \mu}, \quad \mathcal{O}_{W_{\mathrm{L}}^4, 2}^{(p)}=W_{\mathrm{L} \mu \nu}^I W_{\mathrm{L}}^{J \mu \nu} W_{\mathrm{L} \lambda \rho}^I W_{\mathrm{L}}^{J \lambda \rho} \,,\\
& \mathcal{O}_{W_{\mathrm{L}}^2 W_{\mathrm{R}}^2, 1}^{(p)}=W_{\mathrm{L} \mu \nu}^I W_{\mathrm{L} \lambda \rho}^J W_{\mathrm{R}}^{I \nu \rho} W_{\mathrm{R}}^{J \lambda \mu}, \quad \mathcal{O}_{W_{\mathrm{L}}^2 W_{\mathrm{R}}^2, 2}^{(p)}=W_{\mathrm{L} \mu \nu}^I W_{\mathrm{L} \lambda \rho}^I W_{\mathrm{R}}^{J \nu \rho} W_{\mathrm{R}}^{J \lambda \mu} \,,\\
& \mathcal{O}_{W_{\mathrm{R}}^4, 1}^{(p)}=W_{\mathrm{R} \mu \nu}^I W_{\mathrm{R} \lambda \rho}^J W_{\mathrm{R}}^{I \nu \rho} W_{\mathrm{R}}^{J \lambda \mu}, \quad \mathcal{O}_{W_{\mathrm{R}}^4, 2}^{(p)}=W_{\mathrm{R} \mu \nu}^I W_{\mathrm{R}}^{J \mu \nu} W_{\mathrm{R} \lambda \rho}^I W_{\mathrm{R}}^{J \lambda \rho}\,.
\end{aligned}
\end{equation}
The WC space of the $4W$ operators can be defined as 
\begin{equation}
(\mathcal{C}_{W_{\mathrm{L}, 1}^4}^{(p)},\mathcal{C}_{W_{\mathrm{L}}^4,2}^{(p)},\mathcal{C}_{W_{\mathrm{L}}^2 W_{\mathrm{R}}^2, 1}^{(p)},\mathcal{C}_{W_{\mathrm{L}}^2 W_{\mathrm{R}}^2, 2}^{(p)},\mathcal{C}_{W_{\mathrm{R}}^4, 1}^{(p)},\mathcal{C}_{W_{\mathrm{R}}^4, 2}^{(p)})\equiv(C_1,C_2,C_3,C_4,C_5,C_6)\,.
\end{equation}
By applying the UV selection to Table~\ref{table:4HW}, we can check whether some UV states are ruled out or not.
\\
\\
\textbf{1.}~Let's us write out such UV Lagrangian $WWV$ where $W$ is the $W$ boson, and $V$ represents the heavy vector. In the term $WWV$, the indices of the Lorentz and the gauge groups has been omitted for simplification of marking. The first leading contribution would match to $D_2W_4$ which corresponds to dim-$10$. So $WWV$ couplings can be excluded.
\\
\\
\textbf{2.}~Meanwhile, Table~\ref{Table:4HJ} gives the possibility of existing spin-2 UV couplings as 
 $W_{L}W_{L}X$. However, if you calculate the matching of UV state $ W^{\mu\nu}_{I\ L}W_{I\ L\nu\rho}\mathcal{G}^{\rho\mu}$ to the P-Basis,
\begin{equation}
\begin{aligned}
& \wick{W^{\mu\nu}_{I\ L}W_{I\ L\nu\rho}\c{\mathcal{G}^{\rho\mu}} W^{\alpha\beta}_{J\ L}W_{J\ L\beta\gamma}\c{\mathcal{G}^{\gamma\alpha}}}=\\ &W^{\mu\nu}_{I\ L}W_{I\ L\nu\rho} W^{\alpha\beta}_{J\ L}W_{J\ L\beta\gamma}*\frac{1}{M_S^2}(g^{\rho\gamma}g^{\mu\alpha}+g^{\rho\alpha}g^{\mu\gamma}-\frac{2}{3}g^{\rho\mu}g^{\gamma\alpha})\\
&\propto W_{\mathrm{L} \mu \nu}^I W_{\mathrm{L} \lambda \rho}^J W_{\mathrm{L}}^{I \nu \rho} W_{\mathrm{L}}^{J \lambda \mu}=\mathcal{O}_{W_{\mathrm{L}, 1}^4}^{(p)}\,.
\end{aligned}
\end{equation}
Thus, the matching result for this UV state $ W^{\mu\nu}_{I\ L}W_{I\ L\nu\rho}\mathcal{G}^{\rho\mu}$ exists in the ray $(1,0,0,0,0,0)$ of the WC space.
The result violates the J-Basis analysis result $(-4,-3,0,0,0,0)$ for the UV state $(2,1,1,0)$ in the channel $(W_L,W_L,W_L,W_L)$. Besides, Ref.~\cite{Zhang:2020jyn} provides another character of the dispersion relation in Eq.~\ref{eqdis} that
 the amplitude cone is a silent cone. This means there shouldn't exist any other UV state in the negative direction of the UV state with quantum number $(0,1,5,0)$ for the $4W$ scattering case. In Table~\ref{table:4HW}, it shows that in the opposite direction of $(0,1,5,0)$, there exists the UV state with the quantum number $(2,1,1,0)$.
The three result from the J-Basis method, from the UV matching, and from geometry perspective seem incongruous in that case. 
 However, there is no conflict among the three results because the UV states of tensor particle with the form $WW\mathcal{G}$ can be eliminated by the EOM. 
For the tensor coupling $W_{\mathrm{L} \mu \nu}^I W_{\mathrm{L} \rho}^I{ }^\nu \mathcal{G}^{\mu \rho}$, the interaction Lagrangian can be rewritten as follow
\begin{equation}
\begin{aligned}
& W_{\mathrm{L}}^{I \alpha \beta} W_{\mathrm{L} \alpha}^{I\ \gamma} \mathcal{G}_{\beta \gamma \dot{\alpha}}{ }^{\dot{\alpha}} =(\sigma_{\mu\nu})^{\alpha\beta}(\sigma_{mn})^{\gamma}_{\alpha}(\sigma_{u_{1}})^{\dot{\alpha}}_{\beta}(\sigma_{v_{1}})_{\gamma\dot{\alpha}}W_{L}^{I\mu\nu}W_{L}^{Imn}\mathcal{G}_{\mu_{1}\nu_{1}}\,.
\end{aligned}
\label{eq49}
\end{equation}
By applying the characters of the $\sigma$ matrix, Eq.~\ref{eq49} can be expanded as
\begin{equation}
\begin{aligned}
&W_{\mathrm{L}}^{I \alpha \beta} W_{\mathrm{L} \alpha}^{I\ \gamma} \mathcal{G}_{\beta \gamma \dot{\alpha}}{ }^{\dot{\alpha}}=
\left(-2 g^{\mu \mu_{1}} \sigma_{t \dot{\alpha}}^{\nu}+2 g^{\nu \mu_{1}} \sigma_{t\dot{\alpha}}^{\mu}+2 i \varepsilon^{\mu \nu \mu_{1} \lambda} \sigma_{\lambda t\dot{\alpha}}\right) \varepsilon^{t\alpha} * \\
& \left(-2 g^{m \nu_{1}} \sigma_{\alpha \dot{\beta}}^n+2 g^{n \nu_{1}} \sigma_{\alpha \dot{\beta}}^m+2 i \varepsilon^{mn\nu_{1} \lambda} \sigma_{\lambda \dot{\beta}}\right) \varepsilon^{\dot{\alpha} \dot{\beta}}W_{L}^{I\mu\nu}W_{L}^{Imn}\mathcal{G}_{\mu_{1}\nu_{1}}\,.
\end{aligned}
\label{ex1}
\end{equation}
There are many kinds of terms in the expansion of Eq.~\ref{ex1}, but all the terms can be transformed to the form $W_{\mathrm{L} \mu \nu}^I W_{\mathrm{L} \rho}^I{ }^\nu \mathcal{G}^{\mu \rho}$ by using $\operatorname{Tr}\left(\sigma_{\lambda} \bar{\sigma}_\rho\right) = 2 g_{\lambda \rho}$,
\begin{equation}
 \begin{aligned}
     g^{\mu \mu_{1}} \sigma_{t \dot{\alpha}}^{\nu}g^{m \nu_{1}} \sigma_{\alpha\dot{\beta}}^n W_{L}^{I\mu\nu}W_{L}^{Imn}\mathcal{G}_{\mu_{1}\nu_{1}}&=g^{\mu \mu_{1}}g^{m\nu_{1}}Tr\left[\sigma_{t}\bar{\sigma_{n}})\right] W_{L}^{I\mu\nu}W_{L}^{Imn}\mathcal{G}_{\mu_{1}\nu_{1}}\\&=W_{\mathrm{L} \mu \nu}^I W_{\mathrm{L} \rho}^I{ }^\nu \mathcal{G}^{\mu \rho}\,.
 \end{aligned}   
\end{equation}
Likewise,
\begin{equation}
\begin{aligned}
&  \varepsilon^{\mu \nu \nu_{1} \lambda} \varepsilon^{mn \nu \rho} \sigma_{\lambda t\alpha} \sigma_{\rho \dot{\alpha} \dot{\beta}} \varepsilon^{t \alpha} \varepsilon^{\alpha \dot{\beta}} \quad W_L^{\mu \nu} W_L^{m n} \mathcal{G}^{ \mu_1 \nu_1} \\
& = \varepsilon^{\mu \nu \mu_{1} \lambda} \varepsilon^{m n \nu \rho} \operatorname{Tr}\left(\sigma_{\lambda} \bar{\sigma}_\rho\right) W_L^{\mu \nu} W_L^{m n} \mathcal{G}^{\mu \nu_1} \\
& =-4  W_L^{\mu \nu} W_{L\nu \rho} \mathcal{G}^{\rho }_{\mu} \,.
\end{aligned}
\end{equation} 
Finally, Eq.~\ref{trans1}, the transformation relationship can be obtained
\begin{equation}
\begin{aligned}
& W_{\mathrm{L} \mu \nu}^I W_{\mathrm{L} \rho}^I{ }^\nu \mathcal{G}^{\mu \rho}\propto W_{\mathrm{L}}^{I \alpha \beta} W_{\mathrm{L} \alpha}^{I\ \gamma} \mathcal{G}_{\beta \gamma \dot{\alpha}}{ }^{\dot{\alpha}} \\
&= W_{\mathrm{L}}^{I \alpha \beta} W_{\mathrm{L} \alpha}^{I\ \gamma} \mathcal{G}_{\mu \nu}\left(-g^{\mu \nu} \epsilon_{\beta \gamma}-i \sigma_{\beta \gamma}^{\mu \nu}\right)\,.
\end{aligned}
\label{trans1}
\end{equation}
\\
\textbf{3.}~By applying the EOM of the massive spin-2 particles, we can show that the Eq.~\ref{trans1} equals zero.
The free Lagrangian of the massive spin-2 quantum theory \cite{Blasi:2017pkk, Li:2023wdz} is
\begin{equation}
S_{F P}\left[h ; m_1^2\right] \equiv S_{L G}[h]+S_m\left[h ; m_1^2,-m_1^2\right]\,.
\end{equation}
 The expressions $S_{LG}$ (kinetic term) and $S_{m}$ (mass term) are
\begin{equation}
\begin{aligned}
&S_{L G}[h]=\int \mathrm{d}^4 \mathrm{x}\left[\frac{1}{2} h \partial^2 h-h_{\mu \nu} \partial^\mu \partial^\nu h-\frac{1}{2} h^{\mu \nu} \partial^2 h_{\mu \nu}+h^{\mu \nu} \partial_\nu \partial^\rho h_{\mu \rho}\right],\\
&S_m\left[h ; m_1^2, m_2^2\right]=\frac{1}{2} \int \mathrm{d}^4 x\left(m_1^2 h_{\mu \nu} h^{\mu \nu}+m_2^2 h^2\right)\,.
\end{aligned}
\end{equation}
By applying the Euler-Lagrange equation, we can obtain the EOMs and find that the $h_{\mu\nu}$ is traceless.
\begin{equation}
\begin{aligned}
& \left(\partial^2-m_1^2\right) h_{\mu \nu}(x)=0 \,,\\
& \partial^\mu h_{\mu \nu}(x)=0\,, \\
& h_{\mu}^{\mu}=h(x)=0\,.
\end{aligned}
\label{WWG}
\end{equation}
By using the relation $\mathcal{G}_{\beta \gamma \dot{\alpha}}{ }^{\dot{\alpha}}=\mathcal{G}_{\mu \nu} \epsilon^{\dot{\alpha} \dot{\beta}} \sigma_{\beta \dot{\alpha}}^\mu \sigma_{\gamma \dot{\beta}}^\nu =\mathcal{G}_{\mu \nu}\left(-g^{\mu \nu} \epsilon_{\beta \gamma}-i \sigma_{\beta \gamma}^{\mu \nu}\right)$,
the contraction between symmetry tensor $\mathcal{G}_{\mu \nu}$ and anti-symmetry tensor $\sigma_{\beta \gamma}^{\mu \nu}$ is 0. Then according to Eq.~\ref{WWG}, the   $\mathcal{G}_{\mu \nu}g^{\mu\nu}$ is 0. So the couplings with the form $W_{L}W_{L}\mathcal{G}$ and $W_{R}W_{R}\mathcal{G}$ are eliminated by the EOMs.

The above discussions show that not all the amplitude decompositions correspond to the determined UV states in any case. The results of amplitude decomposition require the UV selection by the EOMs, the repeat field and other identities. Finally, we can write out all possible UV states for the $4W$ scattering in Table~\ref{table4w}.

\begin{table}[!ht]\normalsize
    \centering
    \renewcommand\arraystretch{2.2}
    \begin{tabular}{|c|c|c|}
    \hline
(Spin,$SU(3)$,$SU(2)$,$U(1)_y$) & Interaction Lagrangian & $\vec{c}(p)$ \\ \hline
        (0,1,5,0) & \makecell[c]{$W_{\mathrm{L} \mu \nu}^I W_{\mathrm{L}}^{J \mu \nu}\left(T^{A}\right)^{I J} \mathcal{S}^{A}+$\\$x_{1}W_{\mathrm{R} \mu \nu}^I W_{\mathrm{R}}^{J \mu \nu}\left(T^{A}\right)^{I J} \mathcal{S}^{A}$} & $(4,3,-12x_1,4x_1,4x_1^2,3x_1^2)$ \\ \hline
        (0,1,1,0) & $W_{\mathrm{L} \mu \nu}^I W_{\mathrm{L}}^{I \mu \nu} S+x_2W_{\mathrm{R} \mu \nu}^I W_{\mathrm{R}}^{I \mu \nu} S$ & $(-2,0,0,-4x_2,-2x_2^2,0)$ \\ \hline
        (2,1,5,0) & $W_L^{I \mu \nu} W_{R\nu \rho}^J T_K^{I J} G^{K \rho \mu}$ & $(0,0,-1,-3,0,0)$ \\ \hline
        (2,1,3,0) & $W_L^{I \mu \nu} W_{R \nu \rho}^J \varepsilon^{I J K} G^{ {K \rho \mu }}$ & $(0,0,-1,1,0,0)$ \\ \hline
        (2,1,1,0) & $W_L^{I \mu \nu} W_{R \nu \rho}^I G^{\rho \mu}$ & $(0,0,-1,0,0,0)$ \\ \hline
    \end{tabular}
     \caption{Matching Results for the $4W$ scattering.}
     \label{table4w}
\end{table}
Now according to Eq.~\ref{expand0}, Eq.~\ref{expand1} and Eq.~\ref{expand2}, the normal vectors of the $4W$ scattering amplitude cone can be calculated to obtain bounds. The cone has three categories of normal vectors which have the form in the WC space as
\begin{equation}
\left\{ \begin{aligned}
    &(0, 1, 0, 0, 0, 0)\,,\\
    &(1, -\frac{4}{3}, 0, 0, 0, 0)\,,\\
    &(-x_{2}, \frac{4 (x_1 + x_{2})}{3}, 1, 1, -\frac{1}{x_{2}}, \frac{4 (x_1 + x_{2})}{3 x_1 x_{2}})\,(x_1 \leq 0, x_{2} \leq 0))\,.
\end{aligned}\right.
\label{4WWW}
\end{equation}
\\
The EFT amplitude $(C_1,C_2,C_3,C_4,C_5,C_6)$ should exist in the cone. So the product between the EFT WCs and the normal vector above should be positive, which represents the positivity bounds.

\begin{figure}[htbp]
	\subfigure[] 
	{
 \begin{minipage}[t]{6cm}
			\centering          
			\includegraphics[scale=0.4]{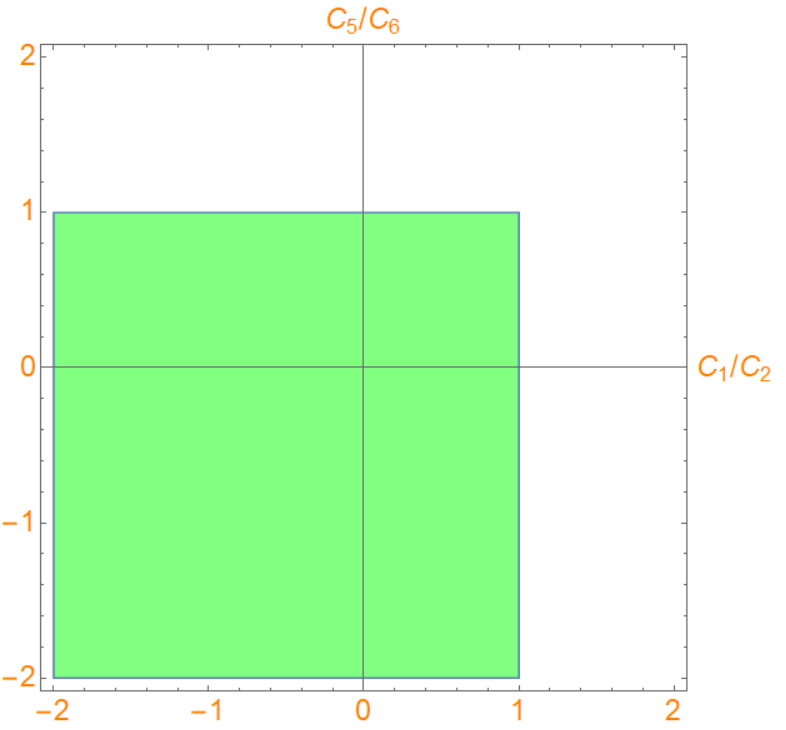}   
		\label{fig:a}\end{minipage}
	}
	\subfigure[] 
	{
		\begin{minipage}[t]{6cm}
			\centering      
			\includegraphics[scale=0.8]{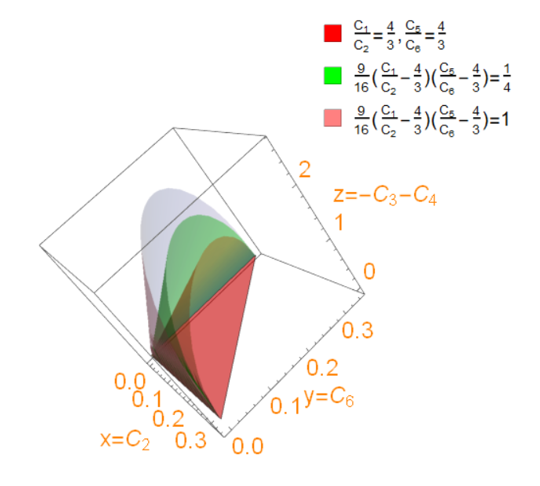}   
		\label{fig:b}\end{minipage}
	}
	\caption{The WC space of the $4W$ scattering process. Fig.~\ref{fig:b} is the slice of 4W scattering cone with different value 
of $\frac{C_{1}}{C_{2}}$ and $\frac{C_{5}}{C_{6}}$. While the red semiconical represents $\frac{C_{1}}{C_{2}}= \frac{4}{3}$, 
$\frac{C_{5}}{C_{6}}= \frac{4}{3}$ point in Fig.~\ref{fig:a};
The green semiconical represents curve $\frac{9}{16} (\frac{C_{1}}{C_{2}} -  \frac{4}{3}) ( 
  \frac{C_{5}}{C_{6}} -\frac{4}{3}) = \frac{1}{4}$ in Fig.~\ref{fig:a}; 
The pink semiconical represents curve $\frac{9}{16} (\frac{C_{1}}{C_{2}} -  \frac{4}{3}) ( 
  \frac{C_{5}}{C_{6}} -\frac{4}{3}) = 1$ in Fig.~\ref{fig:a}. } 
 \label{fig:4W}
\end{figure}

Then the positive argument of the vectors in the Eq.~\ref{4WWW} in the WCs space can be obtained by solving such a system of binary quadratic inequalities:
\begin{equation}
\left\{
\begin{aligned} & C_2 \geq 0\,, \\ & -C_1+\frac{4 C_2}{3} \geq 0\,, \\ & (C_5 -\frac{4 C_6 }{3})x_1-\frac{4 C_6 x_2}{3}-(C_3 +C_4) x_1 x_2-(\frac{4}{3} C_2 x_1-C_1 x_2+\frac{4}{3} C_2 x_2)x_1 x_2\geq 0 \,\left(x_1, x_{2} \leq 0\right)\,.
\end{aligned}
\right.
\end{equation}
To solve the third inequality, there are some tricks, i.e. we can regard $x_1$ as a known number so as to calculate the single quadratic inequality. Then we obtain quadratic inequality of $x_2$ from $b^2 \geq 4ac$.
Finally, we obtain the bounds as
\begin{equation}
\left\{
\begin{aligned} & C_6 \geq 0 \,,\\ & C_2 \geq 0 \,,\\ & C_1-\frac{4C_2}{3} \leq 0 \,,\\ & C_5-\frac{4C_6}{3} \leq 0 \,,\\ & -C_3-C_4-\frac{8 \sqrt{C_2 C_6}}{3} \leq 2 \sqrt{\left[\left(C_1-\frac{4C_2}{3}\right)\left(C_5-\frac{4C_6}{3}\right)\right]}\,.\end{aligned}\right.
\end{equation}
The volume of the allowed WC space is 0.435\% by the Monte Carlo Sampling. Despite that the cone is described by more than 6 WCs, we can still show the structure of the cone in 3D space as in Fig.~\ref{fig:4W} by choosing the specific slicing in the dim-6 WC space. In the scheme of the slice in Fig.~\ref{fig:4W}, the UV state $(2,1,3,0)$ is projected to origin while the UV states $(2,1,5,0)$ and $(2,1,1,0)$ are projected to the $y$ axes. More than that, the circle corresponds to the UV state $(0,1,5,0)$, and the $(0,1,1,0)$ is degenerated to a linear ray $y=4x$. All of them are in the inner or surface of the slice.

In the previous result in Ref.~\cite{Zhang:2020jyn},
the projectors formed by $SO(2)$ and $SU(2)_w$ CG coefficients were used to represent UV states.
the previous work considered the CP-conserving case and reached the results of the 9 possible external rays (UV states) presented by $E_{m,n}$ where $m,n$ are different Irreps of $SU(3)_c$ and $SU(2)_w$. However we reach the conclusion that there are only 5 possible UV states in the tree level completion. For example, $E_{1,2}$ means $(0,1,3,0)$ in  Table~\ref{table:4HW} whose contribution is zero after the decomposition of the Lorentz and the gauge group.
In conclusion, we find that not all irrep projectors can be realized with the UV completion.

\subsubsection{Comment with the 4 Gluon Scattering}\label{subsec432}
According to the detailed discussions about the UV completion of the $4W$ scattering in the Sec.\ref{WWWW}, we find that the number of UV states in the tree level completion to restrict vector boson cones is less than previous results obtained by projection in Ref.~\cite{Zhang:2021eeo}.
Hence, the 4 gluon scattering is similar. More specifically, color group direct product decompositions (projectors) are listed as follows, while the  
\boldmath$\bar{10}$ \unboldmath representations in \boldmath$8\otimes8=1+8+\bar{8}+10+\bar{10}+27$\unboldmath ~ from Ref.~\cite{Bellazzini:2014waa} is eliminated for it doesn't correspond to the inverted symmetry ($ij\rightarrow ji$,~$kl\rightarrow lk$). The projectors corresponding the group decompositions of the direct product of the two $SU(3)_c$ adjoint representations are listed in Eq.~\ref{gluon}.
The $SO(2)$ group decompositions are the same as Eq.~\ref{SO2}. Finally, in Ref.~\cite{Zhang:2021eeo}, it reaches the conclusion that there are 15 possible UV states for the 4 Gluon scattering case,
\begin{equation}
\begin{aligned}
P_{\mathbf{1}}^{a b, c d} &=\frac{\delta^{a b} \delta^{c d}}{N^2-1}, \\
P_{\mathbf{D}}^{a b, c d} &=\frac{N}{N^2-4} d^{a b e} d^{c d e},\\
P_{\mathbf{F}}^{a b, c d}&=\frac{f^{a b e} f^{c d e}}{N},\\
P_{\mathbf{T}}^{a b, c d}&=\frac{N^2-4}{4 N^2}\left(\delta^{a c} \delta^{b d}-\delta^{a d} \delta^{b c}\right)-\frac{1}{2 N}\left(d^{a c e} d^{b d e}-d^{a d e} d^{b c e}\right)-\frac{1}{4}\left(d^{b c e} f_{a d e}+d^{a d e} f_{b c c}\right),\\
P_{\mathbf{X}}^{a b, c d}&=\frac{N+2}{4 N}\left(\delta^{a c} \delta^{b d}+\delta^{a d} \delta^{b c}\right)-\frac{N+2}{2 N(N+1)} \delta^{a b} \delta^{c d}+\frac{1}{4}\left(d^{a c c e} d^{b d e}+d^{a d e} d^{b c e}\right)\\&-\frac{N+4}{4(N+2)} d^{a b c} d^{c d e}\,.
\end{aligned}
\label{gluon}
\end{equation}

However, according to discussions in Sec.~\ref{subsec432}, five spin-1 UV states couldn't exist for their leading contribution correspond to the dim-10 EFT operators. Besides, the UV state $ G^{\mu\nu}_{i}G_{j\nu\rho}f_{ijk}S$ corresponding to the quantum number ($Spin=0$, $SU(3)_c=1$) obviously equals to zero in Lagrangian.
This means that based on the J-Basis method, searching UV states by applying the UV selection can obtain the more reasonable bounds.

\subsection{4 Lepton Scattering}
In this case, the involved P-Basis operators can be divided into four categories based on the their symmetry of the corresponding Young-Tableau in Eq.~\ref{4Lop}.
\begin{equation}
\ytableausetup{smalltableaux}
\begin{aligned} & \mathcal{O}_{D^2 L^2 L^{\dagger 2}, 1}^{(p)}=\frac{1}{4} \mathcal{Y}[
\begin{ytableau}
1 & 2 
\end{ytableau},\begin{ytableau}
3 & 4 \end{ytableau}]_p \delta_{i_3}^{i_1} \delta_{i_4}^{i_2}\left(L_{p_1 i_1} L_{p_2 i_2}\right)\left(D^\mu L_{p_3}^{\dagger i_3} D_\mu L_{p_4}^{\dagger i_4}\right)\,, \\ & \mathcal{O}_{D^2 L^2 L^{\dagger 2}, 2}^{(p)}=\frac{1}{4} \mathcal{Y}[
\begin{ytableau}
1 & 2 
\end{ytableau},\begin{ytableau}
3 & 4 \end{ytableau}]_p\delta_{i_3}^{i_1} \delta_{i_4}^{i_2}\left(L_{p_1 i_1} \sigma_{\mu \nu} L_{p_2 i_2}\right)\left(D^\mu L_{p_3}^{\dagger i_3} D^\nu L_{p_4}^{\dagger i_4}\right)\,, \\ & \mathcal{O}_{D^2 L^2 L^{\dagger 2}, 3}^{(p)}=\frac{1}{4} \mathcal{Y}[\begin{ytableau}
1 \\ 2 
\end{ytableau},\begin{ytableau}
3 \\ 4 \end{ytableau}]_p \delta_{i_3}^{i_1} \delta_{i_4}^{i_2}\left(L_{p_1 i_1} L_{p_2 i_2}\right)\left(D^\mu L_{p_3}^{\dagger i_3} D_\mu L_{p_4}^{\dagger i_4}\right)\,, \\ & \mathcal{O}_{D^2 L^2 L^{\dagger 2}\,, 4}^{(p)}=\frac{1}{4} \mathcal{Y}[
\begin{ytableau}
1 \\ 2 
\end{ytableau},\begin{ytableau}
3 \\ 4 \end{ytableau}]_p\delta_{i_3}^{i_1} \delta_{i_4}^{i_2}\left(L_{p_1 i_1} \sigma_{\mu \nu} L_{p_2 i_2}\right)\left(D^\mu L_{p_3}^{\dagger i_3} D^\nu L_{p_4}^{\dagger i_4}\right) \,.\\ & \end{aligned}\label{4Lop}
\end{equation}
Here $p_{i}$ represents the generation of the particle $i$. Thus, the corresponding Young-Tableau gives the corresponding tensor structure of the generation of operators. So the WC space can be defined as 
\begin{equation}
    (\mathcal{C}_{D^2 L^2 L^{\dagger 2}, 1}^{(p)},\mathcal{C}_{D^2 L^2 L^{\dagger 2}, 2}^{(p)},\mathcal{C}_{D^2 L^2 L^{\dagger 2}, 3}^{(p)},\mathcal{C}_{D^2 L^2 L^{\dagger 2}, 4}^{(p)})\,.
    \label{4oL}
    \end{equation}
By applying the J-Basis method in amplitude decomposition, we can obtain Table~\ref{Table:4L} as a possible list of the UV completion.
\begin{table}[H]
\renewcommand\arraystretch{1.5}
    \centering
\begin{tabular}{|c|c|c|c|c|c|} 
\hline State & Spin & $SU(2)_w$/$U(1)_y$ & Interaction & $\vec{c}(p)$ \\ 
\hline $\mathcal{W}_1$ & 1 & $3_1$ & $g_{p_1 p_2} \epsilon^{i_1 m}\left(\tau^I\right)_m^{i_2} \mathcal{W}_1^{\mu I}\left(L_{p_1 i_1} i \stackrel{\leftrightarrow}{D}_\mu L_{p_2 i_2}\right)+h . c$. & $(0,0,0,-4)$ \\ \hline
$\Xi$ & 0 & $3_1$ & $g_{p_1 p_2} \epsilon^{i_1 m}\left(\tau^I\right)_m^{i_2} \Xi^I\left(L_{p_1 i_1} L_{p_2 i_2}\right)+h . c$. & $(-4,0,0,0)$ \\ \hline
$\mathcal{B}_1$ & 1 & $1_1$ & $g_{p_1 p_2} \epsilon^{i_1 i_2} \mathcal{B}_1^\mu\left(L_{p_1 i_1} i \stackrel{\leftrightarrow}{D}_\mu L_{p_2 i_2}\right)+h . c$. & $(0,-4,0,0)$ \\ \hline
$S$ & 0 & $1_1$ & $g_{p_1 p_2} \epsilon^{i_1 i_2} S\left(L_{p_1 i_1} L_{p_2 i_2}\right)+h . c$. & $(0,0,-4,0)$ \\ \hline
$\mathcal{H}$ & 2 & $3_0$ & $g_{p_1 p_3}\left(\tau^I\right)_{i_3}^{i_1} \mathcal{H}^{\mu \nu I}\left(L_{p_1 i_1} i \sigma_\mu \stackrel{\leftrightarrow}{D}_\nu L_{p_3}^{\dagger i_3}\right)$ & $(-5,9,15,-3)$ \\ \hline
$\mathcal{W}_0$ & 1 & $3_0$ & $g_{p_1 p_3}\left(\tau^I\right)_{i_3}^{i_1} \mathcal{W}_0^{\mu I}\left(L_{p_1 i_1} \sigma_\mu L_{p_3}^{\dagger i_3}\right)$ & $(-1,-3,3,1)$ \\ \hline
$\mathcal{G}$ & 2 & $1_0$ & $g_{p_1 p_3} \delta_{i_3}^{i_1} \mathcal{G}^{\mu \nu}\left(L_{p_1 i_1} i \sigma_\mu \stackrel{\leftrightarrow}{D}_\nu L_{p_3}^{\dagger i_3}\right)$ & $(-5,-3,-5,-3)$ \\ \hline
$\mathcal{B}_0$ & 1 & $1_0$ & $g_{p_1 p_3} \delta_{i_3}^{i_1} \mathcal{B}_0^\mu\left(L_{p_1 i_1} \sigma_\mu L_{p_3}^{\dagger i_3}\right)$ & $(-1,1,-1,1)$\\ \hline
\end{tabular}
\caption{UV completion for the 4 lepton scattering. Here the $g_{pi}g_{pj}$ means coupling constant of fermions between different generations $p_i,p_j$.}
\label{Table:4L}
\end{table}

\subsubsection{One Generation}
In this case, the involved operators become degenerate 
\begin{equation}
\mathcal{O}_1=\partial_\mu\left(\bar{l} \gamma_\nu l\right) \partial^\mu\left(\bar{l} \gamma^\nu l\right), \quad \mathcal{O}_2=\partial_\mu\left(\bar{l} \gamma_\nu \tau^I l\right) \partial^\mu\left(\bar{l} \gamma^\nu \tau^I l\right)\,,
\label{1gf}
\end{equation}
because the last two types of operators in Eq.~\ref{4Lop} are eliminated for the Young Diagram's anti-symmetry character of $\mathcal{O}_{D^2 L^2 L^{\dagger 2}, 3}^{(p)},\mathcal{O}_{D^2 L^2 L^{\dagger 2}, 4}^{(p)}$. So we can obtain the positivity bounds as
\begin{equation}
C_1\leq0\,,\quad 9C_1+5C_2\leq0\,.
\label{eq95}
\end{equation}
\begin{figure}
    \centering
   \includegraphics[width=0.6\textwidth]{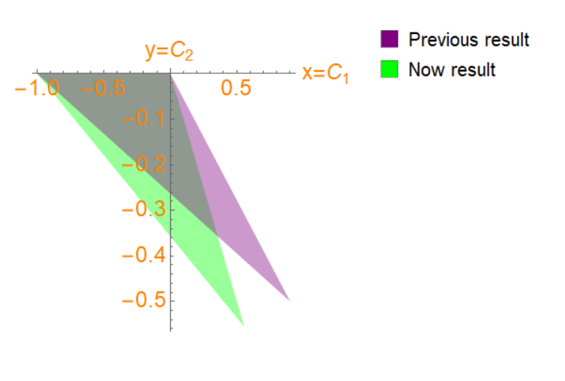}
    \caption{The 2-D cone of the 4 Lepton scattering amplitude in one-generation case.}
    \label{fig:4L1}
\end{figure}
Eq.~\ref{eq95} gives a cone marked by green with external rays respectively representing the UV states $\mathcal{H}$ and $\Xi_1$ in Fig.~\ref{fig:4L1} .
However in Ref.~\cite{Zhang:2021eeo},
it only obtained UV states $\mathcal{B}_1,\mathcal{B},\Xi_1,\mathcal{W}$. Hence, the bounds in Ref.~\cite{Zhang:2021eeo} are $C_1\leq0,~C_1+C_2\leq0$ which shows a looser bounds marked by purple in Fig.~\ref{fig:4L1} than this results in Eq.~\ref{fig:4L1}.
\subsubsection{How to Deal with the Multi-Generation Case}
We need to expand the generation indices of the operators in Eq.~\ref{4oL}, because when we choose different generations $(p_{1}p_{2}p_{3}p_{4})$ in the same type of operators, the coefficients $g_{p_{1},p_{2}},~g_{p_{3},p_{4}}^*$ are different.
We use the UV state $\mathcal{B}_{0}^{\mu}$  as an example to show how to expand generation indices.
For simplification we only consider the lepton coupling with two-generation like
\begin{equation}
\mathcal{L}=g_{1 2} \delta_{i_3}^{i_1} \mathcal{B}_0^\mu\left(L_{1 i_1} \sigma_\mu L_{2}^{\dagger i_3}\right)+g_{21} \delta_{i_3}^{i_1} \mathcal{B}_0^\mu\left(L_{2 i_1} \sigma_\mu L_{1}^{\dagger i_3}\right)\,.
\end{equation}
Next, we use the combination $(p_{i}p_{j}p_{k}p_{l})$ where the index $p_i$ represents the generation of the particle $i$ in the operator to refer to the operators with different generation combinations.
Then based on the permutation group, combination of generation indices $(p_{1}p_{2}p_{3}p_{4})$ can take
$(1212),(1221),(2112),(2121)$. 
For the operator with the type $\mathcal{O}_{D^2 L^2 L^{\dagger 2}, 1}^{(p)}$ or $\mathcal{O}_{D^2 L^2 L^{\dagger 2}, 2}^{(p)}$, we can obtain that $(1212)=(1221)=(2121)=(2112)$.
As for the operators with the form $\mathcal{O}_{D^2 L^2 L^{\dagger 2}, 3}^{(p)}$ or $\mathcal{O}_{D^2 L^2 L^{\dagger 2}, 4}^{(p)}$, $(1212)=-(1221)=(2121)=-(2112)$.

Let's try to write the matching vectors with components of generations tensor $(p_1,p_2,p_3,p_4)$ in the WC space as
\begin{equation}
    \begin{aligned}
&(\mathcal{C}_{D^2 L^2 L^{\dagger 2}, 1}^{(p)(1111)},\mathcal{C}_{D^2 L^2 L^{\dagger 2}, 2}^{(p)(1111)},\mathcal{C}_{D^2 L^2 L^{\dagger 2}, 1}^{(p)(2222)},\mathcal{C}_{D^2 L^2 L^{\dagger 2}, 2}^{(p)(2222)},\mathcal{C}_{D^2 L^2 L^{\dagger 2}, 1}^{(p)(1122)},\mathcal{C}_{D^2 L^2 L^{\dagger 2}, 2}^{(p)(1122)},\\&\mathcal{C}_{D^2 L^2 L^{\dagger 2}, 1}^{(p)(2211)},\mathcal{C}_{D^2 L^2 L^{\dagger 2}, 2}^{(p)(2211)},\mathcal{C}_{D^2 L^2 L^{\dagger 2}, 1}^{(p)(1212)},\mathcal{C}_{D^2 L^2 L^{\dagger 2}, 2}^{(p)(1212)},\mathcal{C}_{D^2 L^2 L^{\dagger 2}, 3}^{(p)(1212)},\mathcal{C}_{D^2 L^2 L^{\dagger 2}, 4}^{(p)(1212)})\,\\&\equiv(C_1,C_2,C_3,C_4,C_5,C_6,C_7,C_8,C_9,C_{10},C_{11},C_{12}).
\end{aligned}\nonumber
\end{equation}
\\
After expanding the generation indices, we could obtain the matching results in Table~\ref{4L2}.
\begin{table}[H]
    \centering
    \begin{tabular}{|c|c|c|c|c|c|c|c|}
    \hline
        (Spin,$SU(2)_w$,$U(1)_y$) & $\vec{c}(p)$ \\ \hline
        (1,3,1) & (0, 0, 0, 0, 0, 0, 0, 0, 0, 0, 0, 0) \\ \hline
        (0,3,1) & $(-1, 0, -y^2, 0, -y, 0, -y, 0, 0, 0, 0, 0)$  \\ \hline
        (1,1,1) & $(0, -1, 0, -y^2, 0, -y, 0, -y, 0, 0, 0, 0)$ \\ \hline
        (0,1,1) & $(0, 0, 0, 0, 0, 0, 0, 0, 0, 0, 0, 0)$ \\ \hline
        (2,3,0) & $(-1, 1, -y^2, y^2, 0, 0, 0, 0, -2 y, 2 y, -2 y, 2 y)$ \\ \hline
        (1,3,0) & $(-5, -3, -5 y^2, -3 y^2, 0, 0, 0, 0, -10 y, -6 y, -10 y, -6 y)$ \\ \hline
        (2,1,0) & $(-1, -3, -y^2, -3 y^2, 0, 0, 0, 0, -2 y, -6 y, 6 y, 2 y)$ \\ \hline
        (1,1,0) & $(-5, 9, -5 y^2, 9 y^2, 0, 0, 0, 0, -10 y, 18 y, 30 y, -6 y)$ \\ \hline
    \end{tabular}
    \caption{Matching results for the form $\mathcal{L}=(g_{12}L_{1}L_{2}^{\dagger}+g_{21}L_{2}L_{1}^{\dagger})X$. Here $X$ is the UV state while we omit the derivative $D^{\mu}$ and $\sigma$ matrix for the simplification of marking.}
    \label{4L2}
\end{table}
We can obtain positivity bounds as
\begin{equation}
\left\{
\begin{aligned}
& C_1+\frac{5C_2}{9}<=0 \,,\\
& C_3+\frac{5C_4}{9}<=0 \,,\\
& \frac{5C_{10}}{9}+2C_5+\frac{10C_7}{9}+C_9 \leq 2 \sqrt{\left[\left(C_1+\frac{5C_2}{9}\right)\left(C_3+\frac{5C_4}{9}\right)\right]}\,.
\end{aligned}\right.
\end{equation}
\subsubsection{The full Flavor Case}
Considering two-generation of fermion, the UV Lagrangian can be written as the form
\begin{equation}
\mathcal{L}=(g_{12}L_{1}L_{2}^{\dagger}+g_{21}L_{2}L_{1}^{\dagger}+g_{11}L_{1}L_{1}^{\dagger}+g_{22}L_{2}L_{2}^{\dagger})X\,. \nonumber
\end{equation}
Here $X$ is the UV state while we omit the derivative $D^{\mu}$, $\sigma$ matrix and other indices for the convenience of marking.
the WC space with the tensor indexed by generations and types of Lorentz structure can be defined as follows,
\begin{equation}
\begin{aligned}
&((1111)_{1},(1111)_{2},(2222)_{1},(2222)_{2},(1112)_{1},(1112)_{2},(1122)_{1},\\&(1122)_{2},(1222)_{1},(1222)_{2},(1212)_{1},(1212)_{2},(1212)_{3},(1212)_{4})\\&\equiv(C_1,C_2,C_3,C_4,C_5,C_6,C_7,C_8,C_9,C_{10},C_{11},C_{12},C_{13},C_{14})\,. \nonumber
\end{aligned}
\end{equation}
Here $p_1,p_2,p_3,p_4$ in $(p_{1}p_{2}p_{3}p_{4})_e$ represent particles' generations of the operator while $e$ represents the serial number in Eq.~\ref{4Lop} which stand for the Lorentz structure and Young-Tableau form of the operator.

The matching results are shown in Table~\ref{4L:2}. The corresponding cone parametrized by the ratios of couplings between different generations $\frac{g_{22}}{g_{11}}=x$ and $\frac{g_{12}}{g_{11}}=y$, exists in a 14-D space. It's very hard to obtain its analytical solutions. However, some analytical constraints can be obtained in some cases like the Minimal Flavor Violation (MFV).
\begin {table}[H]
    \centering
    \begin {tabular} {|c| c|c |c|c|c|c|c|}
    \hline
(Spin, $SU(2)_w$, $U(1)_y$) & $\vec{c}(p)$ \\\hline
(1, 3, 1) & $ (0, 0, 0, 0, 0, 0, 0, 0, 0, 0, 0, 0, 0, -4) $ \\\hline
(0, 3, 1) & $ (-1, 0, -y^2, 0, -2 x, 0, -xy, 0, -2 x y, 0, -4 x^2, 0, 0, 0) $ \\\hline
(1, 1, 1) & $(0, -1, 0, -y^2, 0, -2 x, 0, -xy, 0, -2 x y, 0, -4 x^2, 0, 0)$ \\\hline
(0, 1, 1) & $ (0, 0, 0, 0, 0, 0, 0, 0, 0, 0, 0, 0, -4, 0) $ \\\hline
(2, 3, 0) & $ \begin{aligned}
    &(-5, 9, -5 y^2, 9 y^2, -10 x, 18 x, -5 x^2, 9 x^2, -10 x y, 18 x y,\\ &-5 
(2 x^2 + 2 y), 9 (2 x^2 + 2 y), 15 (-2 x^2 + 2 y), -3 (-2 x^2 + 2 y))\end{aligned}$ \\\hline
(1, 3, 0) & $ \begin{aligned}&(-1, -3, -y^2, -3 y^2, -2 x, -6 x, -x^2, -3 x^2, -2 x y, -6 x y, -2 \\&
x^2 - 2 y, -3 (2 x^2 + 2 y), 3 (-2 x^2 + 2 y), -2 x^2 + 2 y )\end{aligned}$ \
\\\hline
(2, 1, 0) & $\begin{aligned}& (-5, -3, -5 y^2, -3 y^2, -10 x, -6 x, -5 x^2, -3 x^2, -10 x y, -6 x y, \\&
-5 (2 x^2 + 2 y), -3 (2 x^2 + 2 y), -5 (-2 x^2 + 2 y), -3 (-2 x^2 + 
   2 y)) \end{aligned}$ \
\\\hline
(1, 1, 0) & $\begin{aligned}& (-1, 1, -y^2, y^2, -2 x, 2 x, -x^2, x^2, -2 x y, 2 x y,\\& -2 x^2 - 2 y, 
 2 x^2 + 2 y, 2 x^2 - 2 y, -2 x^2 + 2 y)\end{aligned}$ \\\hline
    \end {tabular}\caption{Matching results for the full flavor case of the 4 Lepton scattering involving two generations.}
    \label{4L:2}
\end {table}

\subsubsection{The MFV Case}
MFV is represented that all the flavor violation is generated from Yukawa coupling terms with the form $\mathbb{Y}_{ij}\mathbf{L}_{i}\mathbf{L^{\dagger}}_{j}$ and only EFT operators which are Yukawa singlet can exist
\cite{DAmbrosio:2002vsn,Isidori:2012ts,Bonnefoy:2020yee}.
It give strong constraints both on the EFT and the UV theory. First considering the UV lepton sector in Table~\ref{Table:4L}, the Yukawa matrix is an identity matrix. So it excludes the first four coupling terms with the form $\mathbf{L}_{i}\mathbf{L}_{j}$ so that we need only consider the term of $\mathbf{L}_{i}\mathbf{L^{\dagger}}_{i}$ coupling in Table~\ref{MFV}. Hence, we need to find all operators whose tensors of generation indices $(p_1p_2p_3p_4)$ are singlet to obtain involved operators. For two-generation cases, singlet tensor combinations are
$(1111), (2222), (1212)$.
By defining the WC space as 
\begin{equation}
    \begin{aligned}
&(\mathcal{C}_{D^2 L^2 L^{\dagger 2}, 1}^{(p)(1111)},\mathcal{C}_{D^2 L^2 L^{\dagger 2}, 2}^{(p)(1111)},\mathcal{C}_{D^2 L^2 L^{\dagger 2}, 1}^{(p)(2222)},\mathcal{C}_{D^2 L^2 L^{\dagger 2}, 2}^{(p)(2222)},\mathcal{C}_{D^2 L^2 L^{\dagger 2}, 1}^{(p)(1212)},\mathcal{C}_{D^2 L^2 L^{\dagger 2}, 2}^{(p)(1212)},\mathcal{C}_{D^2 L^2 L^{\dagger 2}, 3}^{(p)(1212)},\mathcal{C}_{D^2 L^2 L^{\dagger 2}, 4}^{(p)(1212)})\\&\equiv(C_1,C_2,C_3,C_4,C_5,C_6,C_7,C_8)\,,
 \end{aligned}\nonumber
\end{equation}
and the previous UV selection, Table~\ref{4L:2} can be reduced to Table~\ref{MFV}.
\begin{table}[ht]
\centering
\begin{tabular}{|c|c|c|c|} 
\hline State & Spin & $SU(2)_w$/$U(1)_y$ &  $\vec{c}(p)$ \\
\hline 
$\mathcal{H}$ & 2 & $3_0$  & $(-5, 9, -5 y^2, 9 y^2, -10 y, 18 y, 30 y, -6 y)$ \\ \hline
$\mathcal{W}_0$ & 1 & $3_0$ & $(-1, -3, -y^2, -3 y^2, -2 y, -6 y, 6 y, 2 y)$ \\ \hline
$\mathcal{G}$ & 2 & $1_0$  & $(-5, -3, -5 y^2, -3 y^2, -10 y, -6 y, -10 y, -6 y)$ \\ \hline
$\mathcal{B}_0$ & 1 & $1_0$  & $(-1, 1, -y^2, y^2, -2 y, 2 y, -2 y, 2 y)$\\ \hline
\end{tabular}
\caption{Matching results for the full flavor case of the 4 Lepton scattering in the MFV case. Here $y=\frac{g_{22}}{g_{11}}$.}
\label{MFV}
\end{table}

According to Table~\ref{MFV}, the corresponding normal vectors can be simply obtained,
\begin{equation}
    \begin{aligned}
       & (y_1^2, \frac{5 y_1^2}{9}, 1, \frac{5}{9}, -y_1, -\frac{5 y_1}{9}, 0, 0)\,,\\
        & (y_1^2, -\frac{y_1^2}{3}, 1, -\frac{1}{3}, -y_1, \frac{y_1}{3}, 0, 0)\,.
    \end{aligned}
\end{equation}
By using Eq.~\ref{expand0}, Eq.~\ref{expand1}, Eq.~\ref{expand2}, we obtain the positivity bounds for the $4$ Lepton scattering with the two-generation under the MFV assumption:
\begin{equation}
\left\{
    \begin{aligned}
& -C_{3}-\frac{5C_4}{9} \geq 0 \,,\\
& -C_1-\frac{5C_2}{9} \geq 0 \,,\\
& -4\left(-C_1-\frac{5C_2}{9}\right)\left(-C_3-\frac{5C_4}{9}\right)+\left(C_5+\frac{5C_6}{9}\right)^2 \leq 0 \,,\\
& -C_1+\frac{C_2}{3} \geq 0 \,,\\
& -C_3+\frac{C_4}{3} \geq 0 \,,\\
& \left(C_5-\frac{C_6}{3}\right) ^2-4\left(-C_1+\frac{C_2}{3}\right)\left(-C_3+\frac{C_4}{3}\right) \leq 0\,.
\end{aligned}\right.
\end{equation}
The allowed area's volume of the WC space is 0.974\% by Monte Carlo sampling in the WC space.

For the three-generation condition, the amplitude cone is an 18-D cone with curved surfaces parameterized by $x=\frac{g_{22}}{g_{11}}$ and $y=\frac{g_{33}}{g_{11}}$. The matching results are listed in Table~\ref{num1}.
\begin{table}[ht]
\centering
    \begin{tabular}{|c|c|c|c|c|} 
\hline State & Spin & $SU(2)_w$/$U(1)_y$ &  $\vec{c}(p)$ \\ \hline 
$\mathcal{H}$ & 2 & $3_0$  & $\begin{aligned}&(-5, 9, -5 x^2, 9 x^2, -5 y^2, 9 y^2, -10 x, 18 x, 30 x, -6 x, -10 y, \\&
18 y, 30 y, -6 y, -10 x y, 18 x y, 30 x y, -6 x y)\end{aligned} $\\ \hline 
$\mathcal{W}_0$ & 1 & $3_0$ &  $\begin{aligned}&(-1, -3, -x^2, -3 x^2, -y^2, -3 y^2, -2 x, -6 x, 6 x, 2 x, -2 y, -6 y,\\& 
6 y, 2 y, -2 x y, -6 x y, 6 x y, 2 x y) \end{aligned}$\\ \hline 
$\mathcal{G}$ & 2 & $1_0$  & $\begin{aligned}&(-5, -3, -5 x^2, -3 x^2, -5 y^2, -3 y^2, -10 x, -6 x, -10 x, -6 x, -10 \\&
y, -6 y, -10 y, -6 y, -10 x y, -6 x y, -10 x y, -6 x y)\end{aligned} $\\ \hline 
$\mathcal{B}_0$ & 1 & $1_0$  & $\begin{aligned}&(-1, 1, -x^2, x^2, -y^2, y^2, -2 x, 2 x, -2 x, 2 x, -2 y, 2 y, -2 y,\\ &2 
y, -2 x y, 2 x y, -2 x y, 2 x y)\end{aligned}$\\ \hline 
\end{tabular}
\caption{Matching results for three generation $4L$ scattering in the MFV case.}
\label{num1}
\end{table}
Despite it being complicated, numerical solutions can be obtained by applying the particle data of SM in Ref.~\cite{ParticleDataGroup:2022pth}:
\begin{equation}
\left\{
    \begin{aligned}
       & C_1 \geq0,\  C_2\geq0,\  33.634 C_3 \geq C_{15},\  33.634 C_4 \geq C_{16},\ 0.1189 C_5\geq C_{15},\ \\ &0.238C_6 \geq C_{16}, \ 3477.22 C_7 \geq C_{15},\  3477.22 C_9\geq C_{17},\  C_{10} \geq 0.000288 C_{18}, \\&   C_{11} \geq 0.00484C_{15},\ C_{12} \geq 0.00484 C_{16},\  C_{13} \geq 0.00484 C_{17}, C_{14} \geq 0.00484 C_{18}\,.
    \end{aligned}\right.
\end{equation}
\subsection{2-to-2 Scattering involving \texorpdfstring{$W$}~ and \texorpdfstring{$B$}~ in the CP-Conservation Case}
For convenience, we only consider for the CP-conserving case. The operators involved in this scattering process are listed as follows,
\begin{equation}
\begin{aligned}
& \mathcal{O}_{W_{\mathrm{L}, 1}^4}^{(p)} =W_{\mathrm{L} \mu \nu}^I W_{\mathrm{L} \lambda \rho}^J W_{\mathrm{L}}^{I \nu \rho} W_{\mathrm{L}}^{J \lambda \mu}, \quad \mathcal{O}_{W_{\mathrm{L}}^4, 2}^{(p)}=W_{\mathrm{L} \mu \nu}^I W_{\mathrm{L}}^{J \mu \nu} W_{\mathrm{L} \lambda \rho}^I W_{\mathrm{L}}^{J \lambda \rho}, \\
&\mathcal{O}_{B_{\mathrm{L}, 1}^4}^{(p)} =B_{\mathrm{L} \mu \nu} B_{\mathrm{L} \lambda \rho} B_{\mathrm{L}}^{ \nu \rho} B_{\mathrm{L}}^{ \lambda \mu} ,\quad
 \mathcal{O}_{W_{\mathrm{L}}^2 B_L^2, 1}^{(p)}=W_{\mathrm{L} \mu \nu}^I W_{\mathrm{L} \lambda \rho}^I B_L^{I \nu \rho} B_{\mathrm{L}}^{ \lambda \mu}, \\&  \mathcal{O}_{W_{\mathrm{L}}^2 B_L^2, 2}^{(p)}=W_{\mathrm{L} \mu \nu}^I W_{\mathrm{L}}^{I \mu \nu } B_{\mathrm{L} }^{ \rho \lambda} B_{\mathrm{L} \rho \lambda} \,.
\end{aligned}
\end{equation}
Now we need to give the accurate UV completion by the UV selection. According to discussion in Sec.~\ref{WWWW}, for the 2-to-2 scattering involving only $B$, it only have two UV terms with $(Spin,SU(3)_c,SU(2)_w,Y)=(0,1,1,0)$ and $(2,1,1,0)$. Then for $W$, $(0,1,5,0)$ and $(0,1,1,0)$ states are left while only $(0,1,1,0)$ is left for $B$ . So we should only consider degeneracy of $(0,1,1,0)$ between $W$ and $B$.
Finally, we can consider UV terms with the form $WBX$ where the Lorentz indices are omitted for simplification of marking. According to Sec.~\ref{WWWW}, the spin-2 term  is also eliminated by the EOMs, while the spin-1 term contributes to dim-10 operators in the lead order. The WC space can be defined as
\begin{equation}
(\mathcal{C}_{W_{\mathrm{L}, 1}^4}^{(p)},\mathcal{C}_{W_{\mathrm{L}}^4, 2}^{(p)},\mathcal{C}_{B_{\mathrm{L}, 1}^4}^{(p)}, \mathcal{C}_{W_{\mathrm{L}}^2 B_L^2, 1}^{p}, \mathcal{C}_{W_{\mathrm{L}}^2 B_L^2, 2}^{p})\equiv(C_1,C_2,C_3,C_4,C_5)\,.
\nonumber
\end{equation}
So we obtain the matching results in Table~\ref{4L2}.
\begin{table}[!ht]
    \centering
    \begin{tabular}{|c|c|c|}
    \hline
(Spin,$SU(3)_c$,$SU(2)_w$,$U(1)_y$) & Interaction Lagrangian & $\vec{c_{p}}$ \\ \hline
        (0,1,5,0) & $W_{\mathrm{L} \mu \nu}^I W_{\mathrm{L}}^{J \mu \nu}\left(T^{A}\right)^{I J} \mathcal{S}^{A}$ & $(4,3,0,0,0)$ \\ \hline
        (0,1,1,0) & $W_{\mathrm{L} \mu \nu}^I W_{\mathrm{L}}^{I \mu \nu} S+x_{1}B_{\mathrm{L} \mu \nu} B_{\mathrm{L}}^{\mu \nu} S$ & $(-2,0,-2x_1^2,0,2x_1)$ \\ \hline
        (0,1,3,0) &$W_{\mathrm{L} \mu \nu}^I B_{\mathrm{L}}^{\mu \nu} S^{I}$  & (0,0,0,0,0)\\ \hline
    \end{tabular}
    \caption{Matching results for the 2-to-2 scattering involving $W$ and $B$. Here we take the slice of the cone as $x=-C_{1}+\frac{4}{3}C_{2}$, $y=-C_{3}$ and $z=-C_{4}$.}
\end{table}
Here the matching result of quantum number $(0,1,3,0)$ is $\mathcal{W}_{I\mu\nu}\mathcal{B}^{\mu\nu}\mathcal{W}_{I\lambda\rho}\mathcal{B}^{\lambda\rho}$ in the m-Basis. However, it is eliminated by the repeat field when it's between the m-Basis and the P-Basis. 

\begin{figure}
    \centering
    \includegraphics[scale=0.6]{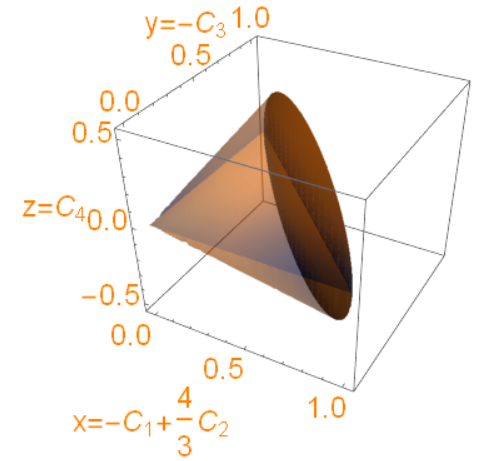}
    \caption{The cone of the 2-to-2 scattering involving $W$ and $B$ boson.}
    \label{fig:W+B}
\end{figure}

After calculation, we obtain the positivity bounds as
\begin{equation}
\left\{
    \begin{aligned}
        &-C_1+\frac{4C_2}{3}\geq0\,,\\
        &-C_3\geq0\,,\\
        &-\sqrt{-C_3\times(-C_1+\frac{4C_2}{3})}\leq C_4\leq\sqrt{-C_3\times(-C_1+\frac{4C_2}{3})}\,.
    \end{aligned}\right.
    \label{1B}
\end{equation}
the slice of the 2D slices of the 3D cone is plotted in Fig.~\ref{fig:W+B}. The last bound in Eq.~\ref{1B} is represented by the circle in Fig.~\ref{fig:W+B} which corresponds to the UV state $(0,1,1,0)$.
\subsection{2-to-2 Scattering involving \texorpdfstring{$W$}~ and Higgs}

For simplification, we limit the involved particles to the $W_{L}$ and the $H$. Considering that $W_{L}$ and $H$ have been discussed separately, in the step of the UV selection, we only need to consider the J-Basis decompositions for type $D^2 HH^{\dagger}W_{L}W_{L}$. In Table~\ref{WH}, we give the possible UV states corresponding to the amplitudes decompositions. First, considering for coupling terms like $W_{L} W_{L}X+xHH^{\dagger}X$ where the indices of the Lorentz and the gauge groups are omitted for the convenience of marking, the $x$ means the coupling constant describing the degeneracy between $W_{L} W_{L}X$ and $BBX$ in the same quantum number, while $X$ mean the heavy state. We have already know the $WWV$ term is impossible in tree level in Sec.~\ref{subsec42}. As a result, we can confirm that the degeneracy coupling only exists in $(0,1,1,0)$. Then 
according to Ref.~\cite{Li:2020gnx},
operators involved $W_L H$ in the P-Basis are
\begin{equation}
\begin{aligned}
&\mathcal{O}_{W_{L}H}^1 = \frac{1}{2} \mathcal{Y} [\begin{ytableau}
p & r \end{ytableau}] W_{Lp}^{I \lambda \nu} W_{Lr}^{I V \lambda}\left(D_\mu H_i\right)\left(D^\mu H^{\dagger^i}\right) \,,\\
&\mathcal{O}_{W_{L}H}^2 =\frac{1}{2} \mathcal{Y} [\begin{ytableau}
p & r \end{ytableau}] \epsilon^{I J K} \tau^{K^i}{ }_j W_{Lp}^{I \lambda \nu} W_{Lr}^{J \lambda \mu}\left(D_\mu H_i\right)\left(D_\nu H^{\dagger j}\right)\,.
\end{aligned}
\label{WH1}
\end{equation}
The matching results of the UV terms with the form $W_{L}HX$ are listed in Table~\ref{WH}.
\begin{table}[!ht]
    \centering
    \begin{tabular}{|c|c|c|}
    \hline
(Spin,$SU(3)_c$,$SU(2)_w$,$U(1)_y$) & Interaction Lagrangian & $\vec{c}(p)$ \\ \hline
        (1,1,4,$\frac{1}{2}$) & $\varepsilon_{i j}\left(\tau^I\right)_i^l W_{L \mu \nu}^I D^\mu H_k V^{i j k}_\nu$ & $(-4I,8)$ \\ \hline
        (1,1,2,$\frac{1}{2}$) & $\tau_i^{I j} W_{L \mu \nu}^I D^\mu H_j V^{\nu i} $ & $(-2I,-8)$ \\ \hline
    \end{tabular}
     \caption{The amplitude's decompositions of the 2-to-2 scattering in the channel $WH\rightarrow WH$  in the P-Basis of Eq.~\ref{WH1}.}
     \label{WH}
\end{table}
Despite that the Table~\ref{WH} give the complex solutions, $\mathcal{O}_{W_{L}H}^2$ has no contribution to the matching results of $W_{L} W_{L}X+xHH^{\dagger}X$, which means we can exclude $C_{W_{L}H}^2$ from the WC space and obtain the real matching results in the WC space. So the WC space can be defined as
\begin{equation}
\begin{aligned}
   & \left(\mathcal{C}_{W_{\mathrm{L}, 1}^4}^{(p)},\mathcal{C}_{W_{\mathrm{L}, 2}^4}^{(p)},C_{H^4}^{(2)},C_{H^4}^{(1)},C_{H^4}^{(3)},\mathcal{C}_{W_{L}H}^1 \right)\equiv\left(C_{1},C_{2},C_{3},C_{4},C_{5},C_{6}\right) \,,
   \end{aligned}
   \nonumber
\end{equation}
Finally we can obtain full matching results in Table~\ref{WH2}.
\begin{table}[ht]
    \centering
    \begin{tabular}{|c|c|}
    \hline
(Spin,$SU(3)_c$,$SU(2)_w$,$U(1)_y$) & $\vec{c}(p)$ \\ \hline
        $(0,1,5,0)$&$\left(4,3,0,0,0,0\right)$\\ \hline
         $(0,1,1,0)$&$\left(-2,0,0,0,x^2,-2x\right)$\\ \hline
         $(2,1,3,1)$&$\left(0,0,3,-2,3,0\right)$\\ \hline
         $(0,1,3,1)$  &$\left(0,0,0,1,0,0\right)$\\ \hline
         $(1,1,1,1)$&$\left(0,0,1,0,-1,0\right)$\\ \hline
         $(2,1,3,0)$&$\left(0,0,-7,3,8,0\right)$\\ \hline
         $(0,1,1,0)$&$\left(0,0,1,1,-2,0\right)$\\ \hline
         $(0,1,3,0)$  &$\left(0,0,2,0,-1,0\right)$\\ \hline
         $(2,1,1,0)$  &$\left(0,0,3,3,-2,0\right)$\\ \hline
         $(1,1,1,0)$  &$\left(0,0,-1,1,0,0\right)$\\ \hline
         $(1,1,4,\frac{1}{2})/(1,1,2,\frac{1}{2})$ &$\left(0,0,0,0,0,-1\right)$\\ \hline
    \end{tabular}
     \caption{Matching results for the 2-to-2 scattering involving $W$ and $H$.}
     \label{WH2}
\end{table}

We obtain the six normal vectors of the cone spanned by the matching results in Table~\ref{WH2} as
\begin{equation}
\begin{aligned}
 &\left(0,1,0,0,0,0\right) , \left(-1,\frac{4}{3},0,0,0,,0\right)   ,\left(-\frac{x^2}{2},\frac{2x^2}{3},1,1,1,x\right) \left(x\leq0\right), \\& \left(-\frac{x^2}{2},\frac{2x^2}{3},1,3,1,x\right) \left(x\leq0\right),\left(-\frac{x^2}{4},\frac{x^2}{3},1,1,\frac{1}{2},\frac{x}{2}\right) \left(x\leq0\right) ,\\& \left(-\frac{x^2}{10},\frac{2x^2}{15},1,\frac{9}{5},\frac{1}{5},\frac{x}{5}\right) \left(x\leq0\right),
\end{aligned}
\end{equation}
we can clearly see that the first and the second normal vectors offer the positivity bounds for the $4W_L$ scattering. When $x$ goes to negative infinity, $W_{L}$ and $H$ would decouple and the last four normal vectors offer bounds for the $4H$ scattering case, similarly in $W$ and quark scattering case.
The total positivity bounds are listed as follows,
\begin{equation}
\left\{
\begin{aligned}
   & C_{2}\geq0, -C_{1}+\frac{4}{3}C_{2}\geq0 \quad\text{(only $4W_{L}$ scattering)}\,,\\
   &C_{3}+C_{4}+C_{5}\geq0,\, C_{3}+3C_{4}+C_{5}\geq0,\, 2C_{3}+2C_{4}+C_{5}\geq0\quad\text{(only $4H$ scattering)}\,,\\&5C_{3}+9C_{4}+C_{5}\geq0,\,|C_5|\leq2\sqrt{\left(C_{3}+C_{4}+C_{5}\right)\left(-\frac{C_{1}}{2}+\frac{2C_{2}}{3}\right)}\,,\\&|C_5|\leq2\sqrt{\left(C_{3}+3C_{4}+C_{5}\right)\left(-\frac{C_{1}}{2}+\frac{2C_{2}}{3}\right)},\\&|C_5|\leq4\sqrt{\left(C_{3}+C_{4}+\frac{C_{5}}{2}\right)\left(-\frac{C_{1}}{4}+\frac{C_{2}}{3}\right)}\,,\\&|C_5|\leq2\sqrt{\left(5C_{3}+9C_{4}+C_{5}\right)\left(-\frac{C_{1}}{2}+\frac{2C_{2}}{3}\right)}\,.
\end{aligned}\right.
\end{equation}
Similarly, the bounds in the first and the second lines can be directly obtained from scattering process involving the same particle while the other bounds represent the degeneracy between $W$ and $H$ particles.
\subsection{2-to-2 Scattering involving \texorpdfstring{$W$}~ and Quark}
For convenience, we only consider $W_{\mathrm{L}}$, $W_{\mathrm{R}}$ and one generation quark. The involved operators in the P-Basis are
\begin{equation}
    \begin{aligned}
    &\mathcal{O}_{W_{\mathrm{L}, 1}^4}^{(p)} =W_{\mathrm{L} \mu \nu}^I W_{\mathrm{L} \lambda \rho}^J W_{\mathrm{L}}^{I \nu \rho} W_{\mathrm{L}}^{J \lambda \mu}, \quad \mathcal{O}_{W_{\mathrm{L}, 1}^4}^{(p)}=W_{\mathrm{L} \mu \nu}^I W_{\mathrm{L}}^{J \mu \nu} W_{\mathrm{L} \lambda \rho}^I W_{\mathrm{L}}^{J \lambda \rho}\,,\\
    & \mathcal{O}_{W_{\mathrm{L}}^2 W_{\mathrm{R}}^2, 1}^{(p)}=W_{\mathrm{L} \mu \nu}^I W_{\mathrm{L} \lambda \rho}^J W_{\mathrm{R}}^{I \nu \rho} W_{\mathrm{R}}^{J \lambda \mu}, \quad \mathcal{O}_{W_{\mathrm{L}}^2 W_{\mathrm{R}}^2, 2}^{(p)}=W_{\mathrm{L} \mu \nu}^I W_{\mathrm{L} \lambda \rho}^I W_{\mathrm{R}}^{J \nu \rho} W_{\mathrm{R}}^{J \lambda \mu}\,, \\
& \mathcal{O}_{W_{\mathrm{R}}^4, 1}^{(p)}=W_{\mathrm{R} \mu \nu}^I W_{\mathrm{R} \lambda \rho}^J W_{\mathrm{R}}^{I \nu \rho} W_{\mathrm{R}}^{J \lambda \mu}, \quad \mathcal{O}_{W_{\mathrm{R}}^4, 2}^{(p)}=W_{\mathrm{R} \mu \nu}^I W_{\mathrm{R}}^{J \mu \nu} W_{\mathrm{R} \lambda \rho}^I W_{\mathrm{R}}^{J \lambda \rho}\,,\\
  & \mathcal{O}_{Q_{1}^4}^{(p)}=\left(D_\mu Q_s^{\dagger a i} D_\nu Q_t^{\dagger b j}\right)\left(Q_{p a i} \sigma^{\mu \nu} Q_{r b j}\right),
  \mathcal{O}_{Q_{2}^4}^{(p)}=\left(D_\mu Q_s^{\dagger a j} D_\nu Q_t^{\dagger b i}\right)\left(Q_{p a i} \sigma^{\mu \nu} Q_{r b j}\right)\,,\\
   &\mathcal{O}_{Q_{3}^4}^{(p)}=\left(Q_{p a i} Q_{r b j}\right)\left(D_\mu Q_s^{\dagger a i} D^\mu Q_t^{\dagger b j}\right),\mathcal{O}_{Q_{3}^4}^{(p)}=\left(Q_{p a i} Q_{r b j}\right)\left(D_\mu Q_s^{\dagger a j} D^\mu Q_t^{\dagger b i}\right)\,,\\&\mathcal{O}_{W_{L}W_{R} Q^2,1}^{(p)}=i \epsilon^{I J K} {W_{L}}_\nu^{I \mu}{W_{R}}_\lambda^{J \nu}\left(Q_{p a i} \sigma^\lambda\left(\tau^K\right)_j^i \overleftrightarrow{D}_\mu Q_r^{\dagger a j}\right), \\&\mathcal{O}_{W_{L}W_{R} Q^2,2}^{(p)}=i W_{L\nu}^{I \mu} {W_{R}}^{I \nu}{ }_\lambda\left(Q_{p a i} \sigma^\lambda \overleftrightarrow{D}_\mu Q_r^{\dagger a i}\right)\,.
    \end{aligned}
\end{equation}
So the WC space can be defined as 
\begin{equation}
\begin{aligned}
    (&\mathcal{C}_{W_{\mathrm{L}, 1}^4}^{(p)},\mathcal{C}_{W_{\mathrm{L}, 1}^4}^{(p)},\mathcal{C}_{W_{\mathrm{L}}^2 W_{\mathrm{R}}^2, 1}^{(p)},\mathcal{C}_{W_{\mathrm{L}}^2 W_{\mathrm{R}}^2, 2}^{(p)},\mathcal{C}_{W_{\mathrm{R}}^4, 1}^{(p)},\mathcal{C}_{W_{\mathrm{R}}^4, 2}^{(p)}, \mathcal{C}_{Q_{1}^4}^{(p)}, \mathcal{C}_{Q_{2}^4}^{(p)}, \mathcal{C}_{Q_{3}^4}^{(p)}, \mathcal{C}_{Q_{4}^4}^{(p)},\mathcal{C}_{W_{L}W_{R} Q^2,1}^{(p)},\mathcal{C}_{W_{L}W_{R} Q^2,2}^{(p)})\\
    &\equiv(C_{1},C_{2},C_{3},C_{4},C_{5},C_{6},C_{7},C_{8},C_{9},C_{10},C_{11},C_{12})\,.
    \end{aligned}
    \nonumber
\end{equation}

The J-Basis analysis for the $4Q$ scattering is listed in Table~\ref{quark}. Given the possible UV resonances from the J-Basis, we select the UV completion by the following steps. First, by assuming that UV states are color singlets to exclude coupling terms with the form $QQX$ where  the indices of the Lorentz and gauge groups are omitted for simplification of marking. We could only discuss the weak sector. As the UV vector boson coupling term in the $\left\{Q_{\dagger}Q\right\} \left\{W_{L}W_{L}\right\}$ channel cannot exist because the $WWV$ coupling's contribution starts at the dim-10 operators at the tree level which are discussed in Sec.~\ref{WWWW}. So we can obtain the conclusion that the degeneracy of $\bar{Q}Q$ and $WW$ exists in the Irreps with quantum numbers $(2,1,3,0)$ and $(2,1,1,0)$.
\begin{table}[ht]
\centering
\renewcommand\arraystretch{1.2}
\begin{tabular}{|c|c|c|}
\hline \multicolumn{3}{|c|}{ group: (Spin, $\left.S U(3)_c, S U(2)_L, U(1)_y\right)$} \\
\hline$\left\{Q_1, Q_3^{\dagger}\right\},\left\{Q_2, Q_4^{\dagger}\right\}$ & $\mathcal{O}_j^{(m)}$ & $\mathcal{O}_j^{(p)}$ \\
\hline$(2,1,3,0)$ & $(5, 3, -10, -6, 0, 0, 0, 0)$ & (-10, 6, 5, -3) \\
\hline$(1,1,3,0)$ & $(-1, 1, 2, -2, 0, 0, 0, 0)$ &  (2, 2, -1, -1)\\
\hline $(2.1.1.0)$ & $(-5, -3, 0, 0, 0, 0, 0, 0)$ &  (0, 0, -5, 3)\\
\hline $(1,1,1,0)$ & $(1, -1, 0, 0, 0, 0, 0, 0)$ &  (0, 0, 1, 1)\\\hline
\end{tabular}
\caption{J-Basis analysis results for the $4Q$ scattering. Here the P-Basis $\mathcal{O}_j^{(p)}$ are ($\mathcal{O}_{Q_{1}^4}^{(p)},\mathcal{O}_{Q_{2}^4}^{(p)},\mathcal{O}_{Q_{3}^4}^{(p)},\mathcal{O}_{Q_{4}^4}^{(p)}$).}
\label{quark}
\end{table}

The matching results are listed in Table~\ref{WQ},
\begin{table}[!ht]
    \centering
    \begin{tabular}{|c|c|}
    \hline
(Spin,$SU(3)_c$,$SU(2)_w$,$U(1)_y$) & \textbf{Matching result in P-Basis} \\ \hline
        (0,1,5,0) &  $(4, 3, -12 x_1, 4 x_1, 4 x_1^2, 3 x_1^2, 0, 0, 0, 0, 0, 0)$ \\ \hline
        (0,1,1,0) &  $(-2, 0, 0, -4 x_2, -2 x_2^2, 0, 0, 0, 0, 0, 0, 0)$ \\ \hline
        (2,1,5,0)  & (0, 0, -1, -3, 0, 0, 0, 0, 0, 0, 0, 0)\\ \hline
         (2,1,3,0)  & $(0, 0, -1, 1, 0, 0, -10 x_3^2, 6 x_3^2, 5 x_3^2, -3 x_3^2, -4 x_3, 0)$\\ \hline
          (2,1,1,0)  & $(0, 0, -1, 0, 0, 0, 0, 0, -5 x_4^2, 3x_4^2, 0, -4 x_4)$\\ \hline
           (1,1,3,0)  & (0, 0, 0, 0, 0, 0, 2, 2, -1, -1, 0, 0)\\ \hline
           (1,1,1,0)  & (0, 0, 0, 0, 0, 0, 0, 0, 1, 1, 0, 0)\\ \hline
    \end{tabular}
     \caption{Matching results for the 2-to-2 scattering involving $W$ and $Q$.}
     \label{WQ}
\end{table}
where we consider that for decompositions with different quantum numbers, the coupling constants $x_i$ of degeneracy between $WWX$ and $QQX$ are different. In the Table~\ref{WQ}, the matching results show that the corresponding cone has curved surface parametrized by four parameters $x_1,x_2,x_3,x_4$ which match to the degeneracy between $W$ and quark. Hence,
obtaining the positivity bounds equals solving a hard quaternion quadratic polynomials problem so only numerical solutions can be obtained.

\section{Summary and Discussion}
\paragraph{Positivity Bounds} Positivity bounds for the EFT operators involving  2-to-2 scattering can be transformed into geometry problems: every UV state contributing the EFT operators corresponds to the possible external ray that form the cone in the WC space of the EFT operators. It means that the more complete UV states we find, the more accurate shape of the cone we can acquire so as to obtain the exact bounds for the WCs. Previously, using the projection method based on the CG coefficients to represent UV or enumerating all possible UV states either provide redundant UV states or omit some UV states so as to obtain a not so strict constraint. Among the results obtained previously, the bounds of the $4W$ scattering show a significant difference. 
\paragraph{the J-Basis method and the UV selection}
We introduce the J-Basis method in Sec.~\ref{sec2}. In fact, the J-Basis takes the Lorentz structure into consideration to provide direct product decompositions of the spin structure and uses the Casimir Operators to give decompositions of gauge structure.
Then according to a quantum number of decompositions, all possible UV Lagrangian in tree-level can be written. After that, we need to process the UV selection to check whether its contribution to tree-level matching is eliminated by the EOMs, the repeat field and other redundancy or not, to give an accurate UV completion. 
We apply the J-Basis method and the UV selection to calculate the bounds of some typical processes, such as the $4H$, $4W$ and $4$ lepton scattering, and present the results in Sec.~\ref{sec4}. 
Despite that the J-Basis can give a systematic scheme to find all the UV states, it's hard to obtain the analytical bounds in some cases. Especially for the 4 fermion scattering with multi-generation we cannot obtain fully analytical solutions due to too many parameters represents couplings between different generations. However, by imposing limitations such as the MFV case, the numerical solution can be obtained. In summary, the J-Basis idea and the UV selection provide a systematic framework to find all the UV states and gives more rigorous limitations in positivity-bound problems.
\paragraph{Discussion}
The positivity bounds based on external rays, by itself, is a powerful tool to determine the exact boundary of the UV-completable EFTs and supersedes bounds from the elastic scattering, and has a better physical interpretation of the relationship between the UV and the SMEFT. Many typical 2-to-2 scattering involving the SM particles are calculated in previous work have been updated in our works by the J-Basis method and the UV selection. However, obtaining the full set of bounds for all the SMEFT operators seems impossible because the degeneracy of two states with the same quantum number turns to obtain bounds to solve corresponding complex multivariate quadratic inequalities. So we should be able to obtain numerical bounds for all the SMEFT operators.


\acknowledgments

Thanks to Hao-Lin Li and Yu-Han Ni's passionate instruction in the J-Basis framework and ABC4EFT Code. Besides, thanks for Hao Sun's help in the basis for the MFV case. 
Thanks to Shuang-Yong Zhou for his valuable comments on the manuscript.
The work is supported in part by the National Science Foundation of China under Grants 
No. 12022514, No. 12375099, No. 12047503
, No.11725520, No.11675002, No.12075257, No.12235001, and National Key Research and Development Program of China Grant No. 2020YFC2201501, and No.2021YFA0718304.

\appendix
\section{Matching Results}

In fact, the J-Basis method only provides the possibility of UV particles' existence and the further UV selection step give all the UV completion. For cross-checking, we need to calculate the UV-EFT matching results to process crosscheck. In this appendix, we list all calculations of tree-level matching for the UV states involved in the $4H$, the $4W$, and the $4$ lepton scattering processes in the P-Basis. The transformation matrices between the different basis can be acquired in Ref.~\cite{Li:2020gnx,Li:2020xlh,Li:2022abx}.

Here we introduce some notations that would be used later in this appendix. The Proca Lagrangian for massive spin-1 particle:
\begin{align}
	\mathcal{L}= -\frac{1}{2} A_{\mu} \mc{G}^{\mu\nu} A_{\nu},
\end{align}
where
\begin{align}
	\mc{G}^{\mu\nu} = -(\square+M^2)g^{\mu\nu} + \partial^{\mu} \partial^{\nu}\;.
\end{align}

\subsection{SM Higgs}
The channel: $\{H_{1}, H_{2}\}, \{H^{\dagger}_{3}, H^{\dagger}_{4}\}$. All the possible UV resonances and the matching results to the SMEFT operators with the form $H^2H^{\dagger 2}D^4$ are listed as follows,

\begin{flalign}
&\textbf{The UV states} ( Spin,SU(2)_w,U(1)_y)=(2,3,1):&\\
&\text{The UV amplitude}&\\
&\ =\quad i \frac{g^2}{M^4}\left(\delta_k^i \delta_l^j+\delta_l^i \delta_k^j\right)\left(s_{13}^2+s_{14}^2-\frac{2}{3} s_{12}^2\right)+\cdots &\\
&\ =i \frac{8 g^2}{3 M^4}\left(\begin{array}{lll}
3 & -2 & 3
\end{array}\right)\left(\begin{array}{l}
\mathcal{M}_{D^4 H^2 H^{\dagger2}}^{(1)} \\
\mathcal{M}_{D^4 H^2 H^{\dagger2}}^{(2)} \\
\mathcal{M}_{D^4 H^2 H^{\dagger2}}^{(3)}
\end{array}\right)+\cdots \,. &\\
\end{flalign}
\begin{flalign}
&\textbf{The UV states} (Spin,SU(2)_w,U(1)_y)=(1,3,1):\\& \mathcal{L}_{int}=g \mathcal{W}_1^{\mu I}(H^T \epsilon \tau^I i \stackrel{\leftrightarrow}{D}_\mu H)+ h.c.\,.&\\
&\text{The UV amplitude}&\\
 &= i \frac{2 g^2}{M^4}\left(e^{j m}\left(\tau^I\right)_m^i\left(\tau^I\right)_k^n \epsilon_{n l}\left(s_{13}^2-s_{14}^2\right)+\epsilon^{j m}\left(\tau^I\right)_m^i\left(\tau^I\right)_l^n \epsilon_{n k}\left(s_{14}^2-s_{13}^2\right)\right)+\cdots &\\
&=  0\,.&
\end{flalign}
\\
\textbf{The UV states}$(Spin,SU(2)_w,U(1)_y)=(0,3,1):\\ \mathcal{L}_{int}=g M \Xi_1^{I \dagger}(H^T \epsilon \tau^I H)+$ h.c..
\\
Considering that,\\
\\
\quad \quad \quad \quad\quad\quad$\left(H_j \epsilon^{j k}\left(\tau^I\right)_k^i H_i\right)^{\dagger}=-  \left(H^{\dagger i}\left(\tau^I\right)_i^k \epsilon_{k j} H^{\dagger j}\right)$,
\\
\\
we have the UV amplitude
  \begin{flalign}
&\ =  -i \frac{2 g^2}{M^4}\left(\epsilon^{j m}\left(\tau^I\right)_m^i\left(\tau^I\right)_l^n \epsilon_{n k}+\epsilon^{j m}\left(\tau^I\right)_m^i\left(\tau^I\right)_k^n \epsilon_{n l}\right) s_{12}^2+\cdots  &\\
&\ =  i \frac{4 g^2}{M^4}\left(\delta_k^i \delta_l^j+\delta_l^i \delta_k^j\right) s_{12}^2+\cdots &\\
 &\ =\quad i \frac{32 g^2}{M^4}\left(\begin{array}{lll}
0 & 1 & 0
\end{array}\right)\left(\begin{array}{l}
\mathcal{M}_{D^4 H^2 H^{\dagger2}}^{(1)} \\
\mathcal{M}_{D^4 H^2 H^{\dagger2}}^{(2)} \\
\mathcal{M}_{D^4 H^2 H^{\dagger2}}^{(3)}
\end{array}\right)+\cdots  \,.\\
\nonumber
\end{flalign}
\begin{flalign}
 &\ \textbf{The UV states}\left(Spin, S U(2)_w, U(1)_y\right)=(1,1,1):&\\&\ \mathcal{L}_{int}= g \mathcal{B}_1^{\mu \dagger}\left(H^T \epsilon i \stackrel{\leftrightarrow}{D}_\mu H\right)+\text {h.c.}&\,. 
 \nonumber
\end{flalign}
 \text{Considering the conjugate relationship}
 \begin{equation}
    \begin{aligned}
& \left(H_j \epsilon^{j i} i \stackrel{\leftrightarrow}{D_\mu} H_i\right)^{\dagger}=i\left(D_\mu H^{\dagger i} \epsilon_{i j} H^{\dagger j}-H^{\dagger i} \epsilon_{i j} D_\mu H^{\dagger j}\right)=-H^{\dagger i} \epsilon_{i j} i \stackrel{\leftrightarrow}{D}_\mu H^{\dagger j}\,, 
 \end{aligned}  
\end{equation}
\begin{flalign}
&\text{we have the UV amplitude}&\\
& =\quad i \frac{4 g^2}{M^4}\left(-\delta_k^i \delta_l^j+\delta_l^i \delta_k^j\right)\left(s_{13}^2-s_{14}^2\right)+\cdots &\\
& =i \frac{32 g^2}{M^4}\left(\begin{array}{lll}
1 & 0 & -1
\end{array}\right)\left(\begin{array}{l}
\mathcal{M}_{D^4 H^2 H^{\dagger2}}^{(1)} \\
\mathcal{M}_{D^4 H^2 H^{\dagger2}}^{(2)} \\
\mathcal{M}_{D^4 H^2 H^{\dagger2}}^{(3)}
\end{array}\right)+\cdots  \,.&\\
\end{flalign}
Now we discuss about the channel $\left\{H_1, H_3^{\dagger}\right\},\left\{H_2, H_4^{\dagger}\right\}$.\\
\begin{flalign}
   & \textbf{The UV states} (Spin,SU(2)_w,U(1)_y)=(2,3,0):&\\& \mathcal{L}_{int}= g M^{-1} \mathcal{H}_0^{\mu \nu I}\left(\partial_\mu H^{\dagger} \tau^I \partial_\nu H\right)\,.&\\
   &\text{The UV amplitude}&\\
& =\quad i \frac{g^2}{4 M^4}\left(\delta_k^i \delta_l^j\left(s_{12}^2-\frac{7}{3} s_{14}^2+\frac{8}{3} s_{13}^2\right)+\delta_l^i \delta_k^j\left(s_{12}^2-\frac{7}{3} s_{13}^2+\frac{8}{3} s_{14}^2\right)\right)+\cdots &\\
& =\quad i \frac{2 g^2}{3 M^4}\left(\begin{array}{lll}
-7 & 3 & 8
\end{array}\right)\left(\begin{array}{l}
\mathcal{M}_{D^4 H^2 H^{\dagger2}}^{(1)} \\
\mathcal{M}_{D^4 H^2 H^{\dagger2}}^{(2)} \\
\mathcal{M}_{D^4 H^2 H^{\dagger2}}^{(3)}
\end{array}\right)+\cdots \,.&
\end{flalign}
\begin{flalign}
    &\textbf{The UV states}(Spin,SU(2)_w,U(1)_y)=(1,3,0):~ \mathcal{L}_{UV}=g \mathcal{W}_0^{\mu I}\left(H^{\dagger} \tau^I i \stackrel{\leftrightarrow}{D_\mu} H\right)\,,&\\
    &\text{The UV amplitude}&\\
& =\quad i \frac{g^2}{M^4}\left(\delta_k^i \delta_l^j\left(s_{12}^2+s_{14}^2-2 s_{13}^2\right)+\delta_l^i \delta_k^j\left(s_{12}^2+s_{13}^2-2 s_{14}^2\right)\right)+\cdots &\\
& =i \frac{8 g^2}{M^4}\left(\begin{array}{lll}
1 & 1 & -2
\end{array}\right)\left(\begin{array}{l}
\mathcal{M}_{D^4 H^2 H^{\dagger2}}^{(1)} \\
\mathcal{M}_{D^4 H^2 H^{\dagger2}}^{(2)} \\
\mathcal{M}_{D^4 H^2 H^{\dagger2}}^{(3)}
\end{array}\right)+\cdots  \,.&
\end{flalign}
\begin{flalign}
\begin{split}
   & \textbf{The UV states}(Spin,SU(2)_w,U(1)_y)=(0,3,0): \\& \mathcal{L}_{int}=g M \Xi_0^I\left(H^{\dagger} \tau^I H\right)\,.\\
   &\text{The UV amplitude}\\
 & =i \frac{g^2}{M^4}\left(\delta_k^i \delta_l^j\left(2 s_{14}^2-s_{13}^2\right)+\delta_l^i \delta_k^j\left(2 s_{13}^2-s_{14}^2\right)\right)+\cdots\\
& =i \frac{8 g^2}{M^4}\left(\begin{array}{lll}
2 & 0 & -1
\end{array}\right)\left(\begin{array}{l}
\mathcal{M}_{D^4 H^2 H^{\dagger2}}^{(1)} \\
\mathcal{M}_{D^4 H^2 H^{\dagger2}}^{(2)} \\
\mathcal{M}_{D^4 H^2 H^{\dagger2}}^{(3)}
\end{array}\right)+\cdots \,.
\end{split}&
\end{flalign}
\begin{flalign}
    \begin{split}
    &\textbf{The UV states}(Spin,SU(2)_w,U(1)_y)=(2,1,0):\\& \mathcal{L}_{int}=g M^{-1} \mathcal{G}^{\mu \nu}\left(\partial_\mu H^{\dagger} \partial_\nu H\right)\,.\\
     &\text{The UV amplitude}\\
& =\quad i \frac{g^2}{8 M^4}\left(\delta_k^i \delta_l^j\left(s_{12}^2+s_{14}^2-\frac{2}{3} s_{13}^2\right)+\delta_l^i \delta_k^j\left(s_{12}^2+s_{13}^2-\frac{2}{3} s_{14}^2\right)\right)+\cdots \\
& =\quad i \frac{g^2}{3 M^4}\left(\begin{array}{lll}
3 & 3 & -2
\end{array}\right)\left(\begin{array}{l}
\mathcal{M}_{D^4 H^2 H^{\dagger2}}^{(1)} \\
\mathcal{M}_{D^4 H^2 H^{\dagger2}}^{(2)} \\
\mathcal{M}_{D^4 H^2 H^{\dagger2}}^{(3)}
\end{array}\right)+\cdots  \,.
\end{split}&
\end{flalign}
\begin{flalign}
& \textbf {The UV states}\left(Spin,SU(2)_w,U(1)_y\right)=(1,1,0):\\& \mathcal{L}_{int}=g \mathcal{B}_0^\mu\left(H^{\dagger} i \stackrel{\leftrightarrow}{D}_\mu H\right) \,.&\\
 &\text{The UV amplitude}&\\
& =\quad i \frac{g^2}{M^4}\left(\delta_k^i \delta_l^j\left(s_{12}^2-s_{14}^2\right)+\delta_l^i \delta_k^j\left(s_{12}^2-s_{13}^2\right)\right)+\cdots \\
& =i \frac{8 g^2}{M^4}\left(\begin{array}{lll}
-1 & 1 & 0
\end{array}\right)\left(\begin{array}{l}
\mathcal{M}_{D^4 H^2 H^{\dagger2}}^{(1)} \\
\mathcal{M}_{D^4 H^2 H^{\dagger2}}^{(2)} \\
\mathcal{M}_{D^4 H^2 H^{\dagger2}}^{(3)}
\end{array}\right)+\cdots  \,.&
\end{flalign}
 \begin{flalign}
     \begin{split}
    & \textbf{The UV states}(Spin, S U(2)_w, U(1)_y)=(0,1,0):\\& \mathcal{L}_{int}=g M \mathcal{S}\left(H^{\dagger} H\right)\,.\\
 &\text{The UV amplitude}\\
& =\quad i \frac{g^2}{M^4}\left(\delta_k^i \delta_l^j s_{13}^2+\delta_l^i \delta_k^j s_{14}^2\right)+\cdots \\
& =i \frac{8 g^2}{M^4}\left(\begin{array}{lll}
0 & 0 & 1
\end{array}\right)\left(\begin{array}{l}
\mathcal{M}_{D^4 H^2 H^{\dagger2}}^{(1)} \\
\mathcal{M}_{D^4 H^2 H^{\dagger2}}^{(2)} \\
\mathcal{M}_{D^4 H^2 H^{\dagger2}}^{(3)}
\end{array}\right)+\cdots \,.
\end{split}&
\end{flalign}
\subsection{\texorpdfstring{$4W$}~ boson}
\subsubsection{Scalar Couplings:}
\begin{flalign}
&\textbf{The UV states}(Spin,SU(3)_c,SU(2)_w,U(1)_y)=(0,1,1,0): &
\end{flalign}
\begin{equation}
    \mathcal{L}_{\mathrm{UV}}=-\frac{1}{2} S\left(\square+M^2\right) S+g W_{\mathrm{L} \mu \nu}^I W_{\mathrm{L}}^{I \mu \nu} S+g^* W_{\mathrm{R} \mu \nu}^I W_{\mathrm{R}}^{I \mu \nu} S \nonumber\,.
\end{equation}
$W^2_{L}W^2_{R}$\\
The EOM of $S$ :
\begin{equation}
    \begin{aligned}
 (\square+M^2) S&=g W_{\mathrm{L} \mu \nu}^I W_{\mathrm{L}}^{I \mu \nu}+g^* W_{\mathrm{R} \mu \nu}^I W_{\mathrm{R}}^{I \mu \nu} \,,\\
\mathcal{L}_{\mathrm{EFT}}&=\frac{g g^*}{M^2} W_{\mathrm{L} \mu \nu}^I W_{\mathrm{L}}^{I \mu \nu} W_{\mathrm{R} \rho \lambda}^J W_{\mathrm{R}}^{J \rho \lambda} \\
& =\frac{4 g g^*}{M^2}\left(\begin{array}{ll}
0 & -1
\end{array}\right)\left(\begin{array}{l}
\mathcal{O}_{W_{\mathrm{L}}^2 W_{\mathrm{R}}^2, 1}^{(p)} \\
\mathcal{O}_{W_{\mathrm{L}}^2 W_{\mathrm{R}}^2, 2}^{(p)}
\end{array}\right) \,.
\end{aligned}
\end{equation}
$W^4_{L}$\\
The EOM of $S$ :
\begin{equation}
    \begin{aligned}
(\square+M^2) S&=g W_{\mathrm{L} \mu \nu}^I W_{\mathrm{L}}^{I \mu \nu} \,,\\
 \mathcal{L}_{\mathrm{EFT}}&= \frac{g^2}{2 M^2} W_{\mathrm{L} \mu \nu}^I W_{\mathrm{L}}^{I \mu \nu} W_{\mathrm{L} \rho \lambda}^J W_{\mathrm{L}}^{J \rho \lambda} \\
&= \frac{2 g^2}{M^2}\left(\begin{array}{ll}
-1 & 0
\end{array}\right)\left(\begin{array}{l}
\mathcal{O}_{W_{\mathrm{L}}^4, 1}^{(p)} \\
\mathcal{O}_{W_{\mathrm{L}}^4, 2}^{(p)}
\end{array}\right)\,.
\end{aligned}
\end{equation}
$\textbf{The UV states}\left(\right.Spin,\left.SU(3)_c, SU(2)_w,U(1)_y\right)=(0, \mathbf{1}, \mathbf{5}, 0): $\\
\begin{equation}
    \mathcal{L}_{\mathrm{UV}}=-\frac{1}{2} S^{\Lambda}\left(\square+M^2\right) S^{A}+g W_{\mathrm{L} \mu \nu}^I W_{\mathrm{L}}^{J \mu \nu}\left(T^{A}\right)^{I J} \mathcal{S}^{A}+g^* W_{\mathrm{R} \mu \nu}^I W_{\mathrm{R}}^{J \mu \nu}\left(T^{A}\right)^{I J} \mathcal{S}^{A}\nonumber\,.
\end{equation}
$W^2_{L}W^2_{R}$\\
The EOM of $S^{A}$:
\begin{equation}
    \begin{aligned}
&(\square+M^2) S^{A}=g W_{\mathrm{L} \mu \nu}^I W_{\mathrm{L}}^{J \mu \nu}\left(T^{A}\right)^{I J}+g^* W_{\mathrm{R}}^I{ }^{\mu \nu} W_{\mathrm{R}}^{J \mu \nu}\left(T^{A}\right)^{I J}\,, \\
\mathcal{L}_{\mathrm{EFT}} & =\frac{g g^*}{M^2} W_{\mathrm{L} \mu \nu}^I W_{\mathrm{L}}^{J \mu \nu} W_{\mathrm{R} \rho \lambda}^K W_{\mathrm{R}}^{L \rho \lambda}\left(T^{A}\right)^{I J}\left(T^{A}\right)^{K L} \\
& =\frac{4 g g^*}{3 M^2}\left(\begin{array}{ll}
-3 & 1
\end{array}\right)\left(\begin{array}{c}
\mathcal{O}_{W_{\mathrm{L}}^2 W_{\mathrm{R}}^2, 1}^{(p)} \\
\mathcal{O}_{W_{\mathrm{L}}^2 W_{\mathrm{R}}^2, 2}^{(p)}
\end{array}\right)\,.
\end{aligned}
\end{equation}
$W^4_{L}$\\
The EOM of $S^{\Lambda}$:
\begin{equation}
    \begin{aligned}
(\square+M^2) S^{A}&=g W_{\mathrm{L} \mu \nu}^I W_{\mathrm{L}}^{J \mu \nu}\left(T^{A}\right)^{I J} \,,\\
 \mathcal{L}_{\mathrm{EFT}}&=\frac{g^2}{2 M^2} W_{\mathrm{L} \mu \nu}^I W_{\mathrm{L}}^{J \mu \nu} W_{\mathrm{L} \rho \lambda}^K W_{\mathrm{L}}^{L \rho \lambda}\left(T^{A}\right)^{I J}\left(T^{A}\right)^{K L} \\
&=\frac{g^2}{3 M^2}\left(\begin{array}{ll}
4 & 3
\end{array}\right)\left(\begin{array}{c}
\mathcal{O}_{W_{\mathrm{L}}^4, 1}^{(p)} \\
\mathcal{O}_{W_{\mathrm{L}}^4, 2}^{(p)}
\end{array}\right)\,.
\end{aligned}
\end{equation}
\subsubsection{Massive Spin-2 Couplings}
We have already discussed, there are only $W_{L}^2W_{R}^2$ terms.
\begin{flalign}
    &\textbf{The UV states}(Spin,SU(3)_c,SU(2)_w,U(1)_y)=(2, 1,5, 0): \mathcal{L}_{\mathrm{int}}=W_{\mathrm{L} \mu \nu}^I W_{\mathrm{R} \rho}^J{ }^\nu\left(T^{A}\right)^{I J} \mathcal{H}_5^{A \mu \rho}\nonumber\,.&
\end{flalign}
\begin{equation}
    \begin{aligned}
\mathcal{L}_{\mathrm{EFT}}= & \frac{g^2}{2 M^2}\left(g^{\mu \lambda} g^{\nu \rho}+g^{\nu \lambda} g^{\mu \rho}-\frac{2}{3} g^{\mu \nu} g^{\lambda \rho}\right)\left(\left(T^{A}\right)^{I J}\left(T^{A}\right)^{K L} W_{\mathrm{L} \mu \xi}^I W_{\mathrm{R} \nu}^J{ }^{\xi} W_{\mathrm{L} \lambda \sigma}^K W_{\mathrm{R} \rho}^L\right) \\
= & \frac{g^2}{3 M^2}(-1-3)\left(\begin{array}{c}
\mathcal{O}_{W_{\mathrm{L}}^2,W_{\mathrm{R}}^2, 1}^{(p)} \\
\mathcal{O}_{W_{\mathrm{L}}^2 W_{\mathrm{R}}^2, 2}^{(p)} \\
\end{array}\right)\,.
\end{aligned}
\end{equation}
\begin{flalign}
    &\textbf{The UV states}(Spin,SU(3)_c,SU(2)_w,U(1)_y)=(2,1,3,0): \mathcal{L}_{\mathrm{int}}=\epsilon_{I J K} W_{\mathrm{L} \mu \nu}^I W_{\mathrm{R} \rho}^J{ }^\nu \mathcal{H}_3^{K \mu \rho}\,.&\nonumber
\end{flalign}
\begin{equation}
    \begin{aligned}
 \mathcal{L}_{\mathrm{EFT}}&=\frac{g^2}{2 M^2}\left(g^{\mu \lambda} g^{\nu \rho}+g^{\nu \lambda} g^{\mu \rho}-\frac{2}{3} g^{\mu \nu} g^{\lambda \rho}\right)\left(\epsilon^{I J M} \epsilon^{K L M} W_{\mathrm{L} \mu \xi}^I W_{\mathrm{R} \nu}^J{ }^{\xi} W_{\mathrm{L}}^K \lambda \sigma W_{\mathrm{R} \rho}^L\right)\\
& =\frac{g^2}{M^2}\left(\begin{array}{ll}
1 & -1
\end{array}\right)\left(\begin{array}{c}
\mathcal{O}_{W_{\mathrm{L}}^2 W_{\mathrm{R}}^2, 1}^{(p)} \\
\mathcal{O}_{W_{\mathrm{L}}^2 W_{\mathrm{R}}^2, 2}^{(p)}
\end{array}\right) \,.
&
\end{aligned}
\end{equation}
\begin{flalign}
    &\textbf{The UV states} (Spin, S U(3)_c, S U(2)_w, U(1)_y)=(2, 1,1, 0): \mathcal{L}_{\mathrm{int}}=W_{\mathrm{L} \mu \nu}^I W_{\mathrm{R} \rho}^I \mathcal{G}^{\mu \rho}\,,&\nonumber
\end{flalign}
\begin{equation}
    \mathcal{L}_{\mathrm{UV}}=-\frac{1}{2}\left(g^{\mu \lambda} g^{\nu \rho}+g^{\nu \lambda} g^{\mu \rho}-\frac{2}{3} g^{\mu \nu} g^{\lambda \rho}\right)^{-1} \mathcal{G}^{\mu \nu}\left(\square+M^2\right) \mathcal{G}^{\lambda \rho}+g W_{\mathrm{L} \lambda \sigma}^I W_{\mathrm{R} \rho}^I \mathcal{G}^{\lambda \rho}\,,\nonumber
\end{equation}
\begin{equation}
    \begin{aligned}
\mathcal{L}_{\mathrm{EFT}} & =\frac{g^2}{2 M^2}\left(g^{\mu \lambda} g^{\nu \rho}+g^{\nu \lambda} g^{\mu \rho}-\frac{2}{3} g^{\mu \nu} g^{\lambda \rho}\right)\left(W_{\mathrm{L} \mu \xi}^I W_{\mathrm{R} \nu}^I{ }^{\xi} W_{\mathrm{L} \lambda \sigma}^J W_{\mathrm{R} \rho}^J{ }^\sigma\right) \\
& =\frac{g^2}{M^2}\left(\begin{array}{ll}
-1 & 0
\end{array}\right)\left(\begin{array}{c}
\mathcal{O}_{W_{\mathrm{L}}^2 W_{\mathrm{R}}^2, 1}^{(p)} \\
\mathcal{O}_{W_{\mathrm{L}}^2 W_{\mathrm{R}}^2, 2}^{(p)}
\end{array}\right)\,.
\end{aligned}
\end{equation}
\subsection{Fermions with the Multi-Generation}
\begin{flalign}
    &\textbf{The UV states} (Spin, S U(3)_c, S U(2)_w, U(1)_y)=(1,1,3,1): &\nonumber
\end{flalign}
\begin{equation}
    \mathcal{L}_{\mathrm{int}}=g_{p_1 p_2} \epsilon^{i_1 m}\left(\tau^I\right)_m^{i_2} \mathcal{W}_1^{\mu I}\left(L_{p_1 i_1} i \stackrel{\leftrightarrow}{D}_\mu L_{p_2 i_2}\right)+ h.c.\,,\nonumber
\end{equation}
\begin{equation}
\begin{aligned}
    \mathcal{L}_{\mathrm{UV}}=&\mathcal{W}_{1 \mu}^{\dagger I}\left(\square+M^2\right) \mathcal{W}_1^{\mu I}+g_{p_1 p_2} \epsilon^{i_1 m}\left(\tau^I\right)_m^{i_2} \mathcal{W}_1^{\mu I}\left(L_{p_1 i_1} i \stackrel{\leftrightarrow}{D}_\mu L_{p_2 i_2}\right)\\&-g_{p_1 p_2}^*\left(\tau^I\right)_{i_2}^m \epsilon_{m i_1} \mathcal{W}_1^{\dagger \mu}\left(L_{p_2}^{\dagger i_2} i \stackrel{\leftrightarrow}{D_\mu} L_{p_1}^{\dagger i_1}\right)\,.\nonumber
    \end{aligned}
\end{equation}
The EOM of $\mathcal{W}^{\mu I}$ :
\begin{equation}
    \begin{aligned}
& \left(\square+M^2\right) \mathcal{W}_{1 \mu}^I=g_{p_1 p_2}^*\left(\tau^I\right)_{i_2}^m \epsilon_{m i_1}\left(L_{p_2}^{\dagger_2} i \stackrel{\leftrightarrow}{D_\mu} L_{p_1}^{\dagger i_1}\right)\,.\nonumber
\end{aligned}
\end{equation}
The EOM of $\mathcal{W}_1^{\dagger \mu I}$ :
\begin{equation}
    \begin{aligned}
    (\square+M^2) \mathcal{W}_{1 \mu}^{\dagger I}=-g_{p_1 p_2} \epsilon^{i_1 m}\left(\tau^I\right)_m^{i_2}\left(L_{p_1 i_1} i \stackrel{\leftrightarrow}{D}_\mu L_{p_2 i_2}\right)\,.\nonumber
\end{aligned}
\end{equation}
\begin{equation}
    \begin{aligned}
 \mathcal{L}_{\mathrm{EFT}}&=\frac{g_{p_1 p_2} g_{p_3 p_4}^*}{M^2} \epsilon^{i_1 m}\left(\tau^I\right)_m^{i_2}\left(\tau^I\right)_{i_4}^n \epsilon_{n i_3}\left(L_{p_1 i_1} i \stackrel{\leftrightarrow}{D}_\mu L_{p_2 i_2}\right)\left(L_{p_4}^{\dagger_4} i \stackrel{\leftrightarrow}{D}{ }^\mu L_{p_3}^{\dagger i_3}\right) \\
& =-\frac{g_{p_1 p_2} g_{p_3 p_4}^*}{M^2} \epsilon^{i_{1 m}}\left(\tau^I\right)_m^{i_2}\left(\tau^I\right)_{{ }_4}^n \epsilon_{n i_3}(\langle 12\rangle\langle 14\rangle[14][34]-\langle 12\rangle\langle 13\rangle[13][34]) \\
& =\frac{g_{p_1 p_2} g_{p_3 p_4}^*}{M^2}\left(\begin{array}{llll}
0 & 0 & 0 & -4
\end{array}\right)\left(\begin{array}{l}
\mathcal{O}_{L^2 L^{\dagger 2} D^2, 1}^{(p)} \\
\mathcal{O}_{L^2 L^{\dagger 2} D^2, 2}^{(p)} \\
\mathcal{O}_{L^2 L^{\dagger 2} D^2, 3}^{(p)} \\
\mathcal{O}_{L^2 L^{\dagger 2} D^2, 4}^{(p)}
\end{array}\right) \,.
\end{aligned}
\end{equation}
\begin{flalign}
    &\textbf{The UV states}(Spin, S U(3)_c, S U(2)_w, U(1)_y)=(0,1,3,1): & \\&\mathcal{L}_{int}=g_{p_1 p_2} \epsilon^{i_1 m}(\tau^I)_m^{i_2} \Xi^I\left(L_{p_1 i_1} L_{p_2 i_2}\right)+ h.c.\,,\nonumber
\end{flalign}
\begin{equation}
    \mathcal{L}_{\mathrm{UV}}=-\Xi^{\dagger I}\left(\square+M^2\right) \Xi^I+g_{p_1 p_2} \epsilon^{i_{1 m}}\left(\tau^I\right)_m^{i_2} \Xi^I\left(L_{p_1 i_1} L_{p_2 i_2}\right)-g_{p_1 p_2}^*\left(\tau^I\right)_{i_2}^m \epsilon_{m i_1} \Xi^{\dagger I}\left(L_{p_2}^{\dagger i_2} L_{p_1}^{\dagger i_1}\right)\,.\nonumber
\end{equation}
The EOM of $\Xi^I$ :
\begin{equation}
\begin{aligned}
-\left(\square+M^2\right) &\Xi^I=g_{p_1 p_2}^*\left(\tau^I\right)_{i_2}^m \epsilon_{m i_1}\left(L_{p_2}^{\dagger i_2} L_{p_1}^{\dagger i_1}\right)\,.\nonumber
\end{aligned}
\end{equation}
The EOM of $\Xi^{\dagger I}$ :
\begin{equation}
    (\square+M^2)\Xi^{\dagger I}=g_{p_1 p_2} \epsilon^{i_1 m}\left(\tau^I\right)_m^{i_2}\left(L_{p_1 i_1} L_{p_2 i_2}\right) \,.\nonumber
\end{equation}
After integrating out $\Xi^I$,
\begin{equation}
    \begin{aligned}
& =-\frac{g_{p_1 p_2} g_{p_3 p_4}^*}{M^4}\left(\begin{array}{llll}
4 & 0 & 0 & 0
\end{array}\right)\left(\begin{array}{l}
\mathcal{O}_{L^2 L^{\dagger 2} D^2, 1}^{(p)} \\
\mathcal{O}_{L^2 L^{\dagger 2} D^2, 2}^{(p)} \\
\mathcal{O}_{L^2 L^{\dagger 2} D^2, 3}^{(p)} \\
\mathcal{O}_{L^2 L^{\dagger 2} D^2, 4}^{(p)}
\end{array}\right)\,.
\end{aligned}
\end{equation}
\begin{flalign}
    &\textbf{The UV states} (Spin, S U(3)_c, S U(2)_w, U(1)_y)=(1,1,1,1): &\\&\mathcal{L}_{int}=g_{p_1 p_2} \epsilon^{i_1 i_2} \mathcal{B}_1^\mu(L_{p_1 i_1} i \stackrel{\leftrightarrow}{D}_\mu L_{p_2 i_2})+ h.c.\,,&\nonumber
\end{flalign}
\begin{equation}
    \begin{aligned}
 \mathcal{L}_{\mathrm{EFT}}&=\frac{g_{p_1 p_2} g_{p_3 p_4}^*}{M^2} \epsilon^{i_1 i_2} \epsilon_{i_4 i_3}\left(L_{p_1 i_1} i \stackrel{\leftrightarrow}{D}_\mu L_{p_2 i_2}\right)\left(L_{p_4}^{i_4} i \stackrel{\leftrightarrow}{D}{ }^\mu L_{p_3}^{\dagger i_3}\right) \\
& =-\frac{g_{p_1 p_2} g_{p_3 p_4}^*}{M^2} \epsilon^{i_1 i_2} \epsilon_{i_4 i_3}(\langle 12\rangle\langle 14\rangle[14][34]-\langle 12\rangle\langle 13\rangle[13][34])\,.
\end{aligned}
\end{equation}
\begin{flalign}
    &\textbf{The UV states}(Spin,S U(3)_c, S U(2)_w, U(1)_y)=(0,1,1,1): \mathcal{L}_{int}=g_{p_1 p_2} \epsilon^{i_1 i_2} S(L_{p_1 i_1} L_{p_2 i_2})+ h.c.\,,&\nonumber
\end{flalign}
\begin{equation}
    \begin{aligned}
 \mathcal{L}_{\mathrm{EFT}}&=\frac{g_{p_1 p_2} g_{p_3 p_4}^*}{M^4} \epsilon^{i_1 i_2} \epsilon_{i_4 i_3} \square\left(L_{p_1 i_1} L_{p_2 i_2}\right)\left(L_{p_4}^{\dagger i_4} L_{p_3}^{\dagger i_3}\right) \\
& =\frac{g_{p_1 p_2} g_{p_3 p_4}^*}{M^4} \epsilon^{i_1 i_2} \epsilon_{i_4 i_3}(\langle 12\rangle\langle 12\rangle[12][34]) \\
& =\frac{g_{p_1 p_2} g_{p_3 p_4}^*}{M^4}\left(\begin{array}{llll}
0 & 0 & -4 & 0
\end{array}\right)\left(\begin{array}{l}
\mathcal{O}_{L^2 L^{+2} D^2, 1}^{(p)} \\
\mathcal{O}_{L^2 L^{+2} D^2, 2}^{(p)} \\
\mathcal{O}_{L^2 \perp^{+2} D^2, 3}^{(p)} \\
\mathcal{O}_{L^2 L^{+2} D^2, 4}^{(p)}
\end{array}\right)\,.
\end{aligned}
\end{equation}
\begin{flalign}
    &\textbf{The UV states}(Spin, S U(3)_c, S U(2)_w, U(1)_y)=(2,1,3,0): \mathcal{L}_{int}=g_{p 1 p_3}\left(\tau^I\right)_{i_3}^{i_1} \mathcal{H}^{\mu \nu I}\left(L_{p 1 i_1} i \sigma_\mu \stackrel{\leftrightarrow}{D}_\nu L_{p_3}^{i_3}\right)\,, &\nonumber
\end{flalign}
\begin{flalign}
    \mathcal{L}_{\mathrm{UV}}&=-\frac{1}{2}\left(g^{\mu \lambda} g^{\nu \rho}+g^{\nu \lambda} g^{\mu \rho}-\frac{2}{3} g^{\mu \nu} g^{\lambda \rho}\right)^{-1} \mathcal{H}^{\mu \nu I}\left(\square+M^2\right) \mathcal{H}^{\lambda \rho I}\\&+g_{p_1 p_3}\left(\tau^I\right)_{i_3}^{i_1} \mathcal{H}^{\mu \nu I}\left(L_{p_1 i_1} i \sigma_\mu \stackrel{\leftrightarrow}{D}_\nu L_{p_3}^{\dagger i_3}\right)\,.\nonumber
\end{flalign}
The EOM of $\mathcal{H}^{\mu \nu I}$ :
\begin{equation}
    \begin{aligned}
 &(\square+M^2)\mathcal{H}^{\mu \nu I}=\left(\tau^I\right)_{i_3}^{i_1}\left(g^{\mu \lambda} g^{\nu \rho}+g^{\nu \lambda} g^{\mu \rho}-\frac{2}{3} g^{\mu \nu} g^{\lambda \rho}\right)\left(L_{p_1 i_1} \sigma_\lambda \stackrel{\leftrightarrow}{D}_\rho L_{p_3}^{\dagger i_3}\right) \,,\\
 \mathcal{L}_{\mathrm{EFT}}=&\frac{g_{p_1 p_3} g_{p_2 p_4}}{2 M^2}\left(\tau^I\right)_{i_3}^{i_1}\left(\tau^I\right)_{i_4}^{i_2}\left(g^{\mu \lambda} g^{\nu \rho}+g^{\nu \lambda} g^{\mu \rho}-\frac{2}{3} g^{\mu \nu} g^{\lambda \rho}\right)*\\&\left(L_{p_1 i_1} i \sigma_\mu \stackrel{\leftrightarrow}{D}_\nu L_{p_3}^{i_3}\right)\left(L_{p_2 i_2} i \sigma_\lambda \stackrel{\leftrightarrow}{D}_\rho L_{p_4}^{\dagger i_4}\right) \\
=&\frac{g_{p_1 p_3} g_{p_2 p_4}}{2 M^2}\left(\tau^I\right)_{i_3}^{i_1}\left(\tau^I\right)_{i_4}^{i_2}[2\langle 12\rangle[34](-\langle 12\rangle[12]+\langle 14\rangle[14]) \\
& -(\langle 14\rangle[43]\langle 23\rangle[34]-\langle 14\rangle[43]\langle 21\rangle[14]-\langle 12\rangle[23]\langle 23\rangle[34]+\langle 12\rangle[23]\langle 21\rangle[14])] \\
=&\frac{g_{p_1 p_3} g_{p_2 p_4}}{M^2}\left(\begin{array}{llll}
-5 & 9 & 15 & -3
\end{array}\right)\left(\begin{array}{l}
\mathcal{O}_{L^2 L^{\dagger+} D^2, 1}^{(p)} \\
\mathcal{O}_{L^2 L^{\dagger 2} D^2, 2}^{(p)} \\
\mathcal{O}_{L^2 L^{\dagger 2} D^2, 3}^{(p)} \\
\mathcal{O}_{L^2 L^{\dagger} D^2, 4}^{(p)}
\end{array}\right) \,.
\end{aligned}
\end{equation}
\begin{flalign}
    &\textbf{The UV states} (Spin, S U(3)_c, S U(2)_w, U(1)_y)=(1,1,3,0): \mathcal{L}_{int}=g_{p_1 p_3}(\tau^I)_{i_3}^{i_1} \mathcal{W}_0^{\mu I}(L_{p_1 i_1} \sigma_\mu L_{p_3}^{\dagger i_3})\,,&\nonumber
\end{flalign}
\begin{equation}
    \mathcal{L}_{\mathrm{UV}}=\frac{1}{2} \mathcal{W}_{0 \mu}^I\left(\square+M^2\right) \mathcal{W}_0^{\mu I}+g_{p_1 p_3}\left(\tau^I\right)_{i_3}^{i_1} \mathcal{W}_0^{\mu I}\left(L_{p_1 i_1} \sigma_\mu L_{p_3}^{\dagger i_3}\right)\,.
\end{equation}
The EOM of $\mathcal{W}_{0 \mu}^I$:
\begin{equation}
    \begin{aligned}
 &(\square+M^2)\mathcal{W}_{0 \mu}^I=-g_{p_1 p_3}\left(\tau^I\right)_{i_3}^{i_1}\left(L_{p_1 i_1} \sigma_\mu L_{p_3}^{\dagger i_3}\right) \,,\\
 \mathcal{L}_{\mathrm{EFT}}&=\frac{g_{p_1 p_3} g_{p_2 p_4}}{2 M^4}\left(\tau^I\right)_{i_3}^{i_1}\left(\tau^I\right)_{i_4}^{i_2}\left(L_{p_1 i_1} \sigma_\mu L_{p_3}^{\dagger i_3}\right) \square\left(L_{p_2 i_2} \sigma^\mu L_{p_4}^{\dagger i_4}\right) \\
& =\frac{g_{p_1 p_3} g_{p_2 p_4}}{M^4}\left(\tau^I\right)_{i_3}^{i_1}\left(\tau^I\right)_{i_4}^{i_2}\langle 12\rangle\langle 24\rangle[24][34] \\
& =-\frac{g_{p_1 p_2} g_{p_3 p_4}^*}{M^2}\left(\begin{array}{llll}
1 & 3 & -3 & -1
\end{array}\right)\left(\begin{array}{l}
\mathcal{O}_{L^2 L^{\dagger} D^2, 1}^{(p)} \\
\mathcal{O}_{L^2 L^{\dagger} D^2, 2}^{(p)} \\
\mathcal{O}_{L^2 L^{\dagger 2} D^2, 3}^{(p)} \\
\mathcal{O}_{L^2 L^{\dagger} D^2, 4}^{(p)}
\end{array}\right) \,.
\end{aligned}
\end{equation}
\begin{flalign}
   &\textbf{The UV states}(Spin, S U(3)_c, S U(2)_w, U(1)_y)=(2,1,1,0): \mathcal{L}_{int}=g_{p_1 p_3} \delta_{i_3}^{i_1} \mathcal{G}^{\mu \nu}(L_{p_1 i_1} i \sigma_\mu \stackrel{\leftrightarrow}{D}_\nu L_{p_3}^{\dagger_3})\,,&\nonumber 
\end{flalign}
\begin{equation}
    \begin{aligned}
 \mathcal{L}_{\mathrm{EFT}}&=\frac{g_{p_1 p_3} g_{p_2 p_4}}{2 M^2} \delta_{i_3}^{i_1} \delta_{i_4}^{i_2}\left(g^{\mu \lambda} g^{\nu \rho}+g^{\nu \lambda} g^{\mu \rho}-\frac{2}{3} g^{\mu \nu} g^{\lambda \rho}\right)\left(L_{p_1 i_1} i \sigma_\mu \stackrel{\leftrightarrow}{D}_\nu L_{p_3}^{\dagger i_3}\right)\left(L_{p_2 i_2} i \sigma_\lambda \stackrel{\leftrightarrow}{D}_\rho L_{p_4}^{\dagger i_4}\right) \\
& =\frac{g_{p_1 p_3} g_{p_2 p_4}}{2 M^2} \delta_{i_3}^{i_1} \delta_{i_4}^{i_2}\left(-8\langle 12\rangle\langle 34\rangle[34]^2+6\langle 13\rangle\langle 24\rangle[34]^2\right) \\
& =\frac{g_{p_1 p_3} g_{p_2 p_4}}{M^2}\left(\begin{array}{llll}
-5 & -3 & -5 & -3
\end{array}\right)\left(\begin{array}{l}
\mathcal{O}_{L^2 L^{\dagger 2} D^2, 1}^{(p)} \\
\mathcal{O}_{L^2 L^{\dagger 2} D^2, 2}^{(p)} \\
\mathcal{O}_{L^2 L^{\dagger 2} D^2, 3}^{(p)} \\
\mathcal{O}_{L^2 L^{\dagger} D^2, 4}^{(p)}
\end{array}\right) \,.
\end{aligned}
\end{equation}
\begin{flalign}
    &\textbf{The UV states}(Spin,SU(3)_c, S U(2)_w, U(1)_y)=(1,1,1,0): \mathcal{L}_{int}=g_{p_1 p_3} \delta_{i_3}^{i_1} \mathcal{B}_0^\mu(L_{p_1 i_1} \sigma_\mu L_{p_3}^{i_3})\,,&\nonumber
\end{flalign}
\begin{equation}
    \begin{aligned}
 \mathcal{L}_{\mathrm{EFT}}&=\frac{g_{p_1 p_3} g_{p_2 p_4}}{2 M^4} \delta_{i_3}^{i_1} \delta_{i_4}^{i_2}\left(L_{p_1 i_1} \sigma_\mu L_{p_3}^{\dagger i_3}\right) \square\left(L_{p_2 i_2} \sigma^\mu L_{p_4}^{\dagger i_4}\right) \\
& =\frac{g_{p_1 p_3} g_{p_2 p_4}}{M^4} \delta_{i_3}^{i_1} \delta_{i_4}^{i_2}\langle 12\rangle\langle 24\rangle[24][34] \\
& =-\frac{g_{p_1 p_3} g_{p_2 p_4}}{M^2}\left(\begin{array}{llll}
1 & -1 & 1 & -1
\end{array}\right)\left(\begin{array}{l}
\mathcal{O}_{L^2 L^{\dagger+2} D^2, 1}^{(p)} \\
\mathcal{O}_{L^2 L^{\dagger 2} D^2, 2}^{(p)} \\
\mathcal{O}_{L^2 L^{\dagger 2} D^2, 3}^{(p)} \\
\mathcal{O}_{L^2 L^{+2} D^2, 4}^{(p)}
\end{array}\right)\,.
\end{aligned}
\end{equation}




\bibliographystyle{JHEP}
\bibliography{ref2.bib}
\end{document}